\documentclass{article}

\usepackage{arxiv}
\pdfoutput=1

\usepackage[utf8]{inputenc} 
\usepackage[T1]{fontenc}    
\usepackage{hyperref}       
\usepackage{url}            
\usepackage{booktabs}       
\usepackage{amsfonts}       
\usepackage{nicefrac}       
\usepackage{microtype}      
\usepackage{doi}

\usepackage{float}
\usepackage{xcolor}

\usepackage{enumerate}

\usepackage{natbib} 

\usepackage{amsmath}
\usepackage{amssymb}
\usepackage{bm}
\usepackage{amsthm}
\newtheorem{theorem}{Theorem}
\newtheorem{corollary}{Corollary}
\theoremstyle{thmstyletwo} %
\newtheorem{proposition}{Proposition}%
\theoremstyle{thmstylethree}%

\usepackage[doublespacing]{setspace}
\usepackage[fontsize=12pt]{fontsize}

\newcommand{\ms}{\scriptscriptstyle} 

\usepackage{graphicx} 
\usepackage[font=scriptsize,labelfont=bf]{caption}
\usepackage{subcaption}

\usepackage{array}
\usepackage{booktabs}

\usepackage{cleveref}

\title{A response-adaptive multi-arm design for continuous endpoints based on a weighted information measure}
\date{}

\author{ \href{https://orcid.org/0000-0001-8765-0003}{\includegraphics[scale=0.06]{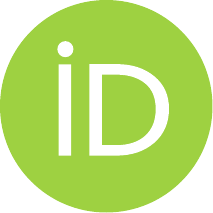}\hspace{1mm}Gianmarco~Caruso}\thanks{corresponding author} \\
	MRC Biostatistics Unit\\
	University of Cambridge\\
	Cambridge, UK CB2 0SR \\
	\texttt{gianmarco.caruso@mrc-bsu.cam.ac.uk} \\
	\And
	\href{https://orcid.org/0000-0001-6810-0284}{\includegraphics[scale=0.06]{orcid.pdf}\hspace{1mm}Pavel~Mozgunov} \\
	MRC Biostatistics Unit\\
	University of Cambridge\\
	Cambridge, UK CB2 0SR \\
	\texttt{pavel.mozgunov@mrc-bsu.cam.ac.uk} \\
}

\renewcommand{\headeright}{Pre-print version}
\renewcommand{\undertitle}{Pre-print version}
\renewcommand{\shorttitle}{}

\hypersetup{
pdftitle={A response-adaptive multi-arm design for continuous endpoints based on a weighted information measure},
pdfauthor={Gianmarco~Caruso, Pavel~Mozgunov}
}

\begin{document}
\maketitle

\def\spacingset#1{\renewcommand{\baselinestretch}%
{#1}\small\normalsize} \spacingset{1}
\spacingset{1.2} 

\begin{abstract}
Multi-arm trials are gaining interest in practice given the statistical and logistical advantages they can offer. The standard approach uses a fixed allocation ratio, but there is a call for making it adaptive and skewing the allocation of patients towards better-performing arms. However, it is well-known that these approaches might suffer from lower statistical power. We present a response-adaptive design for continuous endpoints which explicitly allows to control the trade-off between the number of patients allocated to the "optimal" arm and the statistical power. Such a balance is achieved through the calibration of a tuning parameter, and we explore robust procedures to select it. The proposed criterion is based on a context-dependent information measure which gives greater weight to treatment arms with characteristics close to a pre-specified clinical target. We establish conditions under which the procedure consistently selects the target arm and derive the corresponding limiting allocation ratios. We also introduce a simulation-based hypothesis testing procedure which focuses on selecting the target arm and discuss strategies to effectively control the type-I error rate. The practical implementation of the proposed criterion and its potential advantage over currently used alternatives are illustrated in the context of early Phase IIa proof-of-concept oncology trials.
\end{abstract}


\maketitle

\section{Introduction}
\label{sec:intro}


Clinical trials are medical research studies involving patients with the aim of identifying the best treatment for future recommendation. For many rare diseases, they may be the only chance for patients of receiving a new promising treatment promptly. However, the clinical trial dilemma often requires some participants to receive treatments that may be suboptimal in order to be able to estimate the treatment effect and draw reliable conclusions. In other terms, ethical responsibility in clinical trials is based on the balance of two components: the individual ethics, prioritizing the allocation of participants to better treatments, and the collective ethics, focusing on identifying the best treatment with high probability, ensuring that future patients receive the best possible care \citep{antognini2010compound}. Balanced allocation strategies are the most commonly used in practice, due their easy implementation and their even exploration of all the treatments which generally leads to high statistical power \citep{azriel2012optimal}. Although they comply quite well with the collective objective, their indiscriminate balanced allocation does not allow for a preventive stochastic drop of suboptimal treatments, by often neglecting the benefit of patients enrolled in the trial \citep{thall1989two}. In contrast, response-adaptive designs can skew the allocation of patients towards better-performing treatments, often without a significant sacrifice in terms of statistical power. In fact, rather than using allocation probabilities fixed throughout the trial, these designs sequentially allocates patients based on the responses observed in earlier participants \citep{rosenberger1993use,berry1995adaptive,robertson2023response}.


We consider the case of a multi-armed clinical trial comparing several treatments and where the primary endpoint is continuous. Such endpoints are common in clinical trials and have recently received increasing methodological attention \citep{aslanyan2026sample}. Nevertheless, much of the methodological literature focuses on binary endpoints and, in many real applications, continuous efficacy endpoints tend to be dichotomized. In early phases, common measurements in continuous scale are represented by biomarkers, which consist in measurable indicators of biological conditions and that are often used to evaluate the activity of treatments \citep{zhao2015biomarkers}. While in most applications one targets the highest or the lowest value of the endpoints, in some cases the focus may rather be on an optimal range or a specific target value for the selected biomarker. An example is represented by the levels of glucose in the blood measured through the HbA1c biomarker: values of HbA1c much above a given threshold indicate poor glucose control and risk of complications due to hyperglicemia, while values too low can indicate risk of hypoglycemia \citep{kaiafa2021hba1c}.  
In this context, we consider $K$-armed clinical trials with a clinical target $\gamma\in\mathbb{R}$ set by clinicians in advance, where arm $j^{1*}$ with mean response $\mu_{j^{1*}}$ closest to $\gamma$ is regarded as the true best arm, namely $j^{1*}:=\underset{j=1,\dots,K}{\text{argmin}}\,|\mu_j-\gamma|$. When the goal is to target the highest (or lowest) value of an endpoint, $\gamma$ can be set to the highest (lowest) value which is clinically attainable.


While it is common to focus on comparisons with the control and test hypotheses against it, this approach does not always align with the objectives of some multi-armed trials, such as selecting the best treatment arm. Rather than comparing each arm against the control, we aim at identifying the treatment arm which gives responses on average closer to the target $\gamma$. Under this assumption, competing treatment arms can be ranked from the best to the worst according to the proximity of their true mean responses to $\gamma$, that is $d_{j^{1*}}<d_{j^{2*}}<\dots <d_{j^{K*}}$, with $d_j=|\mu_j-\gamma|$ and $j=\{j^{1*},j^{2*},\dots,j^{K*}\}$, when assuming no ties. Here, arm $j^{l*}$ occupies the $l$-th position of the ranking. We propose a hypothesis testing procedure which compares the smallest and second smallest distances in the ranking, while taking into account the correct identification of these arms. We discuss strategies to control type-I error rate under this hypothesis testing procedure.


In this work, we introduce a class of response-adaptive designs for studies with continuous endpoints aiming at identifying the best treatment arm under patient-benefit constraints (e.g. increasing the number of patients assigned to it). Drawing from the theory of weighted information measures \citep{kelbert2015asymptotic}, we propose to use the information gain (i.e. the difference between the Shannon differential entropy and its weighted version) as a measure for the decision-making in a trial. The use of suitable parametric weight functions allows to comply with patient-benefit constraints by informing entropy measures about which outcomes are more desirable. In this work, we consider weight functions with a particular interest in arms whose mean responses are in the neighbourhood of $\gamma$. The idea of taking into account the ``context" of the experiment directly in the information measure has also been the basis of some recently proposed clinical trial designs with either binary \citep{kasianova2021response} or multinomial responses \citep{mozgunov2019information,mozgunov2020information}. In line with these, we define the information measures based on the posterior distribution of the parameter of interest. We introduce a tuning parameter in the weight function to control the ``exploration vs exploitation" trade-off \citep{azriel2011treatment}, and discuss strategies to select it with respect to the trial's objective. 


The current work can be framed within the broader class of response-adaptive designs (see a recent review by  \cite{robertson2023response}). Multi-Arm Bandit (MAB) approaches are a popular sub-class of response-adaptive designs which can be adopted in multi-armed trials to find a balance between exploration and exploitation. Several MAB designs have been recently proposed both for discrete \citep{villar2015multi,aziz2021multi} and continuous \citep{williamson2020response} responses, but they  focus on the highest (or lowest) response rather than on a specific target $\gamma$. 


This work is motivated by the Phase IIa oncology  trial of a novel inhibitor (called A, the name is masked) in combination with one or several approved immunotherapies. The hypothesis is that, due to its mechanism of action, this inhibitor can enhance the effect of an immunotherapy. However, the challenge is to select which immunotherapy treatment to use. To answer this, a multi-arm trial studying the three combination arms of the novel inhibitor with three approved immunotherapies (called B, C and D) and one arm of A+B+C was proposed. Due to the mechanism of action, the primary endpoint is a PD-marker (at Day 15) that was agreed to be an accurate proxy of the efficacy. A specific value of this marker is being targeted, as lower values would indicate insufficient efficacy, while higher values could lead to severe adverse effects. The aims of the trial are (i) to find the most promising combination among the four, (ii) check whether the difference between the most promising and the second promising arm is ``significant'' and (iii) collect the most information about the most promising arms. Accounting for all three aims, we propose to use the response-adaptive design presented in this work.



\section{Methodology}
\label{sec:methodology}

\subsection{Model assumptions}
\label{subsec:model}
Consider a clinical trial with a continuous endpoint observed for $K$ treatment arms. Each patient is randomized to a single arm, so outcomes are independent across patients. Within arm $j$, the $n_j$ responses $X_{1,\,j},\dots,X_{{n_j},\,j}$ are assumed to be an iid sample from $N(\mu_j,\,\sigma_j^2)$, with $\mu_j\in\mathbb R$ and known $\sigma_j^2\in\mathbb R^+$. We consider an objective Bayesian setting, with independent improper uniform priors over $\mathbb{R}$ on the arm means $\mu_j$, resulting in the following normal posterior density for $\mu_j$:
\begin{equation}
\label{eq:posteriorUniv}
    \pi_{n_j}(\mu_j)=\bigl(2\pi\sigma_j^2\bigr)^{-\frac{1}{2}}\,\exp\Biggl\{-\frac12\biggl(\frac{\bar x_{n_j}-\mu_j}{\sigma_{n_j}}\biggr)^2\Biggr\}\,,
\end{equation}
with $\bar x_{n_j}=\frac{1}{n_j}\sum_{i=1}^{n_j}x_{i,\,j}$ being the sample mean and $\sigma^2_{n_j}=\sigma_j^2/n_j$. 

As $n_j$ grows, $\sigma^2_{n_j}$ shrinks to zero and the posterior density in \eqref{eq:posteriorUniv} concentrates around the true mean $\mu_j\in\mathbb{R}$. By the Law of Large Numbers, $\bar x_{n_j}\rightarrow \mu_j$.
The amount of information needed to estimate $\mu_j$ is given by the Shannon differential entropy of $\pi_{n_j}(\mu_j)$, i.e. $h(\pi_{n_j})=-\int_{\mathbb R}\,\pi_{n_j}(\mu_j)\log\pi_{n_j}(\mu_j)\,d\mu_j=\frac12\log\bigl(2\pi e\,\sigma^2_{n_j}\bigr)$. This quantity only depends on $\sigma^2_{n_j}$, thus it ignores that the information about $\mu_j$ is not uniformly valuable across the entire parametric space of $\mu_j$. 

\subsection{Context-dependent information measure}
\label{subsec:IG}
Along with the amount of information needed to estimate a parameter, in experimental studies it is also relevant to consider the nature of the outcome itself and how desirable it is. Given a pre-specified target mean response $\gamma\in\mathbb{R}$, one can consider a context-dependent measure as the weighted Shannon entropy, i.e. $h^{\phi_\gamma}(\pi_{n_j})=-\int_{\mathbb R}\,\phi_\gamma(\mu_j)\pi_{n_j}(\mu_j)\log\pi_{n_j}(\mu_j)\,d\mu_j\,,$ where the weight function $\phi_\gamma(\mu_j)$ emphasizes the interest on a particular subset of the parametric space (e.g. the neighbourhood of $\gamma$). Following \cite{mozgunov2020information}, the information gain from considering the estimation problem in this sensitive area is given by
\begin{equation}
\label{eq:infoGainGeneral}
    \Delta(\pi_{n_j})=h(\pi_{n_j})-h^{\phi_\gamma}(\pi_{n_j})\,.
\end{equation}
This quantity is the amount of additional information needed to estimate $\mu_j$ when the degree of interest in $\mu_j$ is not uniform across the whole parametric space (i.e. some values of $\mu_j$ are more desirable than others). 
Theorem \ref{th:symmWeight} provides a closed-form expression for the information gain when a family of weight functions centered around the target $\gamma$ is considered, and insights on its asymptotic behaviour.

\begin{theorem}
\label{th:symmWeight}
Let $h(\pi_{n_j})$ and $h^{\phi_\gamma}(\pi_{n_j})$ be the standard and weighted Shannon entropy of the pdf in \eqref{eq:posteriorUniv} of the mean $\mu_j$ for arm $j$, and let $\gamma\in\mathbb R$ denote the clinical target for the continuous endpoint. Consider a family of weight functions of the form
\begin{equation}
\label{eq:phiFormula}
    \phi_\gamma(\mu_j)=C(\bar x_{n_j},\sigma^2_j,\gamma,n_j)\,\exp\biggl\{-\frac12\Bigl(\frac{\mu_j-\gamma}{\sigma_{\phi,\,j}}\Bigr)^2\biggr\}\,,
\end{equation}
\noindent where $\sigma_{\phi,\,j}^2=\sigma_j^p/n_j^\kappa$ is the variance term of the Gaussian kernel, with $\kappa$ and $p$ being real-valued tuning parameters controlling the influence of the sample size and the response variance, respectively, and $C(\bar x_{n_j},\sigma^2_j,\gamma,n_j)$ satisfies the normalization condition $\int_{\mathbb{R}}\,\phi_\gamma(\mu_j)\pi_{n_j}(\mu_j)\,d\mu_j=1$. Then, 
\begin{equation}
    \label{eq:infoGainSymm}
    \Delta_{n_j}=\frac12 \frac{\sigma_j^{2-p} n_j^{\kappa}}{\sigma_j^{2-p} n_j^{\kappa}+n_j}-\frac12\Biggl(\frac{\gamma-\bar x_{n_j}}{\sigma_j}\sqrt{n_j}\Biggr)^2\Biggl(\frac{\sigma_j^{2-p} n_j^{\kappa}}{\sigma_j^{2-p} n_j^{\kappa}+n_j}\Biggr)^2\,.
\end{equation}
And, if $\underset{n_j\rightarrow\infty}{\lim} \bar x_{n_j}=\mu_j\,(\neq \gamma)$, then: (i) $\Delta_{n_j}\sim-\frac12(\gamma- \mu_j)^2\sigma_j^{2(1-p)}n_j^{2\kappa-1}$ for $\kappa<1$; (ii) $\Delta_{n_j}\sim-\frac12(\gamma- \mu_j)^2\Bigl(\frac{\sigma_j^{1-p}}{\sigma_j^{2-p}+1}\Bigr)^2 n_j$ for $\kappa=1$; (iii) $\Delta_{n_j}\sim-\frac12\bigl(\frac{\gamma- \mu_j}{\sigma_j}\bigr)^2 n_j$ for $\kappa>1$.
\end{theorem}
All proofs are provided in Section A of the Supplementary Material (SM).
The weight function in \eqref{eq:phiFormula} has a natural statistical interpretation: it is proportional to a Gaussian density symmetric around its maximum $\gamma$ and therefore downweights regions far from $\gamma$ through a standardized quadratic distance. As a result, values of $\mu_j$ close to $\gamma$ contribute more heavily to the information criterion than those farther away. The variance term $\sigma_{\phi,\,j}^2=\sigma_j^p/n_j^\kappa$ controls the degree of this concentration around the clinically relevant target. The parameters $p$ and $\kappa$ determine the rate at which this concentration increases as $\sigma_j$ decreases and $n_j$ increases, respectively. 
In Section B of the SM, we discuss a weight function that is asymmetric around $\gamma$, allowing for a different weighting of values above and below $\gamma$.

Notably, as $n_j\rightarrow\infty$, the asymptotical behaviour of the information gain in \eqref{eq:infoGainSymm} varies with $\kappa$. For example, this quantity converges to $0$ when $\kappa<0.5$, meaning that the context of the study tends to be neglected as more information is collected on a specific arm. For this reason, we will focus on $\kappa\geq0.5$. When $\kappa>0.5$, the information gain diverges to $-\infty$, while, when $\kappa=0.5$, the information gain converges to a finite quantity which, for $p=1$, depends on the squared distance between $\mu_j$ and $\gamma$ and, for $p=2$, on a standardized (Mahalanobis-type) version of this distance.

\begin{figure}[ht]
    \centering
    \includegraphics[width=0.95\textwidth]{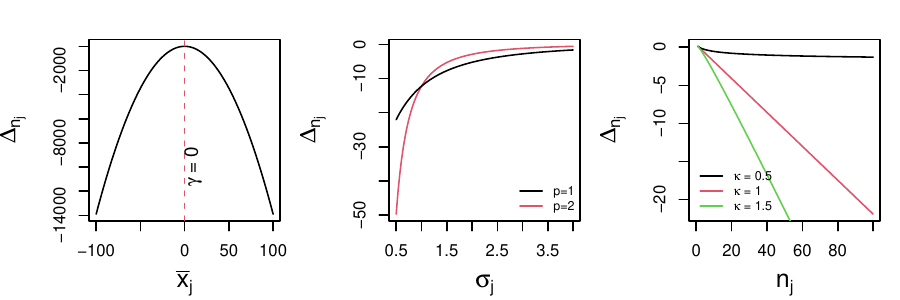}
    \caption{Information gain $\Delta_{n_j}$ in \eqref{eq:infoGainSymm} as function of (a) sample mean $\bar x_{n_j}$ (\textit{left}), (b) standard deviation $\sigma_j^p$ (\textit{center}) and (c) sample size $n_j^\kappa$ of the $j$-th arm (\textit{right}).}
    \label{fig:infoGain}
\end{figure}

The quantity \eqref{eq:infoGainSymm} achieves its maximum at $\bar x_{n_j}=\gamma$ (Figure \ref{fig:infoGain}a). 
Asymptotically, this means that collecting more information about the arm with mean $\mu_j$ close to $\gamma$ implies a maximization of the information gain $\Delta_{n_j}$. This property guarantees that the more valuable the learning about $\mu_j$, the higher the information gain associated to arm $j$. In addition, when $p\geq1$, $\Delta_{n_j}$ is monotonically increasing with $\sigma_j$ (Figure \ref{fig:infoGain}b), reflecting the idea that higher uncertainty about arm $j$ increases the learning value of sampling from it. When $\kappa\geq0.5$, as more observations on arm $j$ are collected, the value of the information gain decreases, with higher values of $\kappa$ leading to a steeper decline of this function (Figure \ref{fig:infoGain}c). For all these desirable properties, we suggest the use of the information gain in \eqref{eq:infoGainSymm} to govern allocation in a trial.

\subsection{Allocation rule}
\label{subsec:simWeight}

Let $N_j$ be the total number of patients allocated to arm $j$ at the end of the trial and $N=\sum_{j=1}^K\,N_j$ the total number of patients treated at the end of the trial. The scheme to allocate the $N$ patients to the $K$ arms works as follows. 

After a burn-in phase where an equal batch of $B\geq1$ patients is allocated to each of the $K$ arms, the following rule is adopted for the $t$-th patient ($t=KB+1,\dots,N$): 
\begin{equation}
    \label{eq:allocRule}
    \texttt{assign patient }t\texttt{ to arm }j^{(t)}\texttt{ such that }j^{(t)}=\underset{j=1,\dots,K}{\text{argmax}}\,\Delta_{n_j}\,.
\end{equation} 
The algorithm proceeds until a total of $N$ patients is treated. Finally, the arm $\widehat j^{1*}$ whose final sample mean $\bar x_{N_j}$ is closest to the target is selected as final recommendation; namely, $\widehat j^{1*}=\underset{j=1,\dots,K}{\text{argmin}}\,|\bar x_{N_j}-\gamma|$. The burn-in phase is necessary to have a first estimate of the sample means of the $K$ arms, even more importantly when the sequential design is based on non-informative prior distributions. 

In what follows, we refer to this class of designs based on the weighted-entropy measure as WE($p$,$\kappa$). This allocation rule may be seen as a deterministic response-adaptive criterion, where the next patient is almost surely allocated to the arm associated with the highest information gain. The ``value" of allocating the next patient to arm $j$ balances the trade-off between reducing uncertainty about arm $j$ and the individual benefit of receiving a treatment whose current average response is at distance $|\bar x_{n_j}-\gamma|$ from the target. Thus, by maximizing $\Delta_{n_j}$, the rule prioritizes learning in the clinically relevant region of the parameter space (i.e., near $\gamma$). 

Notably, smaller values of $p$ tend to mitigate the impact that different values of the standard deviations have on the information gain, thus $p$ can be regarded as a parameter controlling the impact of the response variability on this information measure. 
Moreover, a too greedy algorithm may lead to an insufficient exploration of the alternative arms, resulting in poor performance in terms of statistical power \citep{villar2015multi}. 
For this reason, a key role in the ``exploration vs exploitation" trade-off is played by the parameter $\kappa$ which penalizes an increasing number of observations on the same arm, offering more chances to other arms to be explored. A larger $\kappa$ corresponds to a higher penalization of arms with many patients, favouring a more spread allocation. In Section \ref{sec:kappaSelection}, we show some strategies to select $\kappa$ in accordance with the trial objectives, while, in Section \ref{sec:simStudyUniv}, we offer further insights on how this parameter affects the operating characteristics of the proposed design. 

\subsection{Consistent selection of the ``best" arm}
\label{subsec:consistency}

The consistent selection of the ``best" arm is a desirable property to ensure that the allocation criterion presented in Section \ref{subsec:simWeight} provides a reliable selection of the best arm as the sample size increases. Theorem \ref{th:consistency} establishes conditions under which each arm continues to be sampled indefinitely and the estimated ranking of arms (from closest to $\gamma$ to furthest) converges to the true ranking.

\begin{theorem}
\label{th:consistency}
Consider a $K$-arm trial that is run under the allocation procedure WE$(p,\kappa)$. Let $n_j(t)$ denote the number of observations collected from arm $j$ after $t$ total allocations and let $\bar x_{n_j(t)}$ denote the corresponding sample mean. If $\kappa>0.5$, then each arm is sampled infinitely often, that is $n_j(t)\rightarrow \infty$ as $t\rightarrow\infty$.
Moreover, assume that the true distances from the target, $d_j=|\mu_j-\gamma|$ ($j=1,\dots,K$), are all distinct. Then they admit a unique ordering, which we denote by the ordered vector $(d_{j^{1*}},d_{j^{2*}},\dots,d_{j^{K*}})$ 
where $j^{1*}$ is the arm closest to the target and $j^{K*}$ the one furthest from the target. Similarly, let the estimated distances at time $t$ be $\widehat d_{j(t)}=|\bar x_{n_j(t)}-\gamma|$, and let $(\widehat d_{\widehat j^{1*}(t)},\widehat d_{\widehat j^{2*}(t)},\dots,\widehat d_{\widehat j^{K*}(t)})$ be their order statistics. Suppose the trial ends at time $t=N$. If $\kappa>0.5$, as $N\rightarrow\infty$, the estimated ranking is consistent, that is $P\Bigl((\widehat d_{\widehat j^{1*}(N)},\widehat d_{\widehat j^{2*}(N)},\dots,\widehat d_{\widehat j^{K*}(N)})=(d_{j^{1*}},d_{j^{2*}},\dots,d_{j^{K*}})\Bigr)\rightarrow1\,.$
\end{theorem} 

Theorem \ref{th:consistency} implies that, if $\kappa>0.5$, the posterior mean is a consistent estimator for $\mu_j$ and the selection of the best arm is consistent as the trial size grows. 
    \begin{corollary}
    \label{th:concentr_ineq}
        Consider the WE($p$,$\kappa$) allocation rule under the assumptions of Theorem \ref{th:consistency}, and assume that the true distances $d_j=|\mu_j-\gamma|$, $j=1,\dots,K$, are all distinct, admitting the unique ranking $d_{j^{1*}}<d_{j^{2*}}<\dots<d_{j^{K*}}$. Let $\delta_{\min}:=\min_{i\neq j}|d_i-d_j|>0$ be the smallest gap between any two true distances. Then, for the estimated distances at time $N=\sum_{j=1}^K N_j$, $$P\Bigl((\widehat d_{\widehat j^{1*}(N)},\widehat d_{\widehat j^{2*}(N)},\dots,\widehat d_{\widehat j^{K*}(N)})\neq(d_{j^{1*}},d_{j^{2*}},\dots,d_{j^{K*}})\Bigr)\leq2\sum_{j=1}^K\,\exp\biggl\{-\frac{N_j \delta_{\min}^2}{8\sigma^2_j}\biggr\}\,.$$
    \end{corollary}
Corollary \ref{th:concentr_ineq} provides an upper bound on the probability of incorrectly estimating the arms’ ranking. This bound decays exponentially with the arm-specific sample sizes, with a rate governed by the minimal separation $\delta_{\min}$, which reflects the intrinsic difficulty of the ranking problem. In particular, when some arms are close, the bound becomes less informative, as larger samples are required to reliably distinguish between them. The bound also depends on the variability of the responses, with smaller variances leading to tighter control of the error probability.

Theorem \ref{th:consistency} relies on the assumption of a unique treatment ranking. In the following proposition, we relax this assumption by considering the case where multiple arms may be equally optimal.
\begin{proposition}
\label{th:armselection_ties}
    Consider the WE($p$,$\kappa$) allocation rule under the assumptions of Theorem 2, except that the optimal arm is not necessarily unique. Let $\mathcal{J}^*=\Bigl\{j\in\{1,\dots,K\}:|\mu_j-\gamma|=\min_l|\mu_l-\gamma|\Bigr\}$ denote the set of arms attaining the minimum distance from $\gamma$. Let $\hat j^{1*}$ be the index of the treatment selected at the end of a trial of size $N$. Then, when $\kappa>0.5$, $P(\hat j^{1*}\in\mathcal{J}^*)\rightarrow1$ as $N\rightarrow\infty$; moreover, if the set of sub-optimal arms $\mathcal{J}^-=\{1,\dots,K\}\smallsetminus\mathcal{J}^*$ is non-empty, then $P(\max_{j\in\mathcal{J}^*}\widehat d_j(N)<\min_{i\in\mathcal{J}^-}\widehat d_i(N))\rightarrow1$ as $N\rightarrow\infty$.
\end{proposition} 
Proposition \ref{th:armselection_ties} shows that, when multiple arms attain the same minimal distance from $\gamma$, the procedure is able to asymptotically select one of these optimal treatments with probability tending to one. Moreover, the optimal treatments are asymptotically ranked ahead of all sub-optimal ones (if any), although no unique ordering can be established within the set of optimal treatments.

\subsection{Asymptotic properties of the allocation proportions}
\label{subsec:allocProp}

To evaluate the statistical performance of the WE($p$,$\kappa$) design, it is crucial to understand how patients are allocated across arm. Theorem 3 illustrates the asymptotic behaviour of the allocation ratios, while its corollary provides a sufficient condition under which  the largest proportion of patients is assigned to the best arm.

\begin{theorem}
\label{th:allocProp}
Let $\widehat\rho_j(t)=\frac{n_j(t)}{t}$ denote the allocation ratio of arm $j$ at time $t$, such that $\sum_{j=1}^K\,n_j(t)=t$, and let $\tilde \rho_j=\lim_{t\rightarrow\infty}\widehat\rho_j(t)$ denote its asymptotic version, such that $\sum_{j=1}^K\tilde\rho_j=1$. Then, under the allocation procedure WE$(p,\kappa)$, we have that 
        $$\tilde\rho_j\propto\begin{cases}
            \Bigl[(\gamma-\mu_j)^2\,\sigma_j^{2(1-p)}\Bigr]^{-\frac{1}{2\kappa-1}} & 0.5<\kappa<1 \vspace{0.1cm}\\ 
            \biggl[(\gamma-\mu_j)^2\,\frac{\sigma_j^{1-p}}{1+\sigma_j^{2-p}}\biggr]^{-1} & \kappa=1 \vspace{0.1cm}\\
            \Bigl(\frac{\gamma-\mu_j}{\sigma_j}\Bigr)^{-2} & \kappa>1\\
        \end{cases}\quad.$$
\end{theorem}

Theorem \ref{th:allocProp} shows that, for $\kappa>0.5$, the asymptotic allocation proportion depends on the squared distance between the true mean and the target $\gamma$, on the variability of the response and on the parameters $p$ and $\kappa$.
Notably, for $\kappa>1$, the asymptotic allocations are proportional to the inverse squared standardized distances from the target, independently of $p$ and $\kappa$.

\begin{figure}[ht]
            \centering
            \includegraphics[width=1\linewidth]{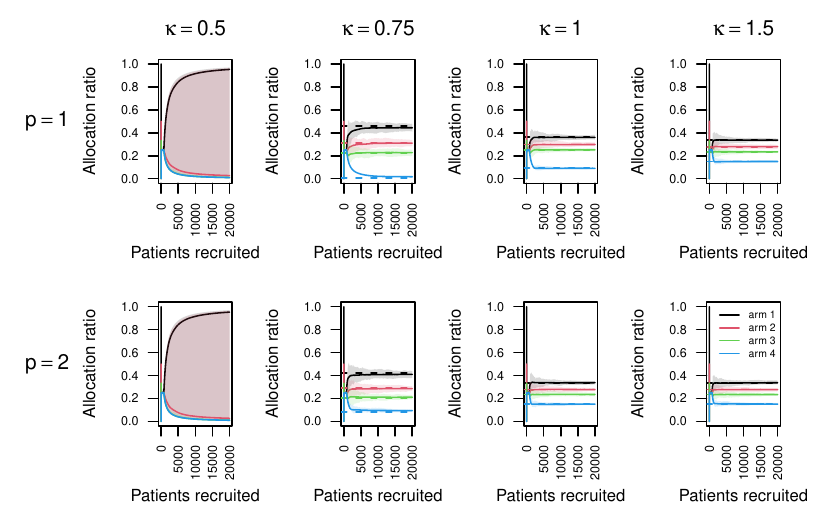}
            \caption{WE($p$,$\kappa$) design's allocation ratios for increasing sample sizes, for four arms and different values of $p$ and $\kappa$: asymptotic allocation ratios (dashed lines); mean allocation ratios across simulations (solid lines); variation observed across replicates (light-shaded areas) under scenario $\bm{\mu}=(1,1.1,1.2,3)$, $\bm{\sigma}=(1,1,1,2)$, $\gamma=0$.}
            \label{fig:allocRatios_WE}
        \end{figure}

Figure \ref{fig:allocRatios_WE} illustrates the allocation ratios as a function of the sample size, for several combinations of $p$ and $\kappa$. 
The solid lines represent the mean allocation ratios across $100$ trial replicates, while the light-shaded areas show the range of variation observed across those. For $\kappa>0.5$, the arm-specific allocation ratios converge to their limiting proportions (dashed lines). In contrast, when $\kappa=0.5$, one arm eventually dominates the others (i.e., its allocation ratio tends to 1), which is consistent with the lack of guarantee that all arms are sampled infinitely often. The choice of $\kappa\in(0.5,1)$ induces a more greedy behaviour of the allocation rule: arms that appear favourable early tends to be over-sampled, and allocation ratios converge much slower to their asymptotic values. Increasing $\kappa$ leads to less imbalanced early allocations,  mitigating the tendency of one arm to dominate the others. This results in faster convergence of allocation ratios when $\kappa\geq1$. The allocation behaviour for competing response-adaptive procedures is provided in SM (Section F).
\begin{corollary}
\label{th:patientGainCondition}
Let $j^{1*}:=\underset{j=1,\dots,K}{\text{argmin}}\,|\mu_j-\gamma|$ denote the best arm in a $K$-armed trial and let $\tilde\rho_j$ denote the asymptotic allocation ratio of arm $j$. Then, under the allocation procedure described in Section \ref{subsec:simWeight}, the choice of $p=1$ and $\kappa\in(0.5,1)$ is sufficient to ensure that $\tilde\rho_{j^{1*}}\geq\tilde\rho_j\,,\forall j\neq j^{1*}$.
\end{corollary}

The proof of Corollary \ref{th:patientGainCondition} follows directly from substituting $p=1$ in the asymptotic allocation ratio of Theorem \ref{th:allocProp}, when $\kappa\in(0.5,1)$. This result shows that, for $p=1$ and $\kappa\in(0.5,1)$, the best arm $j^{1*}$ asymptotically receives the largest share of patients, enhancing the patient-benefit justification for using WE($p$,$\kappa$). However, this limiting share $\tilde\rho_{j^{1*}}\propto[(\gamma-\mu_{j^{1*}})^2]^{-1/(2\kappa-1)}$ decreases as $\kappa$ increases, reflecting a reduced exploitation of the best arm and a greater exploration of the other arms.

\subsection{Multivariate generalization}
\label{subsec:multiWeight}

Consider now the case where $q>1$ endpoints are measured and a vector $\bm \gamma$ of target mean values is considered. Consider $n_j$ responses from arm $j$ as an iid sequence of realizations of $q\times1$ random vectors $\bm X_{i_1,\,j},\dots,\bm X_{i_{n_j},\,j}\overset{iid}{\sim}MVN(\bm \mu_j,\,\bm \Sigma_j)$, with mean vector $\bm \mu_j\in\mathbb R^q$ and known covariance matrix $\bm \Sigma_j$. Under these assumptions, an improper prior uniform on $\mathbb R^q$ for $\bm \mu_j$ yields the normal posterior density given by 
\begin{equation}
\label{eq:posteriorMulti}
    \pi_{n_j}(\bm\mu_j)=(2\pi)^{-\frac{q}{2}}\,\det(\bm\Sigma_{n_j})^{-\frac12}\,\exp\biggl\{-\frac12\Bigl(\bm{\bar x}_{n_j}-\bm\mu_j\Bigr)'\bm\Omega_{n_j}\Bigl(\bm{\bar x}_{n_j}-\bm\mu_j\Bigr)\biggr\}\,,
\end{equation}
where $\bar{\bm x}_{n_j}$ is the vector of sample means for the $j$-th arm, while $\bm\Sigma_{n_j}=\bm\Sigma_j/n_j$ and $\bm\Omega_{n_j}=\bm\Sigma^{-1}_{n_j}$ are  the posterior covariance and precision matrices.

\begin{theorem}
\label{th:multiWeight}
Let $h(\pi_{n_j})$ and $h^{\phi_\gamma}(\pi_{n_j})$ be the standard and weighted Shannon entropy of the density in \eqref{eq:posteriorMulti}. Assume a family of weight functions of the form 
\begin{equation}
\label{eq:multiWeight}
    \phi_\gamma(\bm \mu_j)=C(\bm{\bar x}_{n_j},\bm\Sigma_j,\bm\gamma,n_j)\,\exp\biggl\{-\frac12\Bigl(\bm\mu_j-\bm\gamma\Bigr)'\bm\Omega_{\phi,\,j}\Bigl(\bm\mu_j-\bm\gamma\Bigr)\biggr\}\,,
\end{equation}
where $\bm\Omega_{\phi,\,j}=\bm\Sigma_{\phi,\,j}^{-1}=n_j^\kappa\,\bm\Sigma_j^{-1}$ is the precision matrix term of the weight function and $C(\bm{\bar x}_{n_j},\bm\Sigma_j,\bm\gamma,n_j)$ satisfying $\int_{\mathbb{R}^q}\,\phi_\gamma(\bm\mu_j)\pi_{n_j}(\bm\mu_j)\,d\bm\mu_j=1$.
Then, 
\begin{equation}
\label{eq:infoGainMulti}
    \Delta_{n_j}=\frac{q}{2}\frac{n_j^\kappa}{n_j^\kappa+n_j}-\frac12\bigl(\bm\gamma-\bar{\bm x}_{n_j}\bigr)'\bm\Sigma_j^{-1}\bigl(\bm\gamma-\bar{\bm x}_{n_j}\bigr)'\Biggl(\frac{n_j^{\kappa+\frac12}}{n_j^\kappa+n_j}\Biggr)^2\,.
\end{equation}
\end{theorem}


Theorem \ref{th:multiWeight} provides a multivariate extension of Theorem 1. The bivariate  case ($q=2$) is presented and discussed in Section C.1 of the SM. The information gain in \eqref{eq:infoGainMulti} attains higher values when $\bar{\bm x}_{n_j}$ is close to $\bm \gamma$, is monotonically increasing with the variances (see section C.2 of the SM) and its asymptotical behaviour is similar to the one of the information gain in \eqref{eq:infoGainSymm}.

We adopt the function \eqref{eq:infoGainMulti} as patients' allocation criterion. As for the univariate case, the next patient of the sequential scheme is assigned to the arm with the highest information gain, while the arm selection is still based on the proximity between the final estimates of the means and the respective target mean values, i.e. the estimated best arm is the one whose final estimated mean vector is closest to the target vector. 

The definition of ``best" arm in the multivariate case may highly depend on the adopted distance, and the use of a Manhattan distance may lead to the endpoint with higher magnitude prevailing over the others. For this reason, we adopt the standardized distance between the target vector and the true mean vector, i.e.
\begin{equation}
\label{eq:multivDistance}
\widetilde d(\bm\mu_j,\bm \gamma)=\sum_{l=1}^q\bigl|\mu_j^{\ms (l)}-\gamma^{\ms (l)}\bigr|/\sqrt{\sigma_j^{\ms (ll)}}\,.
\end{equation} 
Therefore, the final recommendation is $\widehat j^{1*}:=\underset{j=1,\dots,K}{\text{argmin}}\,\widetilde d(\bar{\bm{x}}_{N_j},\bm \gamma)$, with $\bar{\bm{x}}_{N_j}$ being the vector of the sample means for arm $j$ at the end of the trial.

\section{Simulation-based hypothesis testing procedure}
\label{sec:test}

\subsection{Hypothesis testing}
\label{subsec:hptest}

We assume a generic context where a $K$-armed trial is considered and no control arm is adopted for comparison. Thus, $\binom{K}{2}$ pairwise comparisons may be necessary to test the superiority of an arm over the others, and this can be a computationally demanding task. Here, following the motivating trial and to be aligned with the final recommendation strategy, we propose a testing procedure which only compares the two best performing arms, by also taking into account their correct identification.

Let $j^{1*}:=\underset{j=1,\dots,K}{\text{argmin}}\,|\mu_j-\gamma|$ be the best treatment arm and $j^{2*}:=\underset{j=1,\dots,K,\,j\neq j^{1*}}{\text{argmin}}\,|\mu_j-\gamma|$ be the second best. In case of tie between two or more treatment arms, the best (or second best) arm is randomly determined among them. If $\bar x_{N_j}$ is the sample mean of arm $j$ at the end of the trial, the estimated best and second best arm are 
$\widehat{j}^*:=\underset{j=1,\dots,K}{\text{argmin}}\,|\bar x_{N_j}-\gamma|$ and $\widehat{j}^{**}:=\underset{j=1,\dots,K,\,j\neq j^{1*}}{\text{argmin}}\,|\bar x_{N_j}-\gamma|$. Denoting $\mu_{j^{1*}}$ and $\mu_{j^{2*}}$ the true mean responses of arm $j^{1*}$ and arm $j^{2*}$, respectively, we propose a test procedure which assesses whether $\mu_{j^{1*}}$ is sufficiently much closer to the target $\gamma$ than $\mu_{j^{2*}}$. Formally, we consider the set of hypotheses $H_0:\,|\mu_{j^{1*}}-\gamma|=|\mu_{j^{2*}}-\gamma|$ and $H_1:\,|\mu_{j^{1*}}-\gamma|<|\mu_{j^{2*}}-\gamma|\,$. 
Under a Bayesian framework, $H_0$ is rejected at a significance level $\alpha$ if $\widehat\pi_{\widehat j^{1*},\widehat j^{2*}}$ exceeds a fixed cut-off probability $\eta_\alpha$, where 
\begin{equation}
\label{eq:BayesTest}
\widehat\pi_{\widehat j^{1*},\widehat j^{2*}}:=\mathbb{P}\bigl(|\mu_{\widehat j^{1*}}-\gamma| < |\mu_{\widehat j^{2*}}-\gamma|\,\bigl|\text{data}\bigr)=\mathbb{E}\Bigl[\mathbb{I} \bigl\{ |\mu_{\widehat j^{1*}}-\gamma| < |\mu_{\widehat j^{2*}}-\gamma| \bigr\} \,\Bigl|\, \text{data} \Bigr]
\end{equation} 
is the posterior probability that the estimated best arm is closer to $\gamma$ than the second best. Notably, the indices $\widehat j^{1*}$ and $\widehat j^{2*}$ are fully determined by the observed data - including both the patients' responses and the realized allocation sequence - and are therefore treated as fixed in \eqref{eq:BayesTest}. This is consistent with the Bayesian principle that inference depends on the current study only through the data actually observed \citep{berry2010bayesian}. Conditional on these data, the joint posterior of $(\mu_{\widehat j^{1*}},\,\mu_{\widehat j^{2*}})$ is bivariate normal with independent marginal components given by \eqref{eq:posteriorUniv}. Further details on the joint posterior of $(\mu_{\widehat j^{1*}},\,\mu_{\widehat j^{2*}})$ are available in Section D.1 of the SM. \\
Once a trial is concluded, the posterior probability $\widehat\pi_{\widehat j^{1*},\widehat j^{2*}}$ is estimated as follows.

\medskip

\noindent\texttt{Monte Carlo procedure A:
\vspace{-0.15cm}
\begin{enumerate}[1)]
        \item estimate the best and second best arm, namely $\widehat{j}^{*}$ and $\widehat{j}^{**}$;
        \vspace{-0.2cm}
        \item choose a number of Monte Carlo replicates $L$. For each $l=1,\dots,L$, 
        \begin{enumerate}[i.]
            \item draw a random vector $(\mu_{\widehat j^{1*}}^{(l)},\mu_{\widehat j^{2*}}^{(l)})$ from $p(\mu_{\widehat j^{1*}},\,\mu_{\widehat j^{2*}}\mid \text{data})$;
            \vspace{-0.2cm}
            \item compute $|\mu^{(l)}_{\widehat j^{1*}}-\gamma|$ and $|\mu^{(l)}_{\widehat j^{2*}}-\gamma|$; 
        \end{enumerate}
        \vspace{-0.2cm}
        \item approximate \eqref{eq:BayesTest} with $\frac1L\sum_{l=1}^L\,\mathbb{I}\biggl\{|\mu^{(l)}_{\widehat j^{1*}}-\gamma|<|\mu^{(l)}_{\widehat j^{2*}}-\gamma|\biggr\}\,.$
    \end{enumerate} 
    }

\subsection{Frequentist properties of $\widehat\pi_{\widehat j^{1*},\widehat j^{2*}}$}

Although conditioning on the realized data is standard in Bayesian trials, it does not generally ensure control of the type-I error rate, so the frequentist properties of the resulting Bayesian test must be assessed separately \citep{spiegelhalter1994bayesian}. The frequentist rejection probability under $H_k$, $k=\{0,1\}$, i.e. $\mathbb{P}(\widehat\pi_{\widehat j^{1*},\widehat j^{2*}}>\eta_\alpha|\,H_k)=\mathbb{E}[\mathbb{I}\{\widehat\pi_{\widehat j^{1*},\widehat j^{2*}}>\eta_\alpha\}|\,H_k]$, can be estimated as follows.

\bigskip

\noindent\texttt{Monte Carlo procedure B:
 \vspace{-0.15cm}
    \begin{enumerate}[1)]
        \item choose a number of Monte Carlo replicates $M$;
         \vspace{-0.2cm}
        \item for each replica $m=1,\dots,M$, 
         \vspace{-0.2cm}
        \begin{enumerate}[i.]
            \item simulate a trial dataset (\text{data}$^{(m)}$) under $H_k$ and the chosen design;
            \item based on \text{data}$^{(m)}$, estimate $\widehat{j}^{*(m)}$ and $\widehat{j}^{**(m)}$;
            \item compute $\widehat\pi_{\widehat j^{1*(m)},\widehat j^{2*(m)}}^{(m)}$ using \texttt{Monte Carlo procedure A};
        \end{enumerate}
         \vspace{-0.2cm}
        \item approximate $\mathbb{P}(\widehat\pi_{\widehat j^{1*},\widehat j^{2*}}>\eta_\alpha|\,H_k)$ with $\frac1M\sum_{m=1}^M\mathbb{I}\bigl\{\widehat\pi_{\widehat j^{1*(m)},\widehat j^{2*(m)}}^{(m)}>\eta_\alpha\bigr\}\,.$
    \end{enumerate}
    }

Comparing distances between the means of the two estimated best arms is not sufficient to make a claim on the true ones, as one needs to guarantee that the ground truth is correctly identified. For this reason, we propose two definitions of power incorporating their correct identification, under the assumption that 
$\mathbb{P}\bigl(\bigl\{\widehat j^{1*}=j^{1*},\,\widehat j^{2*}=j^{2*}\bigr\}\bigl|H_0\bigr)=1$. The first proposal is to consider a conditional power, 
\begin{equation}
\label{eq:condPower}
1-\beta_{C}:=\mathbb{P}\bigl(\bigl\{\widehat\pi_{\widehat j^{1*},\widehat j^{2*}}>\eta_\alpha\bigr\}\bigl|\bigl\{\widehat j^{1*}=j^{1*},\,\widehat j^{2*}=j^{2*}\bigr\},\,H_1\bigr)\,,
\end{equation} 
where rejections of the null are counted in \eqref{eq:condPower} only for trial realizations where the two best arms are correctly identified. Alternatively, we propose a two-components power as a comprehensive measure of the ``selection-and-testing" procedure:
\begin{equation}
\label{eq:twoCompPower}
1-\beta_{TC}:=\mathbb{P}\bigl(\bigl\{\widehat\pi_{\widehat j^{1*},\widehat j^{2*}}>\eta_\alpha\bigr\}\,\cap\,\bigl\{\widehat j^{1*}=j^{1*},\,\widehat j^{2*}=j^{2*}\bigr\}\bigl|H_1\bigr)\,,
\end{equation} 
which jointly accounts for correct identification of the two best arms and the testing component. Notice that $1-\beta_{TC}\leq 1-\beta_{C}$. 

Theorem \ref{th:power} shows that, for $\kappa>0.5$ and under $H_1$, the hypothesis testing procedure based on $\widehat\pi_{\widehat j^{1*},\widehat j^{2*}}$ is consistent as the trial size increases.
\begin{theorem}
\label{th:power}
    Under $H_1:|\mu_{j^{1*}}-\gamma|<|\mu_{j^{2*}}-\gamma|$ and the model assumptions of Section \ref{subsec:model}, for the WE$(p,\kappa)$ design with $\kappa>0.5$ the posterior probability $\widehat\pi_{\widehat j^{1*},\widehat j^{2*}}\rightarrow1$ almost surely as $N\rightarrow\infty$. In addition, for a fixed cut-off probability $\eta_\alpha\in(0,1)$, both $1-\beta_C$ and $1-\beta_{TC}$ tend to 1 almost surely as $N\rightarrow\infty$.
\end{theorem}
Section D.2 of the SM provides an empirical illustration of the result established by Theorem \ref{th:power}. Section \ref{subsec:kappaWithTest} illustrates how the tuning parameter $\kappa$ affects the posterior probability $\widehat\pi_{\widehat j^{1*},\widehat j^{2*}}$ under different finite-sample scenarios.

\subsection{Controlling type-I error rate: mean and strong control}
\label{subsec:typeIcontrol}

Let us consider a wide set $\mathcal{S}_0=\{H_0^{(1)},\,\dots,H_0^{(S_0)}\}$ of plausible null scenarios, where $H_0^{(s)}:=(\mu_1^{(s)},\dots,\mu_K^{(s)})$ is a specific configuration of the true mean vector satisfying the null hypothesis $H_0$ defined in Section \ref{subsec:hptest}. Given a specific design and a maximum tolerated type-I error rate $\alpha$, we define $\eta_\alpha^{(s)}$ as the individual cut-off value calibrated at level $\alpha$ under the null scenario $H_0^{(s)}$ ($s=1,\dots,S_0)$, namely a threshold value $\eta_\alpha^{(s)}$ such that $\mathbb{P}\Bigl(\widehat\pi_{\widehat j^{1*},\widehat j^{2*}}>\eta_\alpha^{(s)}\Bigl|\,H_0^{(s)}\Bigr)=\alpha$. For a fixed $\eta_\alpha^{(s)}$, this probability can be efficiently evaluated using \texttt{Monte Carlo procedure B}. 

To guarantee a strong control of the type-I error rate, one can fix the threshold $\eta_\alpha$ to the maximum of the individual cut-off probabilities, i.e. $\eta_\alpha^\text{max}=\underset{s=1,\dots,S_0}{\max}\,\eta_\alpha^{(s)}$. This is a conservative strategy that ensures that the probability of incorrectly detecting the presence of a superior arm under the null does not exceed $\alpha$ in each of the scenarios in $\mathcal{S}_0$. Ideally, $\mathcal{S}_0$ should contain a wide variety of scenarios in order to control as many situations as possible. In practice, there may be scenarios which are thought to be highly unlikely and can be ruled out. As rule of thumb, we suggest to consider null scenarios with $\mu_j$ not being distant from $\gamma$ more than $10$ times the largest standard deviation. For normal responses it is known that the response $X_j$ will be in the interval $\mu_j\pm 3\sigma_j$ with probability higher than $0.97$. Thus, a distance $|\mu_j-\gamma|$ much higher than $3\sigma_j$ would imply that values in the neighbourhood of $\gamma$ are very unlikely for arm $j$. Furthermore, in some contexts, null scenarios with distances larger than $10\cdot\max_j\sigma_j$ might signify the ineffectiveness of all the treatments and, therefore, having a control on any of them may be out of the trial scope.

In some cases, the strong control of the type-I error rate may be a too stringent requirement which leads to a loss of power. This is even more evident when there are only few scenarios requiring a substantially higher cut-off value than the rest. Similarly to \cite{daniells2024add}, we propose to relax the required type of control, trying to reach an average control of the type-I error rate rather than a strong one. Then the resulting threshold value is $\eta_\alpha^\text{ave}$ such that $\frac{1}{S_0}\sum_{s=1}^{S_0}\,\alpha^{(s)}=\alpha$, where $\alpha^{(s)}$ is the individual type-I error rate associated with scenario $H_0^{(s)}$. In this case, $\alpha$ can be regarded as the average type-I error rate for the set of scenarios in $\mathcal{S}_0$. 

\section{Robust strategies for selecting the tuning parameter $\kappa$}
\label{sec:kappaSelection}

\subsection{Strategy}
\label{subsec:algoKappa}

The choice of $\kappa$ is crucial to control the trade-off between exploration and exploitation. In fact, $\kappa$ has an impact on both the patients' allocation ratios and the distribution of $\widehat\pi_{\widehat j^{1*},\widehat j^{2*}}$ used for the test (see Section \ref{subsec:allocProp} and \ref{subsec:kappaWithTest}, respectively). 

We consider as ``optimal", a value of $\kappa$ which maximizes a specific operating characteristic of interest (e.g. statistical power, percentage of times the best arm is selected in repeated trials). The optimal value of $\kappa$ depends on several parameters: true means, standard deviations, target means, trial size, number of arms, the burn-in size, and the parameter $p$. Apart from the vectors of the true means and true standard deviations, we suppose that all the other parameters are fixed in advance. 

Following \cite{mozgunov2020information}, we discuss a general approach to select the robust optimal value of $\kappa$ which does not require the prior knowledge on the mean responses and the standard deviations. The idea is to consider a large variety of plausible alternative scenarios, each one characterized by  mean responses (and  standard deviations if  assumed unknown), and to select the value of $\kappa$ associated with the average best performance over all the considered scenarios. In practice, one considers a grid of values for $\kappa$ and a set $\mathcal{S}_1=\{H_1^{(1)},\,\dots,H_1^{(S_1)}\}$ of plausible  scenarios, where $H_1^{(s)}$ is characterized by the set of unknown parameters. Then, the following algorithm can be used to find the robust optimal $\kappa$:
\begin{enumerate}
    \item Define an objective function $g(u^{(s)}(\kappa))$, where $u^{(s)}(\kappa)$ is the value assumed by an operating characteristic of interest for a given $\kappa$ and scenario $s$ ($s=1,\dots,S_1$).  \vspace{-3mm}
    \item Compute $u^{(s)}(\kappa)$, for each scenario $s$ and value of $\kappa$ in the considered grid. \vspace{-3mm}
    \item Find the optimal $\kappa$: $\kappa^*=\text{argmin}_\kappa\,\frac{1}{S_1}\sum_{s=1}^{S_1}\,g(u^{(s)}(\kappa))$.
\end{enumerate}

Averaging over a wide set of scenarios offers a robust way to select a suitable value of $\kappa$  regardless of the true values of the parameters: although the optimal value of $\kappa$ may not be convenient for some particular scenario, this procedure ensures that it is nearly optimal. 
Section \ref{subsec:alternativeScenarioCharacteristics} provides further details on how to define a set of plausible scenarios for the proposed strategy.

\subsection{Objective functions}

The objective functions should be chosen in accordance with the operating characteristic that one wants to target in the calibration of $\kappa$. If the main objective is to maximize the patient benefit (PB), we propose to target the average percentage of experimentation of the best arm (evaluated over $M$ trial replicates), namely
\begin{equation}
\label{eq:PBmeasure}
PB=\frac{1}{M}\sum_{m=1}^M\Biggl\{\frac{1}{N}\sum_{t=1}^N\,\mathbb{I}\bigl(a_{tm}=j^{1*}\bigr)\Biggr\}\,,
\end{equation}
where $a_{tm}=j^{1*}$ if and only if patient $t$ is allocated to the true best arm $j^{1*}$ during the trial $m$. Taking $u_1(\kappa)=(PB)_\kappa$ for any $\kappa$, we adopt the  following objective function:
\begin{equation}
\label{eq:PBobjective}
g_1(\kappa)=\frac{1}{S_1}\sum_{s=1}^{S_1}\,\Bigl(u_1^{(s)}(\kappa)-u^{(s)}_\text{max}\Bigr)^2\,,
\end{equation}
quantifying how far each value $u^{(s)}_1(\kappa)$ is from $u^{(s)}_\text{max}=\max_k u_1^{(s)}(\kappa)$.
 
For the objective to achieve a pre-specified level of power, we use 
\begin{equation}
\label{eq:objFunPower}
g_2(\kappa)=\begin{cases}
    \kappa & \frac{1}{S_1}\sum_{s=1}^{S_1}\,\mathbb{I}\Bigl(u_2^{(s)}(\kappa)\geq0.8\times u^{(s)}_{2,FR}\Bigr)\,\geq\,\xi\\
    \infty & \text{otherwise}
\end{cases}\quad,
\end{equation}
where $u_2^{(s)}(\kappa)$ is either the power defined in \eqref{eq:condPower} or in \eqref{eq:twoCompPower}, $u^{(s)}_{2,FR}$ is the power attained using fixed and equal randomization (FR) and $\xi\in[0,1]$. The optimal $\kappa$ will be then the smallest $\kappa$ for which the probability to obtain a power higher than $80\%$ of the power attained by the FR is at least $\xi$. The value $0.8$ relaxes the requirement that the WE design must outperform the FR design in terms of power, a condition that can be overly stringent. Analogous practical compromises can be found in dose-response settings, where the dose achieving $80\%$ of the maximum efficacy is often considered a reasonable target. This value can be modified if closer alignment with the power of FR is required, although targeting a power above that of FR may be unrealistic in the considered context.

Finally, we propose a two-step hierarchical procedure to combine the two objective functions in a single criterion: first, one restricts attention to values of $\kappa$ that ensure a sufficiently high probability of achieving the target power; then, one selects the value that optimizes the patient-benefit objective within this restricted set. Formally, we suggest selecting 
\begin{equation}
\label{eq:objFun_2step}
\kappa^*_\text{2-step}=\arg\min_\kappa g_1(\kappa) \quad \text{subject to} \quad g_2(\kappa)<\infty\,,
\end{equation}
where $g_2(\kappa)<\infty$ is equivalent to satisfying the power constraint in \eqref{eq:objFunPower}.

\section{Simulation study of Phase II clinical trials}
\label{sec:simStudyUniv}

\subsection{Setting}
\label{subsec:simSetting}

Let us consider an early Phase IIa proof-of-concept oncology clinical trial such as the one introduced in the motivating example of Section \ref{sec:intro}. We consider $K=4$ alternative treatments (no control arm) and a total of $N=100$ patients recruited at the end of the trial. We assume that the primary endpoint is PD-marker measured on a continuous scale, and responses from arm $j$ closer to the target $\gamma$ are more desirable. For simplicity, we assume $\gamma=0$, implying that sample means equal to $\epsilon$ and $-\epsilon$ ($\epsilon>0$) are treated equally by WE$(p,\kappa)$. Thus, for a given scenario, the best treatment arm is the one associated with smallest mean response in absolute value. 

We investigate the performance of our method over $S_1=500$ random scenarios under the alternative hypothesis introduced in Section \ref{subsec:hptest}. Each scenario is characterized by a vector of means $(\mu_1,\dots,\mu_K)$, where $\mu_j$ is generated from a $Unif[-4,4]$, $\forall j=1,\dots,K$, while variances are assumed to be known and equal to $\sigma_j^2=2^2$, $j=1,2,3$, $\sigma_4^2=4^2$ for each scenario. Details and a full analysis for the case of unknown variances are presented in Section H of the SM.

Operating characteristics of interest are (i) the type-I error rates in correspondence of specific null scenarios, (ii.a) conditional ($1-\beta_C$) and (ii.b) two-components ($1-\beta_{TC}$) statistical power (see Section \ref{subsec:hptest}), (iii) the average percentage of experimentation of the best arm (see \eqref{eq:PBmeasure}) as measure of patient benefit (PB) and (iv.a) the percentage of trial replicates with correct identification of the best arm (CS$_I(\%)$) and of (iv.b) the two best arms (CS$_{I\&II}(\%)$). For each given scenario, operating characteristics are assessed over $M=10^4$ trial replicates.

In the following, for the class WE($p$,$\kappa$) of designs proposed in Section \ref{subsec:simWeight}, we will only consider the case of $p=\{1,\,2\}$: this will offer insights on how different degrees of discrimination between arm-specific variances affect the resulting operating characteristics. We compare its performance with several alternative allocation procedures which have good properties either in terms of statistical power or patient benefit. For a fair comparison with our proposal and to narrow down the set of competitors, we restrict to response-adaptive designs based on deterministic allocation rule. Additionally, in line with the objective of the trial, we only consider those adaptive designs that try to skew allocations in favour of arms with means closer to the target. Specifically, we compare with the following designs: (i.) Fixed and equal randomization (FR), where each patient can be treated in a specific arm with fixed probability $\frac1K$; (ii.) Current belief (CB), where the next patient is treated in the arm with posterior mean $\bar x_{n_j}$ closest to the target $\gamma$; (iii.) Thompson Sampling (TS), where patient $t$ is treated in the arm which maximizes the adjusted posterior probability of being the best, i.e. $\pi_j=\frac{\mathbb{P}(j=j^{1*}|\texttt{data})^c}{\sum_{l=1}^K\,\mathbb{P}(l=j^{1*}|\texttt{data})^c}\,,$ where $c=\frac{t}{2N}$ helps to stabilize the resulting allocations; (iv.) Symmetric Gittins Index (SGI) and (v.) Targeted Gittins Index (TGI), two modifications of \cite{smith2018bayesian}'s proposal involving $\gamma$ (full details are provided in Section E of the SM). We use objective priors and fix $d=0.99$, in line with the considerations of \cite{wang1991sequential} and \cite{williamson2020response} for a trial of size $N=100$.
Notice that, apart from FR, all the alternative designs can be classified as response-adaptive. FR is taken as benchmark for power comparison, due to its even exploration of all the arms and its wide application in real trials. CB and TS are common and intuitive response-adaptive designs, which are generally associated with good performance in terms of patient benefit. Finally, SGI and TGI are modifications of a forward-looking design which is oriented toward a patient benefit objective \citep{smith2018bayesian,williamson2020response}. In absence of prior knowledge about the treatments, the burn-in size is fixed to $B=5$ for all the response-adaptive designs to help sequential algorithms collecting sufficient preliminary information about all the arms. The same burn-in size is chosen for both the case of known and unknown variances, in order to isolate its effect in the comparison between these two settings. 
Further details are provided in Section G.1 of the SM.

\subsection{Characteristics of randomly generated alternative scenarios}
\label{subsec:alternativeScenarioCharacteristics}

\begin{figure}[ht]
            \centering
            \includegraphics[width=0.9\linewidth]{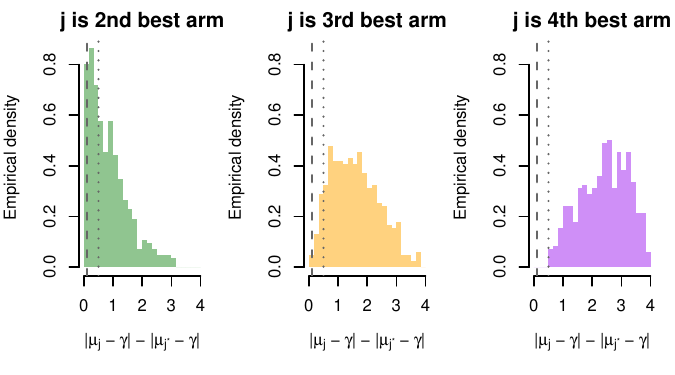}
            \caption{Distribution of differences $d_j-d_{j^{1*}}=|\mu_j-\gamma|-|\mu_{j^{1*}}-\gamma|$ relative to the best arm $j^{1*}$, across the $500$ alternative scenarios randomly generated for the simulation study in Section 5.1. The panels correspond to $j=j^{2*}$ (\textit{left}), $j=j^{3*}$ (\textit{middle}) and $j=j^{4*}$ (\textit{right}). Vertical dashed and dotted lines indicate the thresholds below which all scenarios have differences $d_j-d_{j^{1*}}$ smaller than $\epsilon=0.1$ and $\epsilon=0.5$, respectively.}
            \label{fig:altScenariosGaps}
    \end{figure}

Let $d_j=|\mu_j - \gamma|$ ($j=1,\dots,K$) be the distance of each arm mean from the target. All the 500 scenarios generated in Section \ref{subsec:simSetting} imply a unique best arm $j^{1*}$, namely if $d_j>d_{j^{1*}}$ for all $j \neq j^{1*}$. This allows exploration of a range of deviations of the true arm means $\mu_j$ from $\gamma$. We can characterize these  scenarios by the degree of separation between the best arm and the remaining ones. For a given scenario $H_1^{(s)}$, we can rank the arms from closest to furthest from the target mean: $d_{j^{1*}}<d_{j^{2*}}<d_{j^{3*}}<d_{j^{4*}}$, with the number of asterisks indicating the ranking position. For a threshold $\epsilon > 0$, we compute the proportion of scenarios in which $d_j-d_{j^{1*}}<\epsilon$ for each $j\neq j^{1*}$. A small $\epsilon$ identifies scenarios where the best arm is difficult to distinguish.
Figure \ref{fig:altScenariosGaps} summarizes the distribution of these differences across the $500$ scenarios. As expected, the separation from the best arm increases as we move down the ranking, meaning that the second-best arm is typically only slightly further from $\gamma$, while lower-ranked arms tend to deviate more substantially. The dashed and dotted vertical lines indicate thresholds at $\epsilon = 0.1$ and $\epsilon = 0.5$, respectively, leaving on their left scenarios such that $d_j-d_{j^{1*}}\approx0$. Notably, the scenario most similar to a global null is characterized by $\bm\mu=(-2.98, -2.90,  3.04, -3.40)$ where the difference between the best and worst arms’ distances from the target is only $0.5$, which is small relative to the within-arm standard deviations (each being at least $2$). In such a scenario, all arms lie close to $\gamma$ and are therefore more difficult to distinguish. In contrast, the scenario with the largest separation between the best and second-best arms is characterized by $\bm\mu=(-0.59, 3.73, -3.83,  3.75)$, where one arm is much closer to $\gamma=0$ than the others.

\subsection{Calibration of cut-off probabilities}
\label{subsec:typeIcontrolSim}

For each combination of $p$ and $\kappa$, we calibrate design-specific cut-off probabilities under the set of null configurations $\mathcal{S}_0=\bigl\{(\mu_1^{(s)},\dots,\mu_K^{(s)}):\,\mu_j^{(s)}-\gamma=c_s,\forall j=1,\dots,K,\,c_s\geq0,\,s=1,\dots,S_0\bigr\}$ both considering the strong and the average control of the type-I error rate at a level $\alpha=5\%$. Notice that: (i) the generic element of $\mathcal{S}_0$, denoted by $H_0^{(s)}=(c_s+\gamma,\dots,c_s+\gamma)$, represents the null scenario where all treatment arms are at a common distance $c_s$ from $\gamma$, implying that calibration is done under the least favourable situation where there is no sub-optimal arm; (ii) one only considers $c_s\geq0$ since, for $\gamma=0$, values $c_s$ and $-c_s$ are treated equally in the definition of best arm. Details on the null scenarios are in SM (Section G.2).

\begin{figure}[ht]
    \centering
    \includegraphics[width=0.97\textwidth]{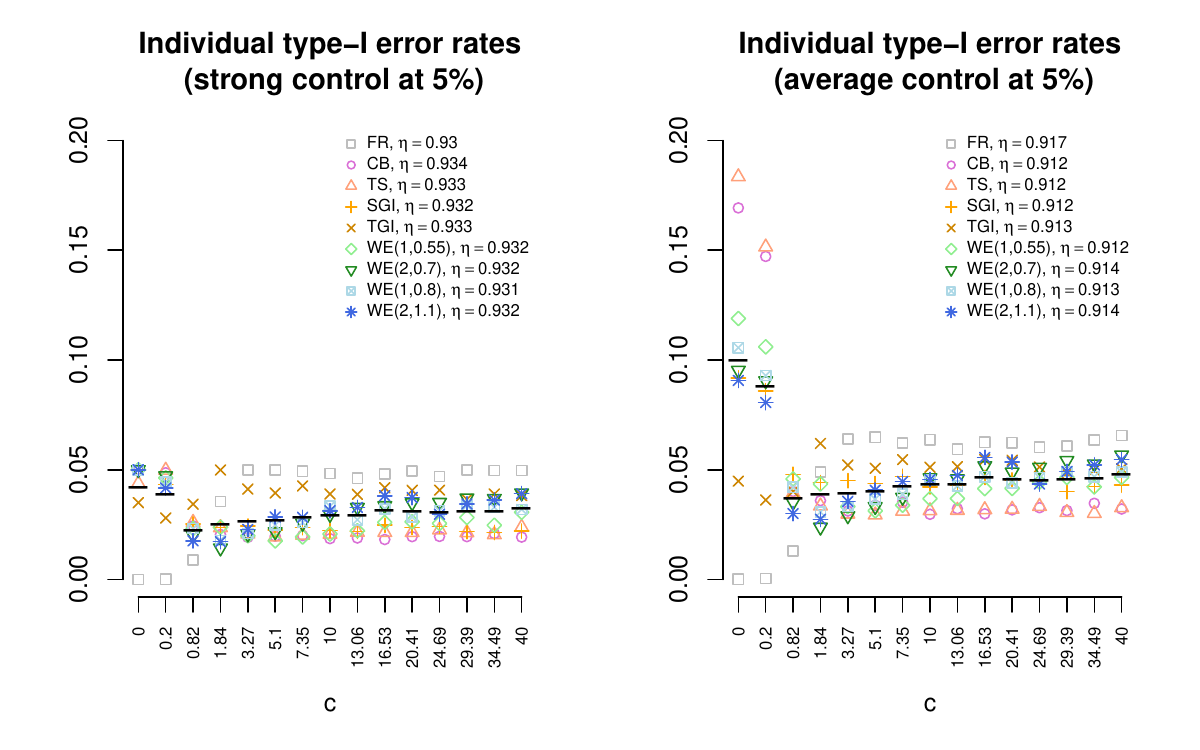}
    \caption{Type-I error rates for a grid of values of $c_s:\,\mu_j^{(s)}-\gamma=c_s,\,\forall j,\,c_s\geq0$ and for several designs. Design-specific cut-off probabilities are calibrated both in the case of strong (\textit{left}) and average (\textit{right}) control of the type-I error rate at level $\alpha=5\%$.}
    \label{fig:alphaRates}
\end{figure}

Results are reported in Figure \ref{fig:alphaRates}, where only four WE designs are shown based on the results of Section \ref{subsec:kappaSelectionSim}. 
Response-adaptive designs tend to be associated with higher type-I error rates when $c_s$ is small, while FR design is associated with type-I error rates which increase with $c_s$ before reaching a plateau. This different behaviour can be explained by the peculiar nature of response-adaptive designs. Indeed, variability around the arm responses can overly penalize some arms at the beginning of the trial in favour of others. Notably, when $c_s$ is very small (i.e. all true means are very close to the target value), most of the patients are likely to be assigned to the arm which, by chance, performs better at the beginning of the trial: as a result, any other arm may not have the chance to be further explored, thus being falsely identified as inferior at the end of the trial. If a practitioner aims at reducing even more the resulting inflation of the type-I error rate associated with response-adaptive designs in this type of scenarios, a possible solution is to consider higher burn-in sizes (see Section G.1 of the SM), possibly at the expense of a lower patient benefit: although a larger $B$ is associated to a more exhaustive exploration of all the arms and to more precise estimates, a too high value may lead to assign a high percentage of patients to sub-optimal treatment arms. 

\subsection{Impact of $\kappa$ on the hypothesis testing procedure}
\label{subsec:kappaWithTest}

\begin{figure}[ht]
            \centering
            \includegraphics[width=1\linewidth]{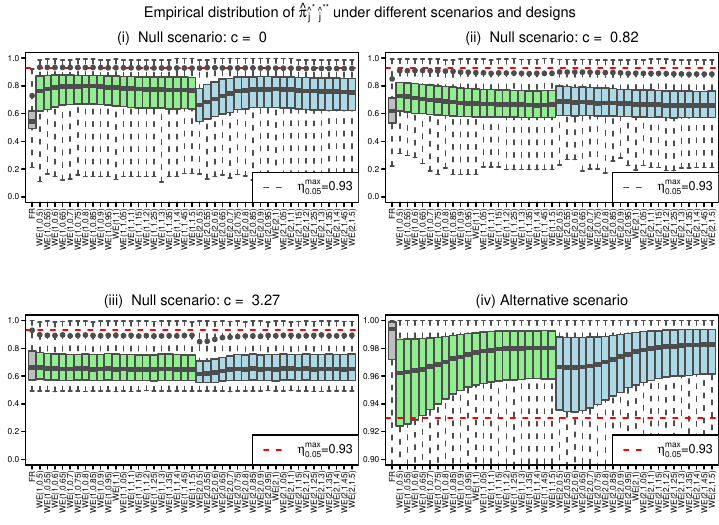}
            \caption{Empirical distribution of $\widehat\pi_{\widehat j^{1*},\widehat j^{2*}}$ for FR and WE$(p,\kappa)$ designs under $H_0:\mu_j-\gamma=c,\,\forall j$, for (i) $c_s=0$, (ii) $c_s=0.82$ and (iii) $c_s=3.27$, and (iv) under the scenario characterized by $\bm\mu=(1.91,-3.36,-0.37,3.99)$, with $\bm\sigma=(2,2,2,4)$, $\gamma=0$, $B=5$; the cut-off $\eta_{0.05}$ (black dot) corresponds to $\alpha=5\%$, with $\eta_{0.05}^{\text{max}}$ (dashed line) approximately controlling type-I error below $5\%$ across all designs and null scenarios.}
            \label{fig:hypothesisTest_byKappa_H0andH1}
    \end{figure}

Figure \ref{fig:hypothesisTest_byKappa_H0andH1} shows the empirical distribution of $\widehat\pi_{\widehat j^{1*},\widehat j^{2*}}$ as the function of $\kappa$. When all means are equal to $\gamma$ [panel (i)], the median value of this quantity increases with $\kappa$ converging to a constant value. Under the other null scenarios [panels (ii-iii)], the distribution of $\widehat\pi_{\widehat j^{1*},\widehat j^{2*}}$ is similar across all $\kappa$ and $p$. The cut-off value $\eta_{0.05}$ for WE designs remains close to $\eta_{0.05}^{\text{max}}=0.93$ across all scenarios and $\kappa$.

Under the alternative scenario [panel (iv)], $\widehat\pi_{\widehat j^{1*},\widehat j^{2*}}$ increases in median value and exhibits reduced variability as $\kappa$ becomes larger, thus increasing the probability of observing a posterior probability $\widehat\pi_{\widehat j^{1*},\widehat j^{2*}}$ above $\eta_{0.05}^{\text{max}}$. This is consistent with the results in Section \ref{subsec:allocProp}, which show that larger values of $\kappa$ lead to less imbalanced allocation ratios. The distribution of $\widehat\pi_{\widehat j^{1*},\widehat j^{2*}}$ under FR shows even less variability across replicates and a larger median value. This is due to the reduced variability in patient allocations with FR, which ensures balanced arm sizes across experiments.

\subsection{Robust selection of the tuning parameter $\kappa$}
\label{subsec:kappaSelectionSim}

\begin{figure}[ht]
    \centering
    \includegraphics[width=0.97\textwidth]{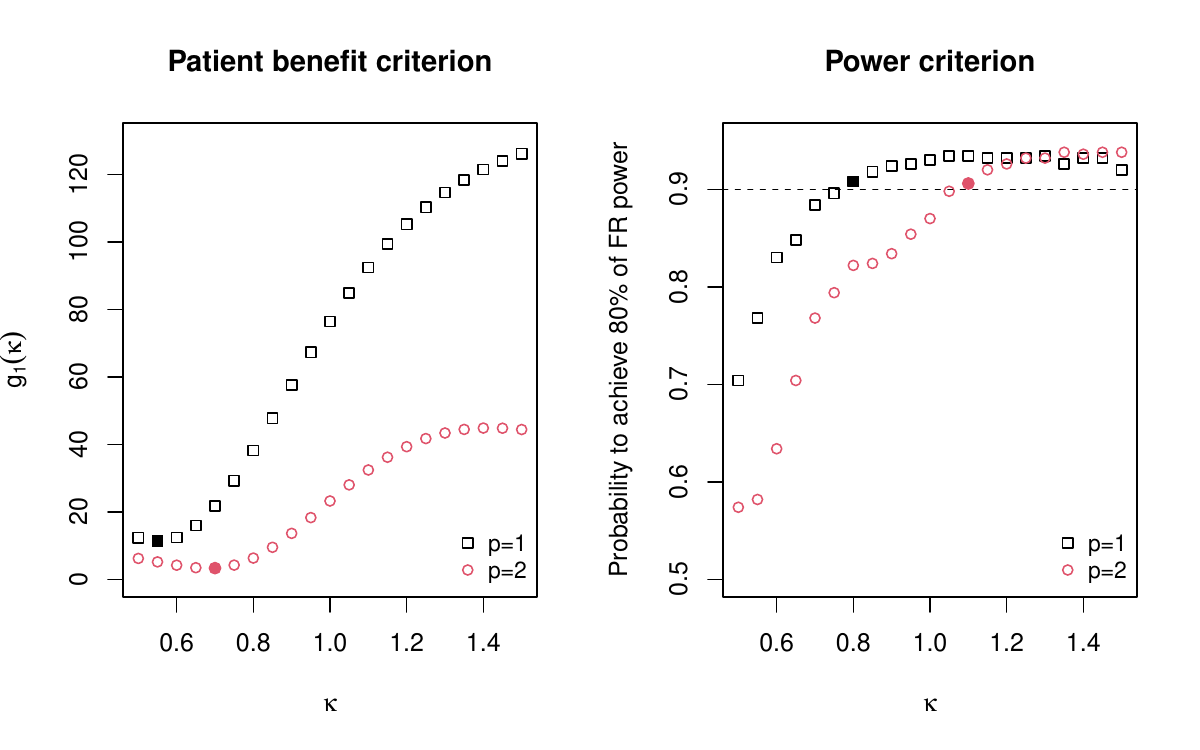}
    \caption{Objective function $g_1$ (\textit{left panel}) and probability to achieve at least $80\%$ of the power of FR (\textit{right panel}) for several values of $\kappa$, and for both $p=1$ (\textit{black square}) and $p=2$ (\textit{red circles}). Filled shapes correspond to robust optimal values of $\kappa$. The target probability to achieve $80\%$ of FR power is $\xi=0.9$ (dashed line).}
    \label{fig:kappaSelection}
\end{figure}

The implementation of WE($p$, $\kappa$) requires to fix $\kappa$ in advance. We apply the procedure presented in Section \ref{sec:kappaSelection} both for the objective function maximizing the patient benefit and the one achieving a specific level of statistical power, conducting separate analyses for $p=1$ and $p=2$. After preliminary checks, the sequence of values for $\kappa$ is restricted in the interval [$0.5$, $1.5$] with step $0.05$. For the power objective function, we require that $80\%$ of the power of the FR is achieved with probability higher or equal to $\xi=0.9$. As a measure for $u_2(\kappa)$, we consider the two-components power defined in \eqref{eq:twoCompPower} with design-specific cut-off probabilities guaranteeing an average control of type-I error rate at $\alpha=5\%$ over the set of scenarios considered in Section \ref{subsec:typeIcontrolSim}. 
Figure \ref{fig:kappaSelection} illustrates the result of the optimization of $g_1$ and $g_2$.
Lower values of $\kappa$ are suggested when the criterion optimizes the patient benefit, while larger values of $\kappa$ are necessary to satisfy the power condition. Higher values of $p$ require higher values of $\kappa$ to attain the specific objective to compensate the higher relevance attributed to the variance. Therefore,  $\kappa=0.55$ and $\kappa=0.8$ will be used for $p=1$, while $\kappa=0.7$ and $\kappa=1.1$ for $p=2$. Interestingly, a closer look to Figure \ref{fig:kappaSelection} shows that the rule \eqref{eq:objFun_2step}, which combines the two objectives, leads to the same optimal solutions as the power criterion.

\subsection{Numerical results}
\label{subsec:numericalResults}

\begin{figure}[ht]
    \centering
    \includegraphics[width=0.95\textwidth]{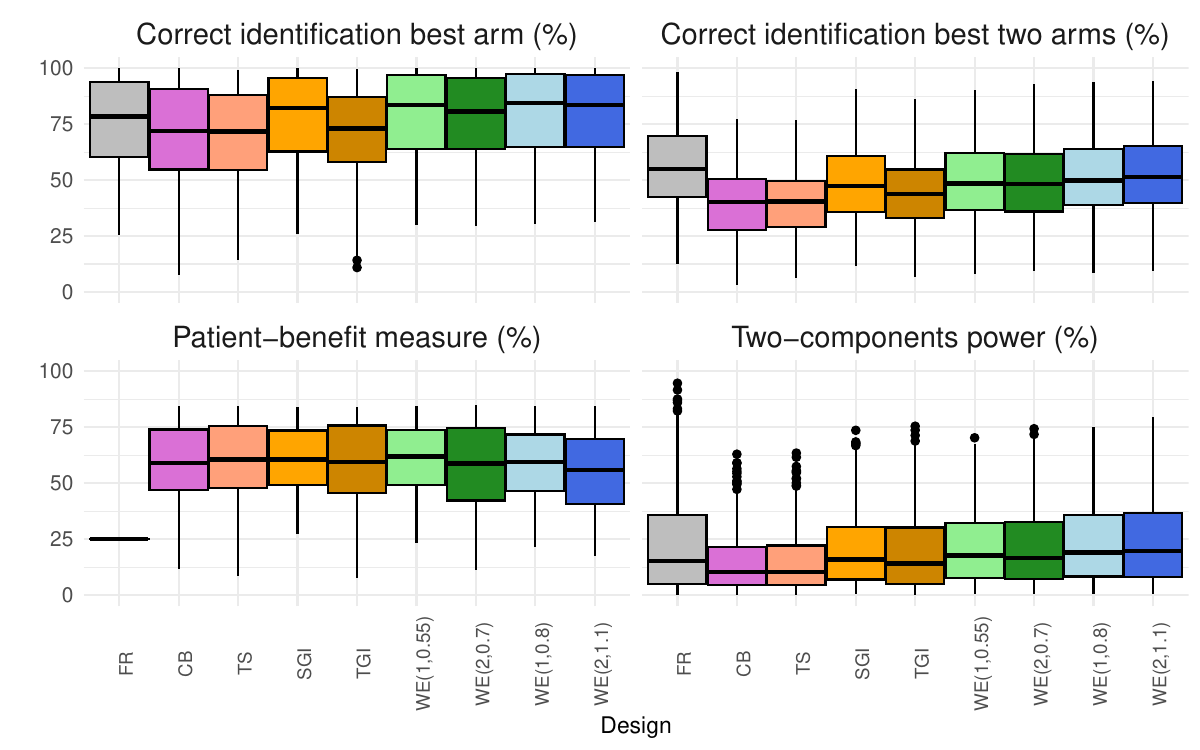}
    \caption{Operating characteristics across $S_1=500$ randomly generated scenarios for the considered set of designs. Each scenario is characterized by a vector of mean responses. For all scenarios, we assume $K=4$, $N=100$, $\bm \sigma=(2,2,2,4)$ (known) and $\gamma=0$. Burn-in size is fixed to $B=5$ for all the response-adaptive designs.}
    \label{fig:OC_Diff}
\end{figure}

A wide overview of the operating characteristics of the considered designs is reported in Figure \ref{fig:OC_Diff}, which summarizes results across all the $500$ randomly generated scenarios.
WE designs achieve good performance in terms of patient benefit and identification of the best arm than their competitors, with WE($1$, $0.8$) and WE($2$, $1.1$) achieving a power almost comparable to FR. The even exploration of all the arms which characterizes FR also leads to a higher percentage of correct identification of the two best arms, but with high variability across the scenarios. CB and TS generally perform worse than the other response-adaptive designs: notably, they both achieve high medians in terms of patient benefit but with high variability around them. In contrast, SGI and WE($1$, $0.55$) achieve comparable patient benefit measures, but with much less variability around their medians.

To further investigate the performance of our proposal, we pick two scenarios from the ones considered in Figure \ref{fig:OC_Diff}. Notably, given $\bm\sigma=(2,2,2,4)$ and $\gamma=0$, we consider: (i) Scenario I, characterized by $\bm\mu=(1.91, -3.36, -0.37, 3.99)$, with the worst treatment arm having the highest variability; (ii) Scenario II, characterized by $\bm\mu=(1.13, -3.48, -3.57, 0.34)$, with the best treatment arm having the highest variability. Table \ref{tab:simUniv_knownVar} illustrates the operating characteristics of the considered designs under Scenario I and II. Under Scenario I, TGI, WE($1$, $0.85$), WE($2$, $1.2$) achieve a conditional power higher or equal to FR, while their two-components power drops down due to a higher difficulty in identifying the true second best arm at the end of the trial: indeed, the difference between CS1 and CS2 is higher than $14\%$ in these three response-adaptive designs, while is lower than $2\%$ for FR. Moreover, FR detects the two best arms and claim the superiority of the best one in $88\%$ of the simulated trials, while this percentage is reduced by $8$ points when WE($2$, $1.2$) is considered. This moderate sacrifice in terms of power allows WE($2$, $1.2$) to assign, on average, more than $3$ out of $4$ patients to the best treatment arm (around $1$ out $4$ with FR), with a large earning in terms of patient benefit. All the other considered response-adaptive designs achieve moderately higher PB at the cost of a much lower two-components power. Notably, the highest PB is achieved by WE($1$, $0.55$), which make this design an appealing alternative whenever one aims at optimizing the benefit of the patients, also due to its low variability (s.e. $=0.06$).
\begin{table}[ht]
\centering

{\footnotesize\textbf{Scenario I:} $(\mu_j,\,\sigma_j)=\bigl\{(1.91,\,2),\,(-3.36,\,2),\,(-0.37,\,2),\,(3.99,\,4)\bigr\}$, $\gamma=0$} \\ 
\small
\setlength{\extrarowheight}{1pt}  
\setlength{\tabcolsep}{11.5pt}  
\begin{tabular}{lccccc}
  \midrule
  Design & PB (s.e.) & CS$_{I}(\%)$ & CS$_{I\&II}(\%)$ & $1-\beta_C$ & $1-\beta_{TC}$ \\ 
  \hline
  FR & 24.99 (0.04) & 99.63 & 97.97 & 0.89 & 0.88 \\ 
  CB & 81.22 (0.14) & 97.10 & 74.49 & 0.72 & 0.53 \\ 
  TS & 81.57 (0.13) & 96.43 & 73.90 & 0.70 & 0.52 \\ 
  SGI & 81.81 (0.07) & 99.67 & 82.75 & 0.85 & 0.70 \\ 
  TGI & 80.99 (0.08) & 94.84 & 75.17 & 0.95 & 0.71 \\ 
  WE(1,0.55) & 82.22 (0.06) & 99.88 & 82.49 & 0.81 & 0.66 \\ 
  WE(2,0.7) & 80.92 (0.07) & 99.85 & 84.46 & 0.86 & 0.73 \\ 
  WE(1,0.8) & 81.12 (0.06) & 99.89 & 83.36 & 0.88 & 0.73 \\ 
  WE(2,1.1) & 77.68 (0.08) & 99.93 & 85.57 & 0.93 & 0.79 \\ 
\end{tabular}

{\footnotesize\textbf{Scenario II:} $(\mu_j,\,\sigma_j)=\bigl\{(1.13,\,2),\,(-3.48,\,2),\,(-3.57,\,2),\,(0.34,\,4)\bigr\}$, $\gamma=0$} \\ 
\begin{tabular}{lccccc}
  \midrule
  Design & PB (s.e.) & CS$_{I}(\%)$ & CS$_{I\&II}(\%)$ & $1-\beta_C$ & $1-\beta_{TC}$ \\ 
  \hline
  FR & 25.05 (0.04) & 75.72 & 75.72 & 0.07 & 0.06 \\ 
  CB & 38.93 (0.37) & 43.31 & 39.31 & 0.52 & 0.21 \\ 
  TS & 35.40 (0.37) & 47.88 & 43.89 & 0.52 & 0.23 \\ 
  SGI & 63.67 (0.28) & 76.94 & 72.49 & 0.39 & 0.28 \\ 
  TGI & 23.31 (0.30) & 78.71 & 76.33 & 0.22 & 0.17 \\ 
  WE(1,0.55) & 67.59 (0.26) & 82.67 & 77.86 & 0.35 & 0.28 \\ 
  WE(2,0.7) & 76.78 (0.14) & 91.99 & 86.67 & 0.29 & 0.25 \\ 
  WE(1,0.8) & 72.12 (0.17) & 88.24 & 83.81 & 0.35 & 0.29 \\ 
  WE(2,1.1) & 76.70 (0.11) & 91.19 & 86.51 & 0.33 & 0.28 \\
\end{tabular}
\caption{Operating characteristics of the WE($p$,$\kappa$) design for different combinations of $\kappa$ and $p$, FR design, CB, SGI and TGI under Scenario I and II. Both $1-\beta_C$ and $1-\beta_{TC}$ are based on the design-specific cut-off probabilities $\eta_{0.05}^\text{ave}$ determined in Section \ref{subsec:typeIcontrolSim}. Results are based on $10^4$ replicated trials.}
\label{tab:simUniv_knownVar}
\end{table}
Under Scenario II, WE($2$, $0.65$) and WE($2$, $1.2$) achieve the highest value of PB, with the latter exhibiting lower variability. CB performs best in terms of conditional power compared to the other designs, but fails in correctly identifying the best treatment more than $50\%$ of the times. In contrast, WE designs correctly identify the two best arms in a high percentage of replicas (i.e. more than $75\%$) and are associated with higher values of two-components power. Notice that FR is associated with a much lower power than the other designs, probably due to the relatively lower exploration ($25$ patients on average) of the best arm.

Overall, the considered WE designs performance are comparable to the ones of the other response-adaptive designs, with small differences in terms of patient benefit but larger gain in terms of correct identification of the best arm and two-components power. In addition, WE designs generally tend to achieve a much higher patient benefit than FR due to their ability to skew allocations according to the accrued data, at the cost of moderately lower power. With a selection of $\kappa$ targeting the power, WE can result in a statistical power closer to the one of FR, but with remarkably larger values of PB. Finally, when the best arm has also the most variable response, the flexibility of WE designs allows it to achieve an even higher gain over FR in terms of power. Further details on how operating characteristics are affected by different choices of $p$ and $\kappa$ are given in Section G.3 of the SM. 

\subsection{Sensitivity analysis under model misspecification}
\label{subsec:modelmisspecif}

We conduct an additional simulation study under skew-normal outcomes to assess the impact of violations of the normality assumption. Details and results are reported in Section G.4 of the SM and show that the operating characteristics remain largely unchanged across the varying degrees of asymmetry considered.

\section{Illustration: co-primary endpoints in Phase II trials}
\label{sec:simStudyBiv}

\subsection{Setting}
\label{subsec:simSettingBiv}

In this section, we illustrate the performance of the novel response-adaptive design in the same trial setting as above but with two co-primary endpoints. The first endpoint is a PD-marker (PDM) targeting a specific value $\gamma^{(1)}=0$, while the second endpoint is the tumour shrinkage rate (TSR, $\%$) whose theoretical target is assumed to be $\gamma^{(2)}=100$. Variance matrices are known and equal to $\bm\Sigma_j=\Big(\begin{smallmatrix}2^2 & 0\\ 0 & 8^2\end{smallmatrix}\Big),\,\forall j$. 

As illustration, we consider the design for bivariate endpoints described in Section \ref{subsec:multiWeight} with $\kappa=\{0.5,\,0.75\}$. We compare its performance to two alternative approaches: (a.) fixed and equal randomization (FR), (b.) current belief (CB). We choose a burn-in size of $B=1$ patient per arm for response-adaptive designs, and we consider a scenario where $\bm\mu^{(1)}=(1,-1,2,-2.5)'$ and $\bm\mu^{(2)}=(10,25,55,60)'$. The four treatment arms can be characterized as follows: the first two treatments have values of the PDM relatively close to the target but associated with a relatively small TSR ($10\%$ and $25\%$, respectively); the third has twice the PDM of the first but it is associated with a remarkable average TSR (i.e. $55\%$); finally, the fourth is associated with a slightly poorer average PDM (e.g. more distant from the target) than the third treatment, but it has the highest average TSR. Adopting the standardized distance in \eqref{eq:multivDistance}, the fourth treatment arm can be regarded as the best one. 

\subsection{Results}
\label{subsec:numericalResultsBiv}

Design-specific cut-off probabilities are calibrated to achieve an average control of the type-I error rate at a level $\alpha=5\%$. The null configurations used for calibration and the type-I error rates for each evaluated design are given in SM (Section I).

Table \ref{tab:simBiv} shows the operating characteristics of the considered designs. FR correctly selects the best two arms in a high number of replicas ($82\%$), but loses power in treating only $50\%$ (on average) patients in these two arms. Both WE(0.5) and CB achieve high patient benefit measure, but the standard error for CB is more than five times larger than for  WE(0.5). Such a high standard error suggests an unpredictable behaviour of CB, which highly depends on the response of the first treated patients. Overall, despite of the low burn-in size ($B=1$), WE designs obtain the best operating characteristics, with WE(0.5) prioritizing the patient benefit objective and WE(0.75) achieving the highest power.
\begin{table}[ht]
\centering 
\small
\setlength{\extrarowheight}{1pt}  
\setlength{\tabcolsep}{11.5pt} 
\begin{tabular}{lccccc}
  \midrule
  Design & PB (s.e.) & CS$_{I}(\%)$ & CS$_{I\&II}(\%)$ & $1-\beta_C$ & $1-\beta_{TC}$ \\ 
  \hline
  FR       &  24.98 (0.04) & 0.82 & 0.82 & 0.41 & 0.34\\
  CB       &  64.41 (0.44) & 0.67 & 0.65 & 0.33 & 0.21\\
  WE(0.5)  &  77.18 (0.08) & 0.89 & 0.89 & 0.51 & 0.45\\
  WE(0.75) &  49.78 (0.03) & 0.89 & 0.89 & 0.52 & 0.47\\
\end{tabular}

\caption{Operating characteristics of FR, CB, WE(0.5) and WE(0.75) under scenario where $\bm\mu^{(1)}=(1,-1,2,-2.5)'$, $\bm\mu^{(2)}=(10,25,55,60)'$, $\bm\gamma=(0,\,100)'$, $\bm\Sigma_j=\Big(\begin{smallmatrix}2^2 & 0\\ 0 & 8^2\end{smallmatrix}\Big),\,j=1,\dots,K$ and $N=100$. Both $1-\beta_C$ and $1-\beta_{TC}$ are based on the design-specific calibrated cut-offs $\eta_{0.05}^\text{ave}$. Results are based on $10^4$ replicated trials.}
\label{tab:simBiv}
\end{table}

\section{Discussion and future directions}
\label{sec:discussion}

In this work, we proposed a class of designs for the selection of arms in a multi-armed setting with continuous outcomes, aiming to identify the arm whose mean lies closest to a pre-specified clinical target $\gamma$. This class of designs - referred to as WE - is based on a weighted entropy measure which retains the information about which outcomes are more desirable, a feature of considerable interest in settings with patient-benefit constraints. Particular emphasis was given on a weight function with Gaussian kernel centered around a pre-specified clinical target $\gamma$. The resulting information gain criterion, WE$(p,\kappa)$, was used to govern patient allocation during the trial. While the working assumption of independent arm-specific parameters may not capture potential cross-arm dependence, it ensures tractability in the construction of the allocation rule and is commonly adopted when reliable information on dependence between treatment effects is unavailable \citep[among others]{mozgunov2020information}.

The allocation rule demonstrates flexibility in addressing the exploration-exploitation trade-off. Empirical results show that WE$(p,\kappa)$ assigns a comparably high average proportion of patients to the best arm during the trial relative to response-adaptive competitors, with reduced variability of this measure, and performs similarly to the FR design in terms of power and correct identification of the best arm when properly calibrated. 
In particular, the penalization parameter $\kappa$ plays a key role for the WE design, and we illustrated a general procedure for its selection, based on a patient benefit criterion or a power criterion. We also discussed a two-step strategy for selecting $\kappa$ that jointly accounts for both criteria. For $\kappa>0.5$, we proved that WE$(p,\kappa)$ leads to a consistent selection of the best arm as the trial size increases and we derived the corresponding asymptotic allocation ratios. Notably, for $\kappa\in(0.5,1)$ and $p=1$, the best arm receives the largest limiting allocation share; although this property is not unique to WE$(p,\kappa)$, it aligns with the within-trial patient-benefit motivation underlying the design, further enhancing the rationale for using this allocation rule in trials with continuous outcomes. In finite-sample applications, small values of $\kappa$ may induce an overly greedy behaviour that excessively favours the apparent best arm early in the trial, while larger values $\kappa$ promote more balanced allocations across arms. The empirical findings of Section \ref{sec:simStudyUniv} suggest that power may still improve for $\kappa\geq1$, although the gains diminish as $\kappa$ approaches 1.5. For these reasons, we recommend exploring $\kappa\in(0.5,1.5]$ in practical applications. 
Regarding the parameter $p\in(1,2)$ - which controls the impact of heterogeneous variances on the allocation strategy - we suggest fixing $p=1$ when heteroscedasticity among different arms is expected, in order to mitigate the effect of a too large variance taking the lead over the others.  

In Section \ref{sec:simStudyUniv}, we chose the burn-in size ($B$) based on considerations regarding the need to control type-I error rates, taking into account both the case of known and unknown variances. 
The burn-in size plays a key role in sequential designs since it can help mitigate the impact of initial random fluctuations that might occur at the start of the trial \citep{pin2025informed}, though it may also reduce the patient-oriented benefit of full adaptiveness when an optimal treatment truly exists. 
 
The choice of the weight function is crucial and should be based on the setting of interest. For example, a multivariate generalization of the symmetric weight function was presented to evaluate multiple endpoints simultaneously. An asymmetric weight function was also presented in the SM (Section B). Although normal outcomes were assumed, the sensitivity analysis in Section G.4 of the SM indicated limited sensitivity to skewed outcomes. When the assumption of normality is not reasonable a priori, the procedure outlined in Section \ref{sec:methodology} can be adapted to accommodate various types of responses and weight functions.

In line with the primary objective of selecting the arm closest to the target $\gamma$, we introduced a hypothesis testing procedure that compares the two arms estimated to be closest to $\gamma$ and assesses their correct identification. This differs from classical frequentist $t$-tests or common Bayesian tests, which typically compare mean responses rather than distances from a target and do not directly accommodate the hypotheses considered here. To assess the performance of our ``selection-and-testing" strategy, we defined two power metrics: conditional power, which quantifies the ability of a design to correctly detect the superiority of the estimated best treatment arm over the second best, conditional on their correct identification; and two-components power, which also considers the ability to correctly identify the two best arms. While the conditional power ignores selection error in the choice of treatment arms to be tested and may therefore lead to overly optimistic assessments \citep{guo2021inference,bibaut2021post}, the two-components power provides a comprehensive measure of performance of the entire selection-and-testing mechanism as it naturally incorporates the uncertainty about the selection of the two best arms. Moreover, a procedure to calibrate cut-off values was illustrated, together with different strategies to control the type-I error rate (either achieving a strong or an average control). In Section \ref{sec:test} we also showed the consistency of the hypothesis testing procedure for the design WE$(p,\kappa)$, when $\kappa>0.5$. 

While this work focuses on the problem of ranking and selection under a unique optimal treatment, a possible extension could consider the identification of a set of tied-optimal treatments or, more generally, ``practically equivalent” treatments, for example by defining optimality via a clinically meaningful tolerance around the target and including all treatments whose expected responses fall within this range.

We adopted a hybrid approach which combines Bayesian and frequentist methodologies. A Bayesian perspective was employed to define information measures based on the posterior distribution of $\mu_j$ derived from an objective prior. While non-informative priors allow equal treatment of all therapies and minimize potential subjective bias, integrating available historical information for some or all treatment arms can be readily implemented within the proposed framework. On the other hand, frequentist properties (such as type-I error rates) of Bayesian hypothesis testing procedures were evaluated. This practice is common, as regulatory bodies often require explicit evidence that frequentist error rates are appropriately controlled \citep{shi2019control}. In addition, in the proposed allocation criterion, unknown variances are updated sequentially using their maximum likelihood estimate, similar to the approach adopted in many sequential designs (e.g., \cite{liu2009sequential}). Although this method has shown good performance in practice, an attempt to make the procedure fully Bayesian is subject of ongoing research.

In this work, we used a deterministic ``select-the-best" rule to allocate patients in the sequential experiments, and we compared our proposed criterion to several allocation procedures which are known to have good properties either in terms of statistical power or patient benefit. When appropriately calibrated through a robust selection of the parameter $\kappa$, the proposed class of designs offers a good balance between patient benefit and statistical power, by making the correct decision in a remarkable proportion of replicated trials. 
The feasibility of such a criterion is shown in \cite{mozgunov2020information}. Other good randomized options can be found in the literature of the ``Top-Two" allocation criteria, a class of allocation rules that, at each iteration, randomly assign patients only to one of the two best performing treatments. Top-Two Probability Sampling and Top-Two Thompson Sampling are two common examples \citep{russo2020simple,jourdan2022top}, but more sophisticated criteria based on weighted information may be explored.

For a fair comparison, we focused on response-adaptive designs based on deterministic rules and grounded in Bayesian considerations. Beyond these rules, the literature on response-adaptive designs is extensive, particularly in the frequentist framework \citep{hu2006theory}. For example, urn-based designs \citep{atkinson2013randomised} are non-parametric and easy to implement but may not adequately penalize heavily unbalanced allocations. Another important class is represented by the doubly-adaptive biased coin designs, which target pre-specified allocation proportions and have optimal asymptotic properties \citep{eisele1994doubly,hu2009efficient}. These procedures do not typically adapt learning toward a clinically relevant target $\gamma$ such in the context that we have considered in this paper. Future research could explore extensions of these designs that are aligned with our motivating example.

\section*{Acknowledgments}

This report is independent research supported by the National Institute for Health Research (NIHR Advanced Fellowship, Dr Pavel Mozgunov, NIHR300576, NIHR Postdoctoral Fellowship, Dr Pavel Mozgunov, NIHR503042) and by UK Medical Research Council (MC\_UU\_00002/19 and MC\_UU\_00040/3). For the purpose of open access, the authors have applied a Creative Commons Attribution (CC BY) licence to any Author Accepted Manuscript version arising. 
The authors would like to thank the three anonymous referees and the editor for their valuable and constructive comments, which helped improve the quality of this manuscript.

\section*{Data Availability Statement}

No new data were generated or analysed in support of this research. All relevant R code to implement the methodology proposed in this paper is available at \href{https://github.com/gcaruso-stats/info-gain-RAD-normal}{https://github.com/gcaruso-stats/info-gain-RAD-normal}.

\bibliographystyle{abbrvnat}
\bibliography{WEdesign_v3}


\renewcommand\thesubsection{\Alph{subsection}}
\renewcommand\thesection{\thesubsection.\arabic{section}}

\renewcommand{\thefigure}{\Alph{subsection}.\arabic{figure}}
\renewcommand{\thetable}{\Alph{subsection}.\arabic{table}}
\counterwithin{figure}{subsection}
\counterwithin{table}{subsection} 

\clearpage
\begin{center}
    \huge\textbf{Supplementary Material}   
\end{center}
\vspace{0.5cm}

\subsection{Proofs}

\subsubsection{Proof of Theorem 1}

\begin{proof}

Consider the generic arm $j$. It is well known that the Shannon entropy of a normal distribution with mean $\mu_j$ and variance $\sigma_{n_j}^2$ is given by 
\begin{equation*}
h(\pi_{n_j})=\frac12+\frac{1}{2}\log\Bigl(2\pi\sigma_{n_j}^2\Bigr)\,,
\end{equation*}
while the weighted Shannon entropy depends on the adopted weight function. 

To begin, one can derive the explicit formula of the (normalized) weight function. For this purpose, the following equation is useful:
\begin{equation}
\label{eqSM:sumOfSquares}
    \biggl(\frac{\mu_j-\gamma}{\sigma_{\phi,\,j}}\biggr)^2 + \biggl(\frac{\mu_j-\bar x_{n_j}}{\sigma_{n_j}}\biggr)^2=\biggl(\frac{\mu_j-\widetilde\mu_j}{\widetilde\sigma_j}\biggr)^2 - \biggl(\frac{\widetilde\mu_j}{\widetilde\sigma_j}\biggr)^2 + \biggl(\frac{\gamma}{\sigma_{\phi,\,j}}\biggr)^2 + \biggl(\frac{\bar x_{n_j}}{\sigma_{n_j}}\biggr)^2\,,
\end{equation}
where $\widetilde{\mu}_j=\frac{\gamma\sigma_{n_j}^2+\bar x_{n_j}\sigma_{\phi,\,j}^2}{\sigma_{n_j}^2+\sigma_{\phi,\,j}^2}$ and $\widetilde{\sigma}^2_j=\frac{\sigma_{n_j}^2\cdot\sigma_{\phi,\,j}^2}{\sigma_{n_j}^2+\sigma_{\phi,\,j}^2}\,,$ and find that the normalization condition $\int_{\mathbb{R}}\,\phi_\gamma(\mu_j)\pi_{n_j}(\mu_j)\,d\mu_j=1$ is satisfied by
\begin{align}
   \phi_\gamma(\mu_j)&=C(\bar x_{n_j},\sigma^2_j,\gamma,n_j)\,\exp\biggl\{-\frac12\Bigl(\frac{\mu_j-\gamma}{\sigma_{\phi,\,j}}\Bigr)^2\biggr\} \nonumber\\
   &=\frac{\sigma_{n_j}}{\widetilde{\sigma}_j}\cdot\exp\Biggl\{-\frac12\Biggl[\biggl(\frac{\mu_j-\gamma}{\sigma_{\phi,\,j}}\biggr)^2+\biggl(\frac{\widetilde{\mu}_j}{\widetilde{\sigma}_j}\biggr)^2-\biggl(\frac{\gamma}{\sigma_{\phi,\,j}}\biggr)^2-\biggl(\frac{\bar x_{n_j}}{\sigma_{n_j}}\biggr)^2\Biggr]\Biggr\}\,. \label{eqSM:phiFormulaSpecific}
\end{align}

Notably,
\begin{equation*}
    \phi_\gamma(\mu_j)\pi_{n_j}(\mu_j)=\bigl(2\pi\widetilde\sigma_j^2\bigr)^{-\frac12}\,\exp\Biggl\{-\frac12\biggl(\frac{\mu_j-\widetilde\mu_j}{\widetilde\sigma_j}\biggr)^2\Biggr\}=\varphi(\mu_j; \widetilde\mu_j,\,\widetilde\sigma_j^2)\,,
\end{equation*}
with $\varphi(\mu_j; \widetilde\mu_j,\,\widetilde\sigma_j^2)$ being the pdf in $\mu_j$ of a normal distribution with mean $\widetilde\mu_j$ and variance $\widetilde\sigma_j^2$.

As a consequence, the weighted Shannon entropy can be written as

\begin{align*}
h^{\phi_\gamma}(\pi_{n_j})&=-\int_{\mathbb R}\,\phi_\gamma(\mu_j)\pi_{n_j}(\mu_j)\log\pi_{n_j}(\mu_j)\,d\mu_j  \\
&=\int_{\mathbb R}\,\varphi(\mu_j; \widetilde\mu_j,\,\widetilde\sigma_j^2)\,\biggl[\frac12\log\Bigl(2\pi\sigma_{n_j}^2\Bigr) + \biggl(\frac{\mu_j-\bar x_{n_j}}{\sigma_{n_j}}\biggr)^2 \biggr]\,d\mu_j \\
&=\frac12\log\Bigl(2\pi\sigma_{n_j}^2\Bigr) + \frac{1}{2\sigma_{n_j}^2}\int_{\mathbb R}(\mu_j-\bar x_{n_j})^2\,\varphi(\mu_j; \widetilde\mu_j,\,\widetilde\sigma_j^2)\,d\mu_j\,.
\end{align*}

Exploiting the fact that $(\mu_j-\bar x_{n_j})^2=(\mu_j-\widetilde\mu_j)^2+2\mu_j(\widetilde\mu_j-\bar x_{n_j})-\widetilde\mu_j^2+\bar x_{n_j}^2$, the weighted Shannon entropy can be finally expressed as 
\begin{equation*}
h^{\phi_\gamma}(\pi_{n_j})=\frac12\log\Bigl(2\pi\sigma_{n_j}^2\Bigr) + \frac{\widetilde\sigma_j^2+(\widetilde\mu_j-\bar x_{n_j})^2}{2\sigma_{n_j}^2}\,.
\end{equation*}

The resulting information gain is then

\begin{equation*}
    \Delta(\mu_j)=h(\pi_{n_j})-h^{\phi_\gamma}(\pi_{n_j})=\frac{\sigma_{n_j}^2-\widetilde \sigma_j^2-(\widetilde\mu_j-\bar x_{n_j})^2}{2\sigma_{n_j}^2}\,.
\end{equation*}

By observing that $\widetilde\mu_j-\bar x_{n_j}=\frac{(\gamma-\bar x_{n_j})\sigma_{n_j}^2}{\sigma_{n_j}^2+\sigma_{\phi,\,j}^2}$, one can easily derive the final expression of the information gain for the weight function in \eqref{eqSM:phiFormulaSpecific}, i.e. 

\begin{equation*}
    \Delta_{n_j}=\frac12 \frac{\sigma_{n_j}^{2}}{\sigma_{n_j}^2+\sigma_{\phi,\,j}^2}-\frac12\Biggl(\frac{\gamma-\bar x_{n_j}}{\sigma_{n_j}}\Biggr)^2\Biggl(\frac{\sigma_{n_j}^{2}}{\sigma_{n_j}^2+\sigma_{\phi,\,j}^2}\Biggr)^2\,.
\end{equation*}

If $\sigma_{\phi,\,j}^2=\frac{\sigma_j^p}{n_j^\kappa}$, then the result immediately follows.

\end{proof}

\subsubsection{Proof of Theorem 2}

\begin{proof}

First, we need to ensure that each arm is sampled infinitely often. For the sake of contradiction, suppose that there exists an arm $j$ such that $n_j(t)$ is bounded by a finite constant $U$, i.e. $n_j(t)\leq U$ as $t\rightarrow\infty$. From Theorem 1, for any arm $i$, $\Delta_{n_i(t)}\rightarrow -\infty$ as $t\rightarrow\infty$ when $\kappa>0.5$. Therefore, there exists a sufficiently large $t'$ such that $\Delta_{n_j(t')}\geq\Delta_{n_i(t')}$, $\forall i\neq j$, with non-zero probability. Hence, arm $j$ must eventually be visited again, implying $n_j(t)\rightarrow\infty$. This contradiction shows that all arms are visited infinitely often when $\kappa>0.5$.\\
Remind that observations within arm $j$ are assumed to be independent and identically distributed, so the Strong Law of Large Numbers ensures that $\bar x_{n_{j}(t)}\rightarrow \mu_{j}$ if $t\rightarrow\infty$. When $\kappa>0.5$, this happens almost surely for all $j$. Now, suppose that the trial ends at time $t=N$. Let denote $\widehat j^{l*}(N)$ the arm placed in the $l$-th position of the estimated ranking at time $t=N$. Then, for each $l=1\dots,K$, $P\Bigl(\widehat d_{\widehat j^{l*}(N)}=d_{j^{l*}}\Bigr)\rightarrow1\,,$ as $N\rightarrow\infty$. That is, when $\kappa>0.5$, each arm eventually occupies its correct position in the ranking as the sample size increases.

\end{proof}

\subsubsection{Proof of Corollary 1}

    \begin{proof}
    Let $N_j$ be the number of observations collected from arm $j$ at the end of the trial, such that $N=\sum_{j=1}^K N_j$. Recall that $\bar X_{N_j}\sim N(\mu_j,\sigma^2_j/N_j)$. Standard Gaussian concentration (see, e.g., \citet[p.~23]{wainwright2019high}) implies that for any $t>0$, $$P(|\bar X_{N_j}-\mu_j|>t)\leq 2\exp\biggl\{-\frac{N_j t^2}{2\sigma^2_j}\biggr\}\,.$$ Now, let $d_j=|\mu_j-\gamma|$ and $\hat d_j=|\bar X_{N_j}-\gamma|$, and assume the true distances are all distinct and ordered as $d_{j^{1*}}<d_{j^{2*}}<d_{j^{3*}}<\dots<d_{j^{K*}}$. Using the reverse triangular inequality, $|\hat d_j-d_j|=\bigl| |\bar X_{n_j}-\gamma|-|\mu_j-\gamma|\bigr|\leq |\bar X_{n_j}-\mu_j|$. Thus, 
    \begin{equation}
    \label{eq:probSingleBound}
    P(|\hat d_j-d_j|>t)\leq P(|\bar X_{n_j}-\mu_j|>t)\leq 2\exp\biggl\{-\frac{n_j t^2}{2\sigma^2_j}\biggr\}\,.
    \end{equation}
    Now we want to show that, if every estimated distance $\hat d_j$ is close enough to the true distance $d_j$, then the entire ranking of the arms must be correct. For any pair of arms $i,j$ such that $d_i<d_j$, we want to have $\hat d_i<\hat d_j$ in order to match the true ranking. If we assume that $$|\hat d_i-d_i|<\frac{d_j-d_i}{2}\quad\text{and}\quad|\hat d_j-d_j|<\frac{d_j-d_i}{2}\,,$$ then the two estimated distances must respect these two bounds: $$\hat d_i<d_i+\frac{d_j-d_i}{2}\quad\text{and}\quad\hat d_j>d_j-\frac{d_j-d_i}{2}\,.$$ Notice that the right side of the previous two inequalities is the same quantity, thus $\hat d_i<\hat d_j$. Notice that the key idea is that the true distances are separated by a positive gap $|d_j-d_i|$, so if the estimation error $|\hat d_k-d_k|$ ($k\in\{i,j\}$) is smaller than half that gap, the two estimated distances cannot swap order (i.e., it must be $\hat d_i<\hat d_j$. \\ To make this hold simultaneously for all pairs $i,j$, we define $$\delta_{\min}=\min_{i\neq j}|d_i-d_j|>0$$ as the smallest gap between any two true distances. Since $d_j-d_i\geq\delta_{\min}$ for every pair with $d_i<d_j$, the condition $|\hat d_k-d_k|<\frac{\delta_{\min}}{2},\,\forall k=1,\dots,K$, is sufficient to guarantee that $\hat d_i<\hat d_j,\,\forall i\neq j$. Thus, if the event $$E=\bigcap_{j=1}^K\,\Bigl\{|\hat d_j-d_j|<\frac{\delta_{\min}}{2}\Bigr\}$$ occurs, then the estimated ranking is correct. Conversely, if the estimated ranking is incorrect, then the complement $E^c=\bigcup_{j=1}^K\,\Bigl\{|\hat d_j-d_j|\geq\frac{\delta_{\min}}{2}\Bigr\}$ occurs. Thus,
    $$\Bigl\{(\widehat d_{\widehat j^{1*}(N)},\widehat d_{\widehat j^{2*}(N)},\dots,\widehat d_{\widehat j^{K*}(N)})\neq(d_{j^{1*}},d_{j^{2*}},\dots,d_{j^{K*}})\Bigr\}\subseteq E^c\,.$$ \\
    If we set $t=\delta_{\min}/2$ in \eqref{eq:probSingleBound}, we can conclude that 
    \begin{align*}
    P\Bigl((\widehat d_{\widehat j^{1*}(N)},\widehat d_{\widehat j^{2*}(N)},\dots,\widehat d_{\widehat j^{K*}(N)}) \neq (d_{j^{1*}},d_{j^{2*}},\dots,d_{j^{K*}})\Bigr) &\leq P(E^c) \\
    &\leq \sum_{j=1}^K P\Bigl(|\hat d_j-d_j|\geq \frac{\delta_{\min}}{2}\Bigr) \\
    &\leq 2\sum_{j=1}^K \exp\biggl\{-\frac{N_j \delta_{\min}^2}{8\sigma_j^2}\biggr\}.
    \end{align*}
    \end{proof}

\subsubsection{Proof of Proposition 1}

\begin{proof}

First, if $\mathcal{J}^*=\{1,\dots,K\}$, every arm is optimal and, trivially, $P(\hat j^{1*}\in\mathcal{J}^*)=1$. Thus, we focus on the case of $\mathcal{J}^*\subset\{1,\dots,K\}$. From Theorem 2, when $\kappa>0.5$, each arm is sampled infinitely often, that is $n_j(t)\rightarrow\infty$, for all $j=1,\dots,K$. Since observations within each arm are independent and identically distributed with finite mean $\mu_j$ the Law of Large Numbers implies that $\bar x_{n_j(t)}\rightarrow\mu_j$ for all $j$. \\
Now, define the suboptimal set $\mathcal{J}^-$ as the complement of $\mathcal{J}^*$, and let $d^{1*}=\underset{l=1,\dots,K}{\min} d_l$. By definition of $\mathcal{J}^*$, we have $d_j=d^{1*},\,\forall i\in\mathcal{J}^*$ and $d_i>d^{1*},\,\forall j\in\mathcal{J}^-$. Since $d^{1*}<\underset{i\in\mathcal{J}^-}{\min} d_i$, the almost sure convergence $\hat d_j(t)$ to $d_j$ implies that there exists $T<\infty$ such that for all $t\geq T$, $$\underset{j\in\mathcal{J}^*}{\max} \hat d_j(t)<\underset{i\in\mathcal{J}^-}{\min} \hat d_i(t)\,,$$ almost surely. Since the allocation rule selects the arm with smallest $\hat d_j(t)$, it follows that $P(\hat j^{1*}\in\mathcal{J}^*)\rightarrow1$.

\end{proof}

\subsubsection{Proof of Theorem 3}

\begin{proof}

By design construction, at least one patient is assigned to each arm, therefore $\widehat\rho_j,\tilde \rho_j\in(0,1)$. For $\kappa>0.5$, Theorem 2 ensures that all arms are visited infinitely often as $t\rightarrow\infty$, thus the information gain must remain of the same order for all arms so that no arm dominates in the long-run. Thus, for a given $\kappa>0.5$, $\Delta_{n_j,\kappa}(t)\sim\Delta_{n_i,\kappa}(t)$ for every pair of arms $i,j$. From Theorem 1, recall that $$\Delta_{n_j,\kappa}(t)\sim -\frac12(\gamma-\mu_j)^2\,Q_j\,n_j^\omega\,,$$ with \begin{itemize}
            \item $Q_j=\sigma_j^{2(1-p)}$ and $\omega=2\kappa-1$, for $0.5<\kappa<1$;
            \item $Q_j=\biggl(\frac{\sigma_j^{1-p}}{1+\sigma_j^{2-p}}\biggr)^2$ and $\omega=1$, for $\kappa=1$;
            \item $Q_j=\sigma_j^{-2}$ and $\omega=1$, for $\kappa>1$.
        \end{itemize}
        Therefore, as $t\rightarrow\infty$, $$\frac{n_j(t)^\omega}{n_i(t)^\omega}\rightarrow\frac{(\gamma-\mu_j)^2 Q_j}{(\gamma-\mu_i)^2 Q_i}$$ $$\iff$$ $$n_j(t)^\omega\rightarrow L\cdot \Bigl[(\gamma-\mu_j)^2 Q_j\Bigr]^{-1/\omega}\,,$$ with $L>0$ being a proportionality factor that does not depend on arm $j$. Since $t=\sum_{j=1}^K\,n_j(t)$, we can conclude that $$\widehat\rho_j(t)=\frac{n_j(t)}{t}\rightarrow\frac{\Bigl[(\gamma-\mu_j)^2 Q_j\Bigr]^{-1/\omega}}{\sum_{l=1}^K\Bigl[(\gamma-\mu_l)^2 Q_l\Bigr]^{-1/\omega}}=\tilde\rho_j\,.$$

\end{proof}

\subsubsection{Proof of Theorem 4}

\begin{proof}

Consider the generic arm $j$. The Shannon entropy of a multivariate normal distribution with $q$-dimensional mean vector $\bm\mu_j$ and $q\times q$ covariance matrix $\bm \Sigma_{n_j}$ is given by $$h(\pi_{n_j})=\frac{q}{2}+\frac{q}{2}\log(2\pi)+\frac{1}{2}\log\det(\bm\Sigma_{n_j})\,.$$

Remind that $\bm\Omega_{\phi,\,j}=\bm\Sigma_{\phi,\,j}^{-1}$ and $\bm\Omega_{n_j}=\bm\Sigma^{-1}_{n_j}$. To derive the (normalized) weight function, the following result is of particular interest:
\begin{align}
    (\bm\mu_j-\bm\gamma)'&\bm\Omega_{\phi,\,j}(\bm\mu_j-\bm\gamma) + (\bm\mu_j-\bar{\bm x}_{n_j})'\bm\Omega_{n_j}(\bm\mu_j-\bar{\bm x}_{n_j}) \nonumber\\
    &=\bm\mu_j'\widetilde{\bm\Omega}_j\bm\mu_j-2\bm\mu_j'\bm b_j+\bm\gamma'\bm\Omega_{\phi,\,j}\bm\gamma+\bar{\bm x}_{n_j}'\bm\Omega_{n_j}\bar{\bm x}_{n_j}\nonumber\\
    &=\bm\mu_j'\widetilde{\bm\Omega}_j\bm\mu_j-2\bm\mu_j'\bm b_j+\bm b_j'\widetilde{\bm\Sigma}_j\bm b_j-\bm b_j'\widetilde{\bm\Sigma}_j\bm b_j+\bm\gamma'\bm\Omega_{\phi,\,j}\bm\gamma+\bar{\bm x}_{n_j}'\bm\Omega_{n_j}\bar{\bm x}_{n_j}\nonumber\\
    &=(\bm\mu_j-\widetilde{\bm \mu}_j)'\widetilde{\bm\Omega}_j(\bm\mu_j-\widetilde{\bm \mu}_j)-\bm b_j'\widetilde{\bm\Sigma}_j\bm b_j+\bm\gamma'\bm\Omega_{\phi,\,j}\bm\gamma+\bar{\bm x}_{n_j}'\bm\Omega_{n_j}\bar{\bm x}_{n_j}\,, \label{eqSM:sumOfSquares_multi}
\end{align}
where $\bm b_j=\bm\Omega_{\phi,\,j}\bm\gamma+\bm\Omega_{n_j}\bar{\bm x}_{n_j}$, $\,\widetilde{\bm \mu}_j=\widetilde{\bm\Sigma}_j\bm b_j$ and $\widetilde{\bm\Omega}_j=\widetilde{\bm\Sigma}_j^{-1}=\bm\Omega_{\phi,\,j}+\bm\Omega_{n_j}$.

Then, using the result in \eqref{eqSM:sumOfSquares_multi}, the normalization condition $\int_{\mathbb{R}^q}\,\phi_\gamma(\bm\mu_j)\pi_{n_j}(\bm\mu_j)\,d\bm\mu_j=1$ is satisfied by
\begin{align*}
   \phi_\gamma(\bm \mu_j)&=C(\bm{\bar x}_{n_j},\bm\Sigma_j,\bm\gamma,n_j)\,\exp\biggl\{-\frac12\Bigl(\bm\mu_j-\bm\gamma\Bigr)'\bm\Omega_{\phi,\,j}\Bigl(\bm\mu_j-\bm\gamma\Bigr)\biggr\}\\
    &=\sqrt{\det{\bigl(\bm\Omega_{n_j}\widetilde{\bm\Sigma}_j\bigr)}}\,\times\,\exp\biggl\{-\frac12\biggl[(\bm\mu_j-\bm\gamma)'\bm\Omega_{\phi,\,j}(\bm\mu_j-\bm\gamma)+\bm{\widetilde\mu}_j'\widetilde{\bm\Sigma}_j\bm{\widetilde\mu}_j-\bm\gamma'\bm\Omega_{\phi,\,j}\bm\gamma-\bar{\bm x}_{n_j}'\Omega_{n_j}\bar{\bm x}_{n_j}\biggr]\biggr\}\,.
\end{align*}

Notably, $$\phi_\gamma(\bm \mu_j)\pi_{n_j}(\bm\mu_j)=(2\pi)^\frac{q}{2}\sqrt{\det\widetilde{\bm\Sigma}_j}\,\exp\biggl\{-\frac12(\bm\mu_j-\widetilde{\bm \mu}_j)'\widetilde{\bm\Omega}_j(\bm\mu_j-\widetilde{\bm \mu}_j)\biggr\}=\varphi(\bm\mu_j;\,\bm{\widetilde\mu}_j,\,\bm{\widetilde\Sigma}_j)\,,$$
with $\varphi(\bm\mu_j;\,\bm{\widetilde\mu}_j,\,\bm{\widetilde\Sigma}_j)$ being the pdf in $\bm\mu_j$ of a normal distribution with mean vector $\widetilde{\bm \mu}_j$ and precision matrix $\widetilde{\bm \Omega}_j$.

As a consequence, the weighted Shannon entropy can be written as
\begin{align*}
h^{\phi_\gamma}(\pi_{n_j})&=-\int_{\mathbb R^q}\,\phi_\gamma(\bm\mu_j)\pi_{n_j}(\bm\mu_j)\log\pi_{n_j}(\bm\mu_j)\,d\bm\mu_j  \\
&=\int_{\mathbb R^q}\,\varphi(\bm\mu_j;\,\bm{\widetilde\mu}_j,\,\bm{\widetilde\Sigma}_j)\,\biggl[\frac{q}{2}\log(2\pi)+\frac12\log\det(\bm\Sigma_{n_j}) +\frac12\Bigl(\bm\mu_j-\bm{\bar x}_{n_j}\Bigr)'\bm\Omega_{n_j}\Bigl(\bm\mu_j-\bm{\bar x}_{n_j}\Bigr)\biggr]\varphi(\bm\mu_j;\,\bm{\widetilde\mu}_j,\,\bm{\widetilde\Sigma}_j) \\
&=\frac{q}{2}\log(2\pi)+\frac12\log\det(\bm\Sigma_{n_j})+\frac12\,\int_{\mathbb R^q}\,\Bigl[(\bm\mu_j-\bm{\bar x}_{n_j})'\bm\Omega_{n_j}(\bm\mu_j-\bm{\bar x}_{n_j})\Bigr]\varphi(\bm\mu_j;\,\bm{\widetilde\mu}_j,\,\bm{\widetilde\Sigma}_j)\,d\bm\mu_j\,.
\end{align*}

To solve the integral in the last line of the previous expression, the following three steps are followed:
\begin{enumerate}
    \item exploiting the properties of the trace of a matrix: $$(\bm\mu_j-\bm{\bar x}_{n_j})'\bm\Omega_{n_j}(\bm\mu_j-\bm{\bar x}_{n_j})=tr\Bigl[\bm\Omega_{n_j}(\bm\mu_j-\bm{\bar x}_{n_j})(\bm\mu_j-\bm{\bar x}_{n_j})'\Bigr]\,.$$

    \item adding and subtracting a same quantity $\bm{\widetilde\mu}_j$:
    \begin{align*}
        (\bm\mu_j-\bm{\bar x}_{n_j})(\bm\mu_j-\bm{\bar x}_{n_j})'&=\bigl[(\bm\mu_j-\bm{\widetilde\mu}_j)(\bm{\widetilde\mu}_j-\bm{\bar x}_{n_j})\bigl]\,\bigl[(\bm\mu_j-\bm{\widetilde\mu}_j)(\bm{\widetilde\mu}_j-\bm{\bar x}_{n_j})\bigl]'\\
        &=(\bm\mu_j-\bm{\widetilde\mu}_j)(\bm\mu_j-\bm{\widetilde\mu}_j)'+2\bm\mu_j(\bm{\widetilde\mu}_j-\bm{\bar x}_{n_j})'+\bm{\bar x}_{n_j}\bm{\bar x}_{n_j}'-\bm{\widetilde\mu}_j\bm{\widetilde\mu}_j'\,.
    \end{align*}

    \item combining the previous results to obtain a sum of three tractable integrals: $$\int_{\mathbb R^q}\,\Bigl[(\bm\mu_j-\bm{\bar x}_{n_j})'\bm\Omega_{n_j}(\bm\mu_j-\bm{\bar x}_{n_j})\Bigr]\varphi(\bm\mu_j;\,\bm{\widetilde\mu}_j,\,\bm{\widetilde\Sigma}_j)\,d\bm\mu_j=\sum_{i=1}^3\,I_i\,,$$
    where
    
    \begin{align*}
        (i)\quad I_1=tr\biggl[\bm\Omega_{n_j}\int_{\mathbb R^q}\,(\bm\mu_j-\bm{\widetilde\mu}_j)(\bm\mu_j-\bm{\widetilde\mu}_j)'\varphi(\bm\mu_j;\,\bm{\widetilde\mu}_j,\,\bm{\widetilde\Sigma}_j)\,d\bm\mu_j\biggr]=tr\bigl[\bm\Omega_{n_j}\bm{\widetilde\Sigma}_j\bigr]
    \end{align*}

    \begin{align*}
        (ii)\quad I_2&=tr\biggl[2\,\bm\Omega_{n_j}\biggl(\int_{\mathbb R^q}\,\bm\mu_j\varphi(\bm\mu_j;\,\bm{\widetilde\mu}_j,\,\bm{\widetilde\Sigma}_j)\,d\bm\mu_j\biggr)\,(\bm{\widetilde\mu}_j-\bm{\bar x}_{n_j})'\biggr]\\
        &=tr\bigl[2\,\bm\Omega_{n_j}\bm{\widetilde\mu}_j(\bm{\widetilde\mu}_j-\bm{\bar x}_{n_j})'\bigr]
    \end{align*}

    \begin{align*}
        (iii)\quad I_3&=tr\biggl[\bm\Omega_{n_j}(\bm{\bar x}_{n_j}\bm{\bar x}_{n_j}'-\bm{\widetilde\mu}_j\bm{\widetilde\mu}_j')\int_{\mathbb R^q}\,\varphi(\bm\mu_j;\,\bm{\widetilde\mu}_j,\,\bm{\widetilde\Sigma}_j)\,d\bm\mu_j\biggr]\\
        &=tr\bigl[\bm\Omega_{n_j}(\bm{\bar x}_{n_j}\bm{\bar x}_{n_j}'-\bm{\widetilde\mu}_j\bm{\widetilde\mu}_j')\bigr]
    \end{align*}
\end{enumerate}

Then,
\begin{align*}
    h^{\phi_\gamma}(\pi_{n_j})&=\frac{q}{2}\log(2\pi)+\frac12\log\det(\bm\Sigma_{n_j})+\frac12\sum_{i=1}^3\,I_i\\
    &=\frac{q}{2}\log(2\pi)+\frac12\log\det(\bm\Sigma_{n_j})+\frac12 tr\bigl[\bm\Omega_{n_j}\bm{\widetilde\Sigma}_j\bigr]+\frac12 tr\bigl[(\bm{\widetilde\mu}_j-\bm{\bar x}_{n_j})'\bm\Omega_{n_j}(\bm{\widetilde\mu}_j-\bm{\bar x}_{n_j})\bigr]\\
    &=\frac{q}{2}\log(2\pi)+\frac12\log\det(\bm\Sigma_{n_j})+\frac12 tr\bigl[\bm\Omega_{n_j}\bm{\widetilde\Sigma}_j\bigr]+\frac12 (\bm{\widetilde\mu}_j-\bm{\bar x}_{n_j})'\bm\Omega_{n_j}(\bm{\widetilde\mu}_j-\bm{\bar x}_{n_j})
\end{align*}
and, after replacing the values of $\bm{\widetilde\mu}_j$ and $\bm{\widetilde\Sigma}_j$,
\begin{equation*}
\begin{split}
    \Delta_{n_j}&=\frac{q}{2}-\frac12\text{tr}\Bigl[\bm\Omega_{n_j}\bigl(\bm\Omega_{n_j}+\bm\Omega_{\phi,\,j}\bigr)^{-1}\Bigr]\\
    &-\frac12\Bigl[\bigl(\bm\gamma-\bar{\bm x}_{n_j}\bigr)'\bm\Omega_{\phi,\,j}\bigl(\bm\Omega_{n_j}+\bm\Omega_{\phi,\,j}\bigr)^{-1}\bm\Omega_{n_j}\bigl(\bm\Omega_{n_j}+\bm\Omega_{\phi,\,j}\bigr)^{-1}\bm\Omega_{\phi,\,j}\bigl(\bm\gamma-\bar{\bm x}_{n_j}\bigr)\Bigr]\,.
\end{split}
\end{equation*}

If $\bm\Omega_{\phi,\,j}=\bm\Sigma_{\phi,\,j}^{-1}=n_j^\kappa\,\bm\Sigma_j^{-1}$, the result immediately follows.

\end{proof}

\subsubsection{Proof of Theorem 5}

\begin{proof}
By Theorem 1, for $\kappa>0.5$, the marginal posteriors for $\mu_j$'s concentrate at the true parameter values and the two best arms (i.e., $j^{1*}$ and $j^{2*}$) are correctly identified as $N\rightarrow\infty$. Thus, as $N\rightarrow\infty$, $$\widehat\pi_{\widehat j^{1*},\widehat j^{2*}}=\mathbb{P}\bigl(|\mu_{\widehat j^{1*}}-\gamma| < |\mu_{\widehat j^{2*}}-\gamma|\,\bigl|\text{data}\bigr)\rightarrow \mathbb{P}\bigl(|\mu_{j^{1*}}-\gamma|<|\mu_{j^{2*}}-\gamma|\,\bigl|\text{data}\bigr)=1\,,$$ since, under $H_1$, the event $\{|\mu_{j^{1*}}-\gamma|<|\mu_{j^{2*}}-\gamma|\}$ is almost sure. Consequently, for any fixed cut-off probability $\eta_\alpha\in(0,1)$, $\mathbb{P}\bigl(\bigl\{\widehat\pi_{\widehat j^{1*},\widehat j^{2*}}>\eta_\alpha\bigr\}\bigl)\rightarrow1\,,$ and both the conditional power and two-components power tend to $1$.
\end{proof}

\subsection{Asymmetric weight function}
\label{subsec:asymmWeight}

The allocation rule for single endpoints presented in the main text has the property to give the same importance to values of the sample mean being above or below the target mean of a same constant value $x\in\mathbb R$. More formally, if $\bar x_{n_j}^-=\gamma-x$ and $\bar x_{n_j}^+=\gamma+x$, then $\phi(x_{n_j}^-)=\phi(x_{n_j}^+)$ and $\Delta_{n_j}(x_{n_j}^-)=\Delta_{n_j}(x_{n_j}^+)$, \textit{ceteris paribus}. However, the assumption of symmetric behaviour around the target may not be tenable whenever the investigator desires to weight values above and below the target in a different way.

To comply with this different context of the study, we propose the following family of piece-wise asymmetric weight functions:
\begin{equation}
\label{eq:asymmWeight}
\phi_\gamma(\mu_j)=C(\bar x_{n_j},\sigma^2_j,\gamma,n_j)\,\cdot\,\begin{cases}
    \exp\biggl\{-\frac12\Bigl(\frac{\mu_j-\gamma}{\sigma_{\phi_a,\,j}}\Bigr)^2\biggr\}\quad\mu_j\leq\gamma\\ 
    \exp\biggl\{-\frac12\Bigl(\frac{\mu_j-\gamma}{\sigma_{\phi_b,\,j}}\Bigr)^2\biggr\}\quad\mu_j>\gamma\\
\end{cases}\,,
\end{equation}
where both pieces are proportional to truncated normal distributions with mean $\gamma$, whereas the left piece is truncated in $(-\infty,\gamma]$ with variance $\sigma^2_{\phi_a,j}=a^2\frac{\sigma_j^2}{n}$ while the right piece is truncated in $(\gamma,+\infty)$ with variance $\sigma^2_{\phi_b,j}=b^2\frac{\sigma_j^2}{n}$, with $a,b\in\mathbb R^+$, and with $C(\bar x_{n_j},\sigma^2_j,\gamma,n_j)$ such that $\int_{\mathbb{R}}\,\phi_\gamma(\mu_j)\pi_{n_j}(\mu_j)\,d\mu_j=1$. Notice that the symmetric weight function can be seen as a special case of the asymmetric one (cfr. \eqref{eq:asymmWeight}) when $a=b=1$. In Theorem \ref{th:asymmWeight}, we illustrate the derivation of the corresponding information gain, $\Delta_{n_j}^{a,b}$.

\setcounter{theorem}{5}
\begin{theorem}
\label{th:asymmWeight}
Consider the same assumption as in Theorem 1 (cfr. main text), but assuming a family of weight functions as the one in \eqref{eq:asymmWeight}. Then the corresponding expression of the information gain is given by 
\begin{equation}
\label{eq:infoGainAsymm}
    \Delta_{n_j}^{a,b}=\frac12-\frac{\dot\sigma_{a,\,j}^2+(\dot\mu_{a,\,j}-\bar x_{n_j})^2}{2\sigma_{n_j}^2}\cdot\frac{D_{a,\,j}}{D_{a,\,j}+D_{b,\,j}}-\frac{\dot\sigma_{b,\,j}^2+(\dot\mu_{b,\,j}-\bar x_{n_j})^2}{2\sigma_{n_j}^2}\cdot\frac{D_{b,\,j}}{D_{a,\,j}+D_{b,\,j}}\,,
\end{equation}
with
\begin{equation*}
    D_{a,\,j}=\tilde\sigma_{a,\,j}\cdot\exp\Biggl\{-\frac12 \Bigl(\frac{\gamma}{\sigma_{\phi_a,\,j}}\Bigr)^2+\frac12 \Bigl(\frac{\tilde\mu_{a,\,j}}{\tilde\sigma_{a,\,j}}\Bigr)^2\Biggr\}\cdot\Phi(z_{a,\,j})
\end{equation*}
\begin{equation*}
    D_{b,\,j}=\tilde\sigma_{b,\,j}\cdot\exp\Biggl\{-\frac12 \Bigl(\frac{\gamma}{\sigma_{\phi_b,\,j}}\Bigr)^2+\frac12 \Bigl(\frac{\tilde\mu_{b,\,j}}{\tilde\sigma_{b,\,j}}\Bigr)^2\Biggr\}\cdot\biggl[1-\Phi(z_{b,\,j})\biggr]
\end{equation*}
\begin{equation*}
    z_{i,\,j}=\frac{\gamma-\tilde\mu_{i,\,j}}{\tilde\sigma_{i,\,j}}
    \,, \quad
    \tilde\mu_{i,\,j}=\frac{\bar x_{n_j}\sigma_{\phi_i,\,j}^2\,+\,\gamma\sigma_{n_j}^2}{\sigma_{\phi_i,\,j}^2\,+\,\sigma_{n_j}^2}\,, \quad \tilde\sigma_{i,\,j}^2=\frac{\sigma_{\phi_i,\,j}^2\,\cdot\,\sigma_{n_j}^2}{\sigma_{\phi_i,\,j}^2\,+\,\sigma_{n_j}^2}\,,\quad i=\{a,\,b\}
\end{equation*}
\begin{equation*}
    \dot\mu_{a,\,j}=\tilde\mu_{a,\,j}-\tilde\sigma_{a,\,j}\frac{\varphi(z_{a,\,j})}{\Phi(z_{a,\,j})}\,, \qquad \dot\sigma^2_{a,\,j}=\tilde\sigma^2_{a,\,j}\cdot\Biggl[1-\frac{z_{a,\,j}\cdot\varphi(z_{a,\,j})}{\Phi(z_{a,\,j})}-\Bigl(\frac{\varphi(z_{a,\,j})}{\Phi(z_{a,\,j})}\Bigr)^2\Biggr]\,,
\end{equation*}
\begin{equation*}
    \dot\mu_{b,\,j}=\tilde\mu_{b,\,j}+\tilde\sigma_{b,\,j}\frac{\varphi(z_{b,\,j})}{1-\Phi(z_{b,\,j})}\,, \qquad \dot\sigma^2_{b,\,j}=\tilde\sigma^2_{b,\,j}\cdot\Biggl[1+\frac{z_{b,\,j}\cdot\varphi(z_{b,\,j})}{1-\Phi(z_{b,\,j})}-\Bigl(\frac{\varphi(z_{b,\,j})}{1-\Phi(z_{b,\,j})}\Bigr)^2\Biggr]\,,
\end{equation*}
and $\Phi(\cdot)$ and $\varphi(\cdot)$ being, respectively, the cdf and the pdf of a standard normal distribution.

\end{theorem}

\begin{figure}[ht]
    \centering
    \includegraphics[width=0.75\textwidth]{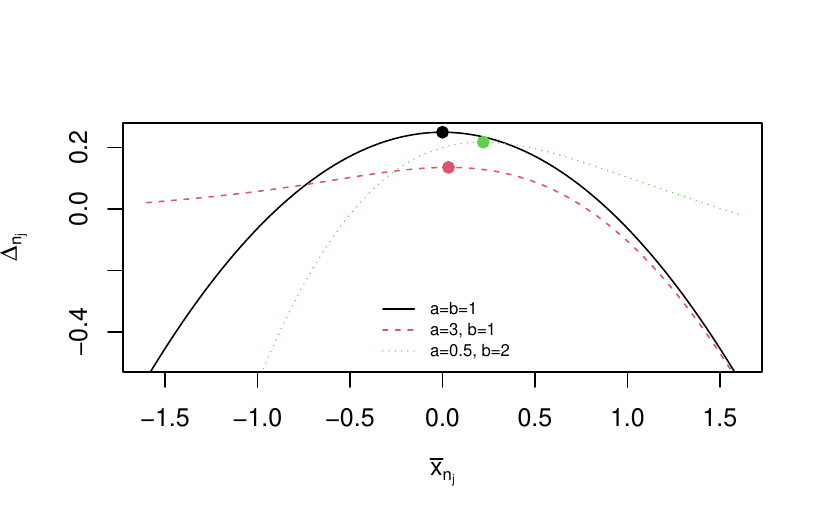}
    \vspace{-0.5cm}
    \caption{Behaviour of the asymmetric information gain as function of the current sample mean when $\gamma=0$. Filled dots show the maximum value assumed by each curve.}
    \label{fig:infoGainAsymm_noOptimal}
\end{figure}

The proof is provided in Section \ref{SM:proofAsymm}. When $a\neq b$, the information gain in \ref{eq:infoGainAsymm} does not generally attain its maximum in correspondence of $\bar x_{n_j}=\gamma$ (cfr. Figure \ref{fig:infoGainAsymm_noOptimal}). Therefore, one needs to find combinations of values $a$ and $b$ that lead to an information gain having this desirable property (henceforth, referred to as optimal combinations). 

\subsubsection{Optimal combinations of a and b}

\begin{figure}[ht]
    \centering
    \includegraphics[width=0.75\textwidth]{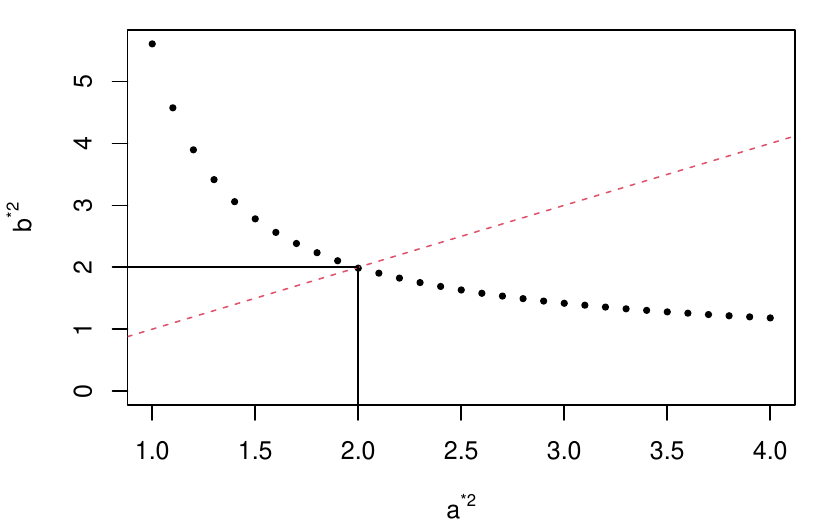}
    \caption{Optimal values $b^{*^2}$ for each considered value $a^{*^2}$. Combinations $(a^*,b^*)$ below the red dashed line are the only admissible values under the condition $a*>b*$.}
    \label{fig:optimalBs}
\end{figure}

Here, we focus on the case where $a>b$ is required. Once we fix $a=a^*$, the corresponding value $b^*$ leading to an information gain which attains its maximum in $\bar x_{n_j}=\gamma$ is such that
\begin{equation}
\label{eq:optimCond}
    b^*=\underset{0<b<a^*}{\text{argmin}}\,\biggl|\underset{\bar x_{n_j}}{\text{max}}\,\Delta_{n_j}^{a^*,b}-\gamma\biggr|\,.
\end{equation}
The closed-form derivation of $\frac{\partial}{\partial \bar x_{n_j}}\,\Delta^{a,b}_{n_j}$ is cumbersome, thus we solve \eqref{eq:optimCond} via numerical optimization.  
Figure \ref{fig:optimalBs} suggests that under the constraint $a>b$, the value $b^*$ does exist and is unique if and only if $a^*>\sqrt{2}$. 

\begin{figure}[ht]
    \centering
    \includegraphics[width=0.7\textwidth]{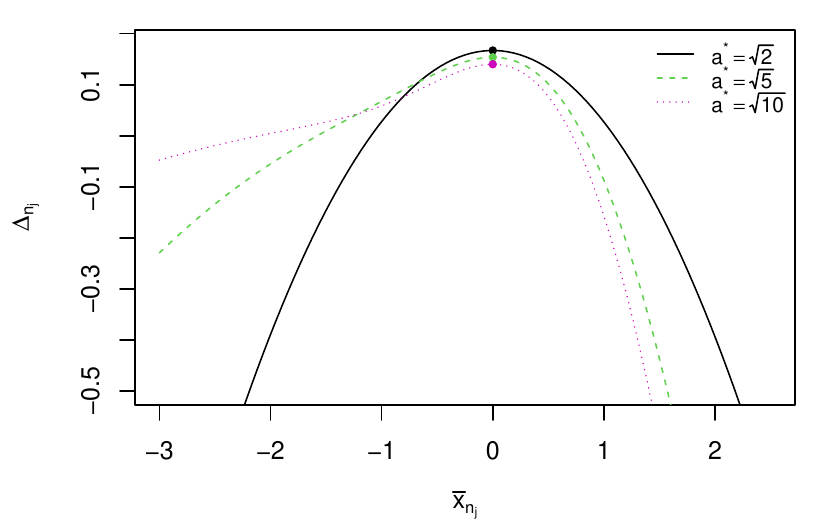}
    \caption{Behaviour of the asymmetric information gain for different values of $a^*$. The other parameters are fixed to $\gamma=0$, $n_j=10$, $\sigma_j=2$ and $\kappa=1$.}
    \label{fig:infoGainAsymm}
\end{figure}

The parameters $a^*$ and $b^*$ can be thus characterized as \textit{asymmetry parameters}. For example, the larger $a^*>\sqrt{2}$ (analogously, $b^*>\sqrt{2}$), the more skewed to the left (right) the resulting information gain and the higher the penalization of values of $\bar x_{n_j}$ higher (lower) than $\gamma$ (cfr. Figure \ref{fig:infoGainAsymm}).
The practical selection of $a^*$ (or $b^*$, when $b>a$) should depend on the objective of the trial.

\begin{figure}[ht]
    \centering
    \includegraphics[width=1\textwidth]{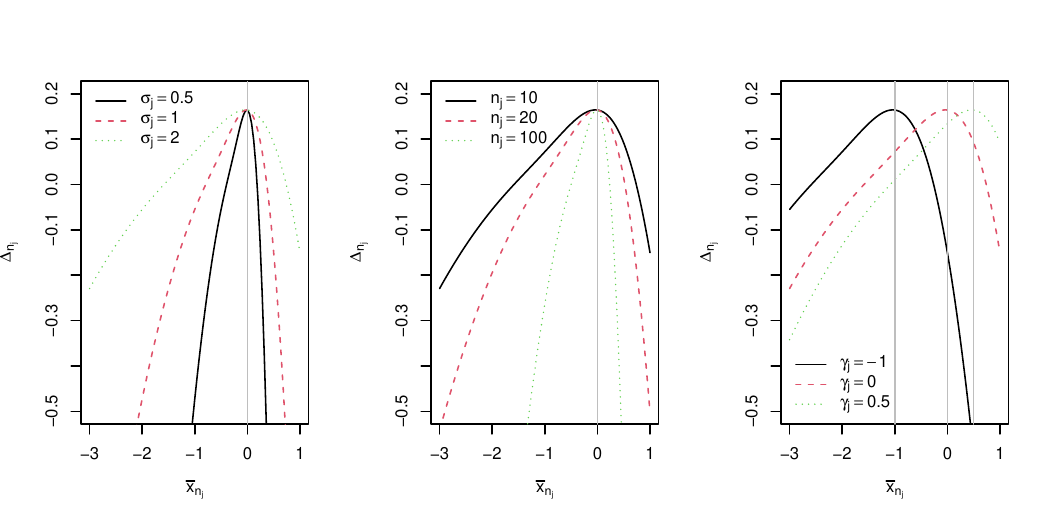}
    \vspace{-0.7cm}
    \caption{Asymmetric information gain $\Delta_{n_j}^{a,b}$ ($a^*=2.236$, $b^*=0.906$) as function of the sample mean for different values of either $\sigma_j$ (\textit{left panel}), $n_j$ (\textit{center panel}) or $\gamma$ (\textit{right panel}). Grey vertical lines help visualizing the maxima of the curves.}
    \label{fig:invarianceOptimalAB}
\end{figure}

Finally, Figure \ref{fig:invarianceOptimalAB} shows that the optimal combination of $a$ and $b$ remains the same regardless the value of $\sigma_j$, $n_j$ and $\gamma$.
\vspace{-0.7cm}

\subsubsection{Proof of Theorem \ref{th:asymmWeight}}
\label{SM:proofAsymm}

\begin{proof}

Consider the generic arm $j$. First, one needs to derive the normalization constant $C(\bar x_{n_j},\sigma^2_j,\gamma,n_j)$ such that $$\int_{\mu_j\leq\gamma}\,\phi_\gamma(\mu_j)\pi_{n_j}(\mu_j)\,d\mu_j + \int_{\mu_j>\gamma}\,\phi_\gamma(\mu_j)\pi_{n_j}(\mu_j)\,d\mu_j = 1$$ $$\iff$$ $$C(\bar x_{n_j},\sigma^2_j,\gamma,n_j)\cdot\bigl[I_{a,\,1} + I_{b,\,1}\bigr]=1\,,$$
where 
\begin{itemize}
    \item $I_{a,\,1}=\displaystyle\int_{\mu_j\leq\gamma}\,\exp\biggl\{-\frac12\Bigl(\frac{\mu_j-\gamma}{\sigma_{\phi_a,\,j}}\Bigr)^2\biggr\}\pi_{n_j}(\mu_j)\,d\mu_j\,;$
    \item $I_{b,\,1}=\displaystyle\int_{\mu_j>\gamma}\,\exp\biggl\{-\frac12\Bigl(\frac{\mu_j-\gamma}{\sigma_{\phi_b,\,j}}\Bigr)^2\biggr\}\pi_{n_j}(\mu_j)\,d\mu_j\,.$
\end{itemize}

Exploiting the equation in \eqref{eqSM:sumOfSquares}, the first integral can be re-written as
\begin{align*}
    I_{a,\,1}&=\bigl(2\pi\sigma_{n_j}^2\bigr)^{-\frac12}\int_{\mu_j\leq\gamma}\,\exp\biggl\{-\frac12\biggl[\Bigl(\frac{\mu_j-\gamma}{\sigma_{\phi_a,\,j}}\Bigr)^2+ \Bigl(\frac{\mu_j-\bar x_{n_j}}{\sigma_{n_j}}\Bigr)^2 \biggr]\biggr\} \,d\mu_j\\
    &=\frac{\tilde\sigma_{a,\,j}}{\sigma_{n_j}}\cdot\exp\biggl\{-\frac12 \biggl[\Bigl(\frac{\bar x_{n_j}}{\sigma_{n_j}}\Bigr)^2+\Bigl(\frac{\gamma}{\sigma_{\phi_a,\,j}}\Bigr)^2- \Bigl(\frac{\widetilde\mu_{a,\,j}}{\tilde\sigma_{a,\,j}}\Bigr)^2\biggr]\biggr\}\int_{\mu_j\leq\gamma}\,\bigl(2\pi\widetilde\sigma_{a,\,j}^2\bigr)^{-\frac12}\exp\biggl\{-\frac12\Bigl(\frac{\mu_j-\widetilde\mu_{a,\,j}}{\widetilde\sigma_{a,\,j}}\Bigr)^2\biggr\}\,d\mu_j\\
    &=\frac{\widetilde\sigma_{a,\,j}}{\sigma_{n_j}}\cdot\exp\biggl\{-\frac12 \biggl[\Bigl(\frac{\bar x_{n_j}}{\sigma_{n_j}}\Bigr)^2+\Bigl(\frac{\gamma}{\sigma_{\phi_a,\,j}}\Bigr)^2- \Bigl(\frac{\tilde\mu_{a,\,j}}{\tilde\sigma_{a,\,j}}\Bigr)^2\biggr]\biggr\}\cdot\Phi\bigl(z_{a,\,j}\bigr)\,,
\end{align*}
and, similarly,
\begin{equation*}
    I_{b,\,1}=\frac{\tilde\sigma_{b,\,j}}{\sigma_{n_j}}\cdot\exp\biggl\{-\frac12 \biggl[\Bigl(\frac{\bar x_{n_j}}{\sigma_{n_j}}\Bigr)^2-\Bigl(\frac{\gamma}{\sigma_{\phi_b,\,j}}\Bigr)^2- \Bigl(\frac{\tilde\mu_{b,\,j}}{\tilde\sigma_{b,\,j}}\Bigr)^2\biggr]\biggr\}\cdot\biggl[1-\Phi\bigl(z_{b,\,j}\bigr)\biggr]\,,
\end{equation*}
where $\Phi(\cdot)$ is the cdf of a standard normal distribution, $z_{i,\,j}=\frac{\gamma-\tilde\mu_{i,\,j}}{\tilde\sigma_{i,\,j}}$, $\tilde\mu_{i,\,j}=\frac{\bar x_{n_j}\sigma_{\phi_i,\,j}^2\,+\,\gamma\sigma_{n_j}^2}{\sigma_{\phi_i,\,j}^2\,+\,\sigma_{n_j}^2}$ and $\tilde\sigma_{a,\,j}^2=\frac{\sigma_{\phi_i,\,j}^2\,\cdot\,\sigma_{n_j}^2}{\sigma_{\phi_i,\,j}^2\,+\,\sigma_{n_j}^2}$, $i=\{a,\,b\}$.

After some computations, the normalization constant of the weight function $\phi_\gamma(\mu_j)$ can be expressed as $$C(\bar x_{n_j},\sigma^2_j,\gamma,n_j)=\frac{\sigma_{n_j}\exp\Bigl\{\frac12 \Bigl(\frac{\bar x_{n_j}}{\sigma_{n_j}}\Bigr)^2\Bigr\}}{D_{a,\,j}+D_{b,\,j}}\,,$$
where $D_{a,\,j}=\tilde\sigma_{a,\,j}\cdot\exp\biggl\{-\frac12 \Bigl(\frac{\gamma}{\sigma_{\phi_a,\,j}}\Bigr)^2+\frac12 \Bigl(\frac{\tilde\mu_{a,\,j}}{\tilde\sigma_{a,\,j}}\Bigr)^2\biggr\}\cdot\Phi(z_{a,\,j})$ and $D_{b,\,j}=\tilde\sigma_{b,\,j}\cdot\exp\biggl\{-\frac12 \Bigl(\frac{\gamma}{\sigma_{\phi_b,\,j}}\Bigr)^2+\frac12 \Bigl(\frac{\tilde\mu_{b,\,j}}{\tilde\sigma_{b,\,j}}\Bigr)^2\biggr\}\cdot\biggl[1-\Phi(z_{b,\,j})\biggr]$.

\bigskip
\bigskip

Now, let's derive the weighted Shannon entropy:
\begin{equation*}
h^{\phi_\gamma}(\pi_{n_j})=-\int_{\mathbb R}\,\phi_\gamma(\mu_j)\pi_{n_j}(\mu_j)\log\pi_{n_j}(\mu_j)\,d\mu_j=-\bigl[I_{a,\,2} + I_{b,\,2}\bigr]\,,
\end{equation*}
where 
\begin{itemize}
    \item $I_{a,\,2}=\displaystyle\int_{\mu_j\leq\gamma}\,\phi_\gamma(\mu_j)\pi_{n_j}(\mu_j)\log\pi_{n_j}(\mu_j)\,d\mu_j\,;$
    \item $I_{b,\,2}=\displaystyle\int_{\mu_j>\gamma}\,\phi_\gamma(\mu_j)\pi_{n_j}(\mu_j)\log\pi_{n_j}(\mu_j)\,d\mu_j\,.$
\end{itemize}

To solve the two integrals, the following four results are particularly useful:
\begin{enumerate}
    \item Once again, exploiting the equation in \eqref{eqSM:sumOfSquares} and assuming that $\varphi(\mu_j;\widetilde\mu_{i,\,j},\,\widetilde\sigma_{i,\,j}^2)$ is the pdf in $\mu_{i,\,j}$ of a normal distribution with mean $\widetilde\mu_{i,\,j}$ and variance $\widetilde\sigma_{i,\,j}^2$ ($i=\{a,b\}$), one obtains that
    \begin{align*}
    \phi_\gamma(\mu_j)\pi_{n_j}(\mu_j)&=\frac{1}{\sqrt{2\pi}(D_{a,\,j}+D_{b,\,j})}\cdot
    \begin{cases}
    \exp\biggl\{-\frac12 \biggl[\Bigl(\frac{\mu_j-\widetilde\mu_{a,\,j}}{\widetilde\sigma_{a,\,j}}\Bigr)^2+\Bigl(\frac{\gamma}{\sigma_{\phi_a,\,j}}\Bigr)^2- \Bigl(\frac{\widetilde\mu_{a,\,j}}{\tilde\sigma_{a,\,j}}\Bigr)^2\biggr]\biggr\}\quad\mu_j\leq\gamma \\ \\
    \exp\biggl\{-\frac12 \biggl[\Bigl(\frac{\mu_j-\widetilde\mu_{b,\,j}}{\widetilde\sigma_{b,\,j}}\Bigr)^2-\Bigl(\frac{\gamma}{\sigma_{\phi_b,\,j}}\Bigr)^2- \Bigl(\frac{\widetilde\mu_{b,\,j}}{\tilde\sigma_{b,\,j}}\Bigr)^2\biggr]\biggr\}\quad\mu_j>\gamma \\
    \end{cases}\\ \\
    &=\frac{1}{D_{a,\,j}+D_{b,\,j}}\cdot
    \begin{cases} 
    D_{a,\,j}\cdot\frac{\varphi(\mu_j;\,\widetilde\mu_{a,\,j},\,\widetilde\sigma_{a,\,j}^2)}{\Phi(z_{a,\,j})}\quad\mu_j\leq\gamma \\ \\
    D_{b,\,j}\cdot\frac{\varphi(\mu_j;\,\widetilde\mu_{b,\,j},\,\widetilde\sigma_{b ,\,j}^2)}{1-\Phi(z_{b,\,j})}\quad\mu_j>\gamma \\
    \end{cases}\,.
    \end{align*}

    \item  Let's define two probability density functions:
    \begin{enumerate}
        \item $\varphi_{(-\infty,\,\gamma]}(\mu_j;\,\widetilde\mu_{a,\,j},\,\widetilde\sigma_{a,\,j}^2)=\frac{\varphi(\mu_j;\,\widetilde\mu_{a,\,j},\,\widetilde\sigma_{a,\,j}^2)}{\Phi(z_{a,\,j})}$, which is the pdf of a $N(\widetilde\mu_{a,\,j},\,\widetilde\sigma_{a,\,j}^2)$ truncated in $(-\infty,\,\gamma]$ and computed in $\mu_j$;
        \item $\varphi_{[\gamma,\,\infty)}(\mu_j;\,\widetilde\mu_{b,\,j},\,\widetilde\sigma_{b ,\,j}^2)=\frac{\varphi(\mu_j;\,\widetilde\mu_{b,\,j},\,\widetilde\sigma_{b ,\,j}^2)}{1-\Phi(z_{b,\,j})}$, which is the pdf of a $N(\widetilde\mu_{b,\,j},\,\widetilde\sigma_{b,\,j}^2)$ truncated in $[\gamma,\,\infty)$ and computed in $\mu_j$.
    \end{enumerate}

    \item In addition, let's define
    \begin{enumerate}
        \item the truncated means: $$\dot\mu_{a,\,j}=\int_{\mu_j\leq\gamma}\,\mu_j\,\cdot\,\varphi_{(-\infty,\,\gamma]}(\mu_j;\,\widetilde\mu_{a,\,j},\widetilde\sigma_{a,\,j}^2)\,d\mu_j=\tilde\mu_{a,\,j}-\tilde\sigma_{a,\,j}\frac{\varphi(z_{a,\,j})}{\Phi(z_{a,\,j})}$$
        $$\dot\mu_{b,\,j}=\int_{\mu_j>\gamma}\,\mu_j\,\cdot\,\varphi_{[\gamma,\,\infty)}(\mu_j;\,\widetilde\mu_{b,\,j},\widetilde\sigma_{b,\,j}^2)\,d\mu_j=\tilde\mu_{b,\,j}+\tilde\sigma_{b,\,j}\frac{\varphi(z_{b,\,j})}{1-\Phi(z_{b,\,j})}\,.$$
        
        \item the truncated variances:
        \begin{align*}
        \dot\sigma_{a,\,j}^2&=\int_{\mu_j\leq\gamma}\,(\mu_j-\dot\mu_{a,\,j})^2\,\cdot\,\varphi_{(-\infty,\,\gamma]}(\mu_j;\,\widetilde\mu_{a,\,j},\widetilde\sigma_{a,\,j}^2)\,d\mu_j\\
        &=\tilde\sigma^2_{a,\,j}\cdot\Biggl[1-\frac{z_{a,\,j}\cdot\varphi(z_{a,\,j})}{\Phi(z_{a,\,j})}-\Bigl(\frac{\varphi(z_{a,\,j})}{\Phi(z_{a,\,j})}\Bigr)^2\Biggr]
        \end{align*}
        
        \begin{align*}
        \dot\sigma_{b,\,j}^2&=\int_{\mu_j>\gamma}\,(\mu_j-\dot\mu_{b,\,j})^2\,\cdot\,\varphi_{[\gamma,\,\infty)}(\mu_j;\,\widetilde\mu_{b,\,j},\widetilde\sigma_{b,\,j}^2)\,d\mu_j\\
        &=\tilde\sigma^2_{b,\,j}\cdot\Biggl[1+\frac{z_{b,\,j}\cdot\varphi(z_{b,\,j})}{1-\Phi(z_{b,\,j})}-\Bigl(\frac{\varphi(z_{b,\,j})}{1-\Phi(z_{b,\,j})}\Bigr)^2\Biggr]
        \end{align*}
    \end{enumerate}

    \item For $i=\{a,b\}$: $(\mu_j-\bar x_{n_j})^2=(\mu_j-\dot\mu_{i,\,j})^2+2\mu_j(\dot\mu_{i,\,j}-\bar x_{n_j})-\dot\mu_{i,\,j}^2+\bar x_{n_j}^2\,.$
\end{enumerate}

Combining the previous four results, one obtains that
\begin{align*}
    I_{a,\,2}&=-\frac{D_{a,\,j}}{(D_{a,\,j}+D_{b,\,j})}\int_{\mu_j\leq\gamma}\,\varphi_{(-\infty,\,\gamma]}(\mu_j;\,\widetilde\mu_{a,\,j},\,\widetilde\sigma_{a,\,j}^2)\cdot\log\pi_{n_j}(\mu_j)\,d\mu_j\\
    &=-\frac{D_{a,\,j}}{D_{a,\,j}+D_{b,\,j}}\cdot\biggl[\frac12\log(2\pi\sigma_{n_j}^2) + \frac{1}{2\sigma_{n_j}^2}\int_{\mu_j\leq\gamma}\,(\mu_j-\bar x_{n_j})^2\cdot\varphi_{(-\infty,\,\gamma]}(\mu_j;\,\widetilde\mu_{a,\,j},\,\widetilde\sigma_{a,\,j}^2)\,d\mu_j\biggr]\\
    &=-\frac{D_{a,\,j}}{D_{a,\,j}+D_{b,\,j}}\cdot\biggl[\frac12\log(2\pi\sigma_{n_j}^2) + \frac{\dot\sigma^2_{a,\,j}+(\dot\mu^2_{a,\,j}-\bar x_{n_j})^2}{2\sigma_{n_j}^2}\biggr]\,.
\end{align*}
and, similarly,
\begin{equation*}
    I_{b,\,2}=-\frac{D_{b,\,j}}{D_{a,\,j}+D_{b,\,j}}\cdot\biggl[\frac12\log(2\pi\sigma_{n_j}^2) + \frac{\dot\sigma^2_{b,\,j}+(\dot\mu^2_{b,\,j}-\bar x_{n_j})^2}{2\sigma_{n_j}^2}\biggr]\,.
\end{equation*}

Finally, 
\begin{align*}
    h^{\phi_\gamma}(\pi_{n_j})&=-\bigl[I_{a,\,2} + I_{b,\,2}\bigr]\\
    &=\frac12\log(2\pi\sigma_{n_j}^2)+\frac{\dot\sigma_{a,\,j}^2+(\dot\mu_{a,\,j}-\bar x_{n_j})^2}{2\sigma_{n_j}^2}\cdot\frac{D_{a,\,j}}{D_{a,\,j}+D_{b,\,j}}+\frac{\dot\sigma_{b,\,j}^2+(\dot\mu_{b,\,j}-\bar x_{n_j})^2}{2\sigma_{n_j}^2}\cdot\frac{D_{b,\,j}}{D_{a,\,j}+D_{b,\,j}}\\
\end{align*}

The resulting information gain immediately follows.

\end{proof}

\subsection{Multivariate generalization: how information gain changes as function of the main parameters (case of $q=2$)}

\subsubsection{Information gain in the bivariate case}

In the bivariate case ($q=2$), the information gain can be expressed in scalar terms as
$$\Delta_{n_j}=\frac{q}{2}\frac{n_j^\kappa}{n_j^\kappa+n_j}-\frac{\bigl(\gamma^{\ms(1)}-\bar x_{n_j}^{\ms(1)}\bigr)^2\sigma_j^{\ms(11)}-2\bigl(\gamma^{\ms(1)}-\bar x_{n_j}^{\ms(1)}\bigr)\bigl(\gamma^{\ms(2)}-\bar x_{n_j}^{\ms(2)}\bigr)\sigma_j^{\ms(12)}+\bigl(\gamma^{\ms(2)}-\bar x_{n_j}^{\ms(2)}\bigr)^2\sigma_j^{\ms(22)}}{\sigma_j^{\ms(11)}\sigma_j^{\ms(22)}-\sigma_j^{{\ms(12)}^2}}\Biggl(\frac{n_j^{\kappa+\frac12}}{n_j^\kappa+n_j}\Biggr)^2\,,$$
where $\bm\gamma=\bigr(\gamma^{\ms(1)},\gamma^{\ms(2)}\bigl)'$, $\bar {\bm x}_{n_j}=\bigr(\bar x_{n_j}^{\ms(1)},\bar x_{n_j}^{\ms(2)}\bigl)'$ and $\bm\Sigma_j=\biggl(\begin{smallmatrix}\sigma_j^{\ms(11)} & \sigma_j^{\ms(12)}\\ \\ \sigma_j^{\ms(12)} & \sigma_j^{\ms(22)}\end{smallmatrix}\biggr)$ with variances in the diagonal and covariance in the off-diagonal.

\subsubsection{Means and variances}

\begin{figure}[ht]
    \centering
    \includegraphics[width=0.7\textwidth]{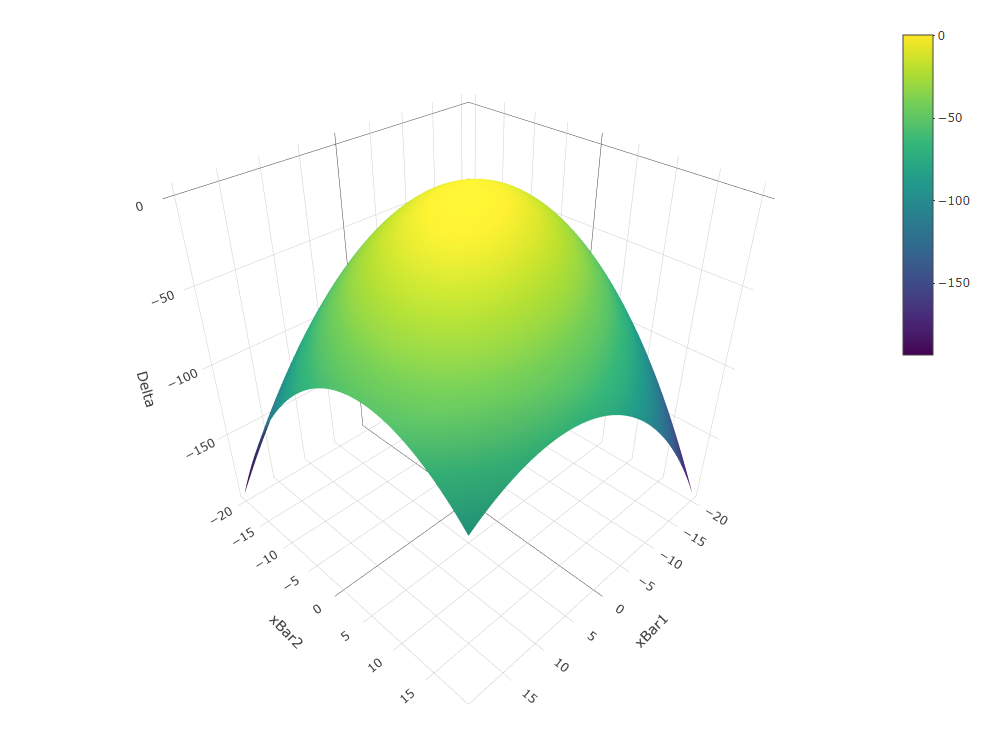}
    \caption{Information gain as function of the two sample means. Other parameters are set as follows: $\sigma_j^{(11)}=4$, $\sigma_j^{(22)}=5$, $\rho_j=0.4$, $n_j=10$ and $\kappa=0.75$. Both target means are equal to $0$.}
    \label{fig:3dmeans}
\end{figure}

\begin{figure}[ht]
    \centering
    \includegraphics[width=0.7\textwidth]{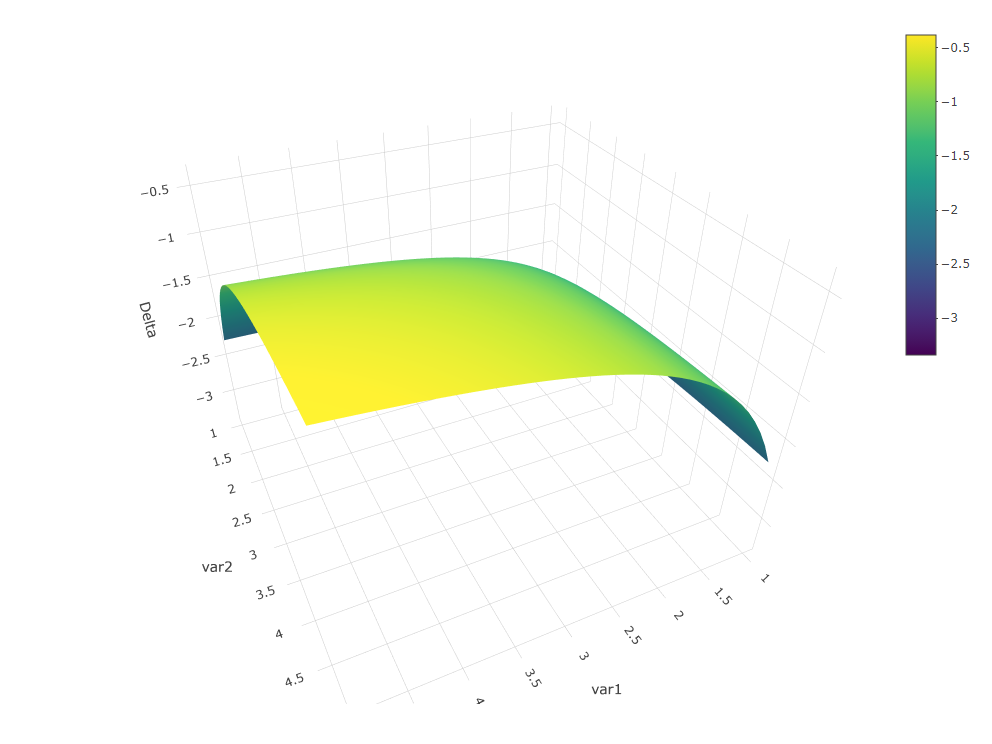}
    \caption{Information gain as function of the two variances. Other parameters are set as follows: $\bar x_{n_j}^{(1)}=\bar x_{n_j}^{(2)}=2$, $\rho_j=0.4$, $n_j=10$ and $\kappa=0.75$. Both target means are equal to $0$.}
    \label{fig:3dvariances}
\end{figure}

Figure \ref{fig:3dmeans} and \ref{fig:3dvariances} show the information gain (bivariate case) as function of the vector of means and the vector of variances, respectively. Similarly to the univariate case, the information gain achieves its maximum when sample means are equal to their corresponding target values. Additionally, the larger the variances, the higher the information gain.

\subsubsection{Correlation}

\begin{figure}[ht]
    \centering
    \includegraphics[width=1\textwidth]{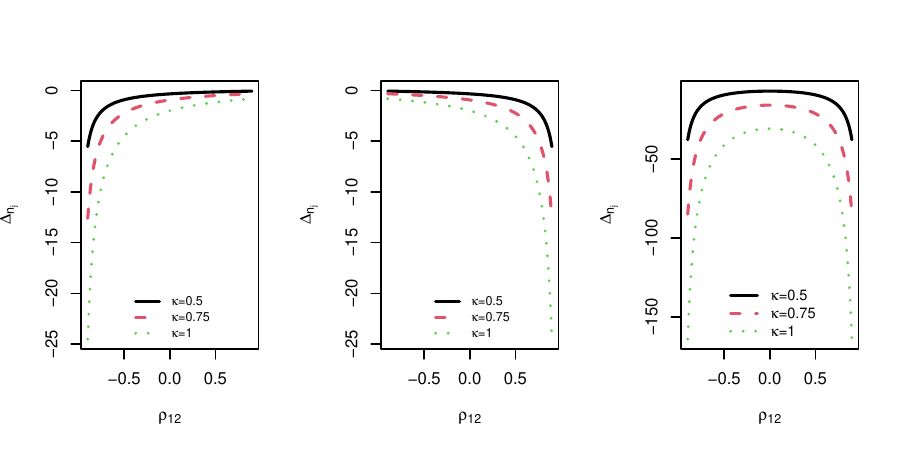}
    \caption{Information gain as function of $\rho_j$ when $\sigma_j^{(11)}=\sigma_j^{(22)}=4$ and $n_j=10$. In each panel, different vectors of sample means are considered: a) $(2,2)$, b) $(2,-2)$ and c) $(0,-10)$.}
    \label{fig:infoGaincorrelation}
\end{figure}

The maximum of the information gain as function of $\rho_j$ - \textit{ceteris paribus} - is attained in $$\rho_j=\text{sign}\Bigl\{\bigl(\gamma^{\ms(1)}-\bar x_{n_j}^{\ms(1)}\bigr)\bigl(\gamma^{\ms(2)}-\bar x_{n_j}^{\ms(2)}\bigr)\Bigr\}\times\min\biggl\{\biggl|\frac{\gamma^{\ms(1)}-\bar x_{n_j}^{\ms(1)}}{\gamma^{\ms(2)}-\bar x_{n_j}^{\ms(2)}}\sqrt{\frac{\sigma_j^{\ms(22)}}{\sigma_j^{\ms(11)}}}\biggl|,\,\biggl|\frac{\gamma^{\ms(2)}-\bar x_{n_j}^{\ms(2)}}{\gamma^{\ms(1)}-\bar x_{n_j}^{\ms(1)}}\sqrt{\frac{\sigma_j^{\ms(11)}}{\sigma_j^{\ms(22)}}}\biggl|\biggr\}\,,$$
with the convention that $\frac{0}{0}=0$. In other words, for the $j$-th arm, the maximum of the information gain is attained in correspondence of the correlation being equal to the ratio of the standardized differences between the endpoint-specific target and sample mean. If variances of the two endpoints are equal and the sample means are at the same distance from the target, then the maximum is attained in $\rho_j=+1$ when the two endpoint-specific sample means are equal and $\rho_j=-1$ when they are unequal (cfr. Figure \ref{fig:infoGaincorrelation}a and \ref{fig:infoGaincorrelation}b, respectively). Intuitively, in the first case responses of the two endpoints tend to be on the same side with respect to their target means (e.g. on average, both above or both below the corresponding target): this would suggest that a positive correlation may exist between the two endpoints and, coherently, higher positive correlations lead to a higher information gain. Conversely, when responses of the two endpoints tend to be on different sides with respect to their target means (e.g. on average, one above and one below the corresponding target), higher negative correlations lead to a higher information gain. Finally, when at least one of the two sample means equals the corresponding target, the information gain is symmetric around the null correlation point (cfr. Figure \ref{fig:infoGaincorrelation}c). 

\subsection{Further details on the proposed hypothesis testing procedure}
\label{subsec:hypothesisTest}

\subsubsection{Joint posterior of $(\mu_{\widehat j^{1*}},\,\mu_{\widehat j^{2*}})$}

Once a trial is completed, the Bayesian analysis treats the collected data - namely, the patients' responses and the realized allocation sequence - as fixed. As described in Section 3.1 of the paper, given the final dataset and the estimated best and second best arms ($\widehat j^{1*},\,\widehat j^{2*}$), the joint posterior of the pair $(\mu_{\widehat j^{1*}},\,\mu_{\widehat j^{2*}})$ is bivariate normal with independent marginal posteriors. Specifically, for $j=\bigl\{\widehat j^{1*},\,\widehat j^{2*}\bigr\}$, the marginal posterior distributions are given by $p(\mu_j \mid \text{data}) = N(\bar x_{N_j}, \sigma_j^2/N_j)$. Under the assumptions made, this result holds irrespective of whether the data were collected under a fixed or a response-adaptive design.

\begin{figure}[ht]
            \centering
            \includegraphics[width=0.9\linewidth]{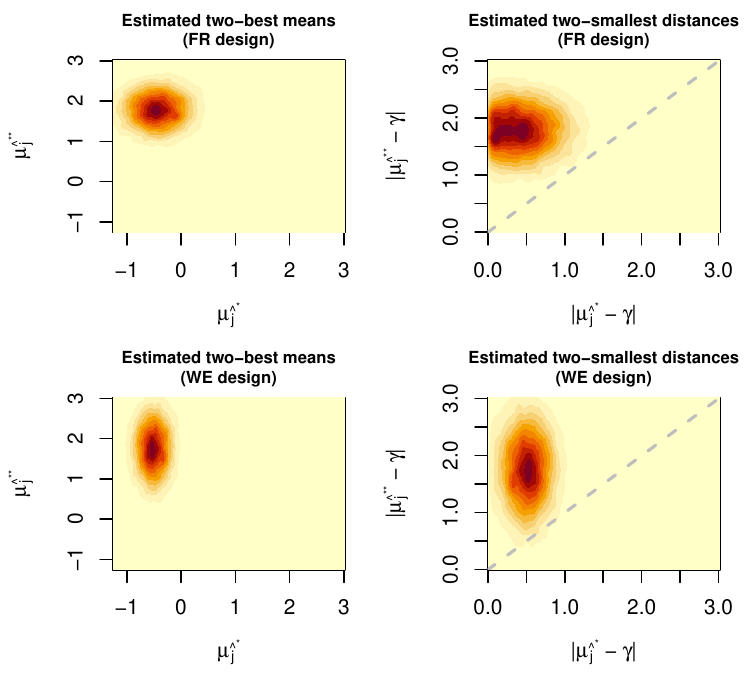}
            \caption{Joint posterior distribution of the estimated two-best arm means (\textit{left panels}) and of the estimated two-smallest distances from the target (\textit{right panels}), given the data collected at the end of a single trial based on either a FR design (\textit{top panels}) or WE$(1,0.75)$ design (\textit{bottom panels}). The grey dashed line identifies the boundary where both arms are equally distant from the target: points above it correspond to posterior draws where the best arm is indeed closer to the target than the second-best. The estimated best and second best arms ($\widehat j^{1*}$ and $\widehat j^{2*}$, respectively) are the same under both designs. The simulation study assumes $\bm \mu=(1.91,-3.36,-0.37,3.99)$, $\bm \sigma=(2,2,2,4)$, $\gamma=0$ and a burn-in $B=5$ under WE designs.}
            \label{fig:jointPostBestTwo}
\end{figure}

Figure \ref{fig:jointPostBestTwo} illustrates the joint posterior distribution of the pair $(\mu_{\widehat j^{1*}},\,\mu_{\widehat j^{2*}})$ (\textit{left panels}) and the joint posterior distribution of $(|\mu_{\widehat j^{1*}}-\gamma|,\,|\mu_{\widehat j^{2*}}-\gamma|)$ (\textit{right panels}), given the data collected at the end of a single trial realization based on either the FR (\textit{top}) or WE($1$,$0.75$) (\textit{bottom}) design. In the right panels, the dashed diagonal marks the region where both arms are equally distant from the target. Accordingly, the quantity $\widehat\pi_{\widehat j^{1*},\widehat j^{2*}}$ - the posterior probability that the estimated best arm is truly closer to $\gamma$, i.e. $\mathbb{P}\bigl(|\mu_{\widehat j^{1*}}-\gamma| < |\mu_{\widehat j^{2*}}-\gamma|\,\bigl|\text{data}\bigr)$ - corresponds to the probability mass lying above the dashed line. For this specific trial realization, we obtain $\widehat\pi_{\widehat j^{1*},\widehat j^{2*}}\approx0.99$ under the FR design and $0.97$ under the WE design.

\subsubsection{Asymptotic behaviour of the frequentist distribution of $\widehat\pi_{\widehat j^{1*},\widehat j^{2*}}$}
        
\begin{figure}[ht]
            \centering
            \includegraphics[width=0.7\linewidth]{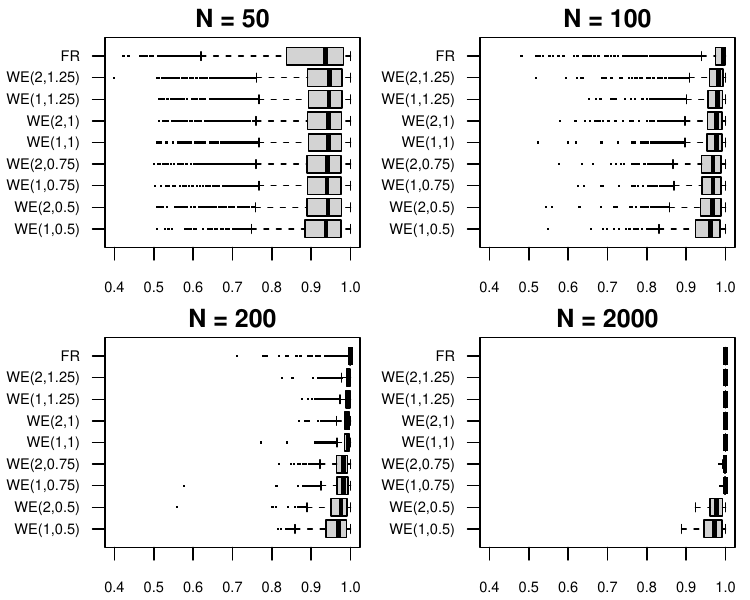}
            \caption{Empirical distribution of the posterior probability $\widehat\pi_{\widehat j^{1*},\widehat j^{2*}}$ ($x$-axis) under FR and WE($p$,$\kappa$) - with $p=\{1,2\}$ and $\kappa=\{0.5,0.75,1,1.25\}$ - designs for different trial sizes. The simulation study assumes $\bm \mu=(1.91,-3.36,-0.37,3.99)$, $\bm \sigma=(2,2,2,4)$, $\gamma=0$ and a burn-in $B=5$ under WE designs.}
            \label{fig:postProbBest_asymptotic_H1}
\end{figure}

Figure \ref{fig:postProbBest_asymptotic_H1} displays the empirical distribution of $\widehat\pi_{\widehat j^{1*},\widehat j^{2*}}$ resulting from FR and WE($p$,$\kappa$) designs for increasing trial sizes $N$, in a scenario where the best and second best arms can be uniquely defined (i.e., under $H_1$). For WE($p$,$\kappa$), all the possible combinations of $p=\{1,2\}$ and $\kappa=\{0.5,0.75,1,1.25\}$ are considered. For both FR and WE designs with $\kappa>0.5$, the empirical distribution of $\widehat\pi_{\widehat j^{1*},\widehat j^{2*}}$ increasingly concentrates its mass near 1 as $N$ increases. In contrast, when $\kappa=0.5$ under the WE design, the distribution of $\widehat\pi_{\widehat j^{1*},\widehat j^{2*}}$ does not concentrate near $1$. This behaviour is consistent with Theorem 5, which guarantees that $\widehat\pi_{\widehat j^{1*},\widehat j^{2*}}$ converges to 1 almost surely for $\kappa>0.5$.

\subsubsection{Additional insights on the impact of $\kappa$ on the hypothesis testing procedure}

\begin{figure}[ht]
            \centering
            \includegraphics[width=1\linewidth]{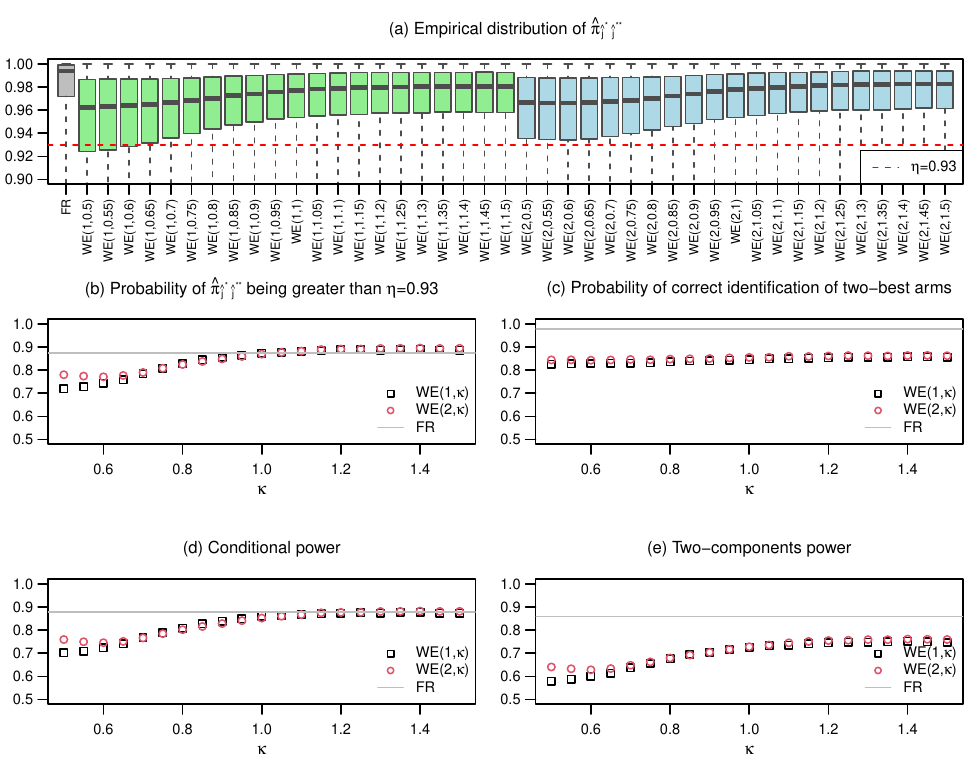}
            \caption{(a) Empirical distribution of the posterior probability $\widehat\pi_{\widehat j^{1*},\widehat j^{2*}}$, (b) estimated probability that $\widehat\pi_{\widehat j^{1*},\widehat j^{2*}}>\eta$, (c) estimated probability of correctly identifying the best-two arms, (d) conditional power and (e) two-components power under FR and WE$(p,\kappa)$. The value $\eta_{0.05}^{\text{max}}$ (dashed line) approximately controls the type-I error rate below $5\%$ across all designs and null scenarios. The simulation study assumes $\bm \mu=(1.91,-3.36,-0.37,3.99)$, $\bm \sigma=(2,2,2,4)$, $\gamma=0$ and a burn-in $B=5$ under WE designs.}
            \label{fig:hypothesisTest_byKappa}
    \end{figure}

Figure \ref{fig:hypothesisTest_byKappa} shows that the posterior probability $\widehat\pi_{\widehat j^{1*},\widehat j^{2*}}$ tends to increase in median value and exhibit reduced variability as $\kappa$ becomes larger [panel (a)], in a scenario where the best arm is truly closer to the target than the second best. A similar increasing trend can be observed for the probability that $\widehat\pi_{\widehat j^{1*},\widehat j^{2*}}$ is greater than the cut-off probability $\eta$ [panel (b)], for the probability of correct identification of the two-best arms [panel (c)] and for the power [panels (d)--(e)]. This improvement is driven by the stronger exploratory behaviour induced by larger values of $\kappa$. However, for values of $\kappa>1.2$, these operating characteristics seem to reach a plateau (no real improvement after this point). Overall, this aligns with the average performance observed across the 500 alternative scenarios in the simulation study (see, for example, the right side of Figure 6 in the manuscript), and justify the choice to explore the range $[0.5,1.5]$ for $\kappa$.

    These results show that when the WE$(p,\kappa)$ design encourages wider exploration across the $K$ arms (i.e., larger $\kappa$), the uncertainty about the suboptimal arms decreases more quickly. Intuitively, this increases the chance of concluding that the estimated best arm is indeed closer to the target than the estimated second best.
        
\subsection{Target-oriented modified Gittins index rules}

\cite{smith2018bayesian}'s design is oriented toward a patient benefit objective and, thus, it is worth to be considered as competitor to our proposal. In \cite{smith2018bayesian}'s formulation, the allocation rule proceeds as follows: if larger responses are desirable, the next patient is assigned to the arm $j$ ($j=1,\dots,K$) with largest Gittins index (GI), $\mathcal{G}_j^+=\bar x_{n_j}+\sigma_j\,\bar{\mathcal{G}}_j(n_j,d)$, where the values of the standardized index $\bar{\mathcal{G}}_j(n_j,d)$ can be interpolated from the tables printed in \citet[pp. 261--262]{gittins2011multi} for several possible combinations of current sample size $n_j$ and $d\in(0,1)$. Vice versa, if smaller responses are desirable, the next patient is assigned to the arm $j$ with smallest quantity $\mathcal{G}_j^-=\bar x_{n_j}-\sigma_j\,\bar{\mathcal{G}}_j(n_j,d)$. The quantity $\sigma_j\,\bar{\mathcal{G}}_j(n_j,d)$ is a measure of the learning reward associated with keep sampling an arm which has already been sampled $n_j$ times, with $d$ being a tuning parameter quantifying the value of learning. However, this design cannot be directly adopted to our setting where desirable values are the ones as close as possible to $\gamma$. Thus, we propose two heuristic extensions of the rule above: the symmetric Gittins index criterion (SGI) and the targeted Gittins index criterion (TGI).

The first proposal is to use an allocation rule which assigns the next patient to the arm $j$ which minimizes the quantity 
\begin{equation}
\label{eq:SGI}
\mathcal{G}^{\gamma}_j=|\bar x_{n_j}-\gamma|-\sigma_j\,\bar{\mathcal{G}}_j(n_j,d)\,,
\end{equation}
which has the property of being symmetrical around its maximum $\bar x_{n_j}=\gamma$ (i.e. it takes the same value in correspondence of $x_{n_j}=\gamma-c_s$ and $x_{n_j}=\gamma+c_s$, $c_s\in\mathbb{R}$, \textit{ceteris paribus}. The basic idea dwells in discounting the learning component $\sigma_j\,\bar{\mathcal{G}}_j(n_j,d)$ from the distance between the current sample mean and the target mean. 

The second proposal arises from considering a targeted Gittins index for arm $j$ ($j=1,\dots,K$), namely the Gittins index that would be associated to arm $j$ if its response were $N(\gamma,\,\sigma_j^2)$: as $n$ goes to infinity, the value of learning goes to $0$ and the targeted Gittins index would tend to $\gamma$ by the Law of Large Numbers. For this reason, our proposal is to adopt the following rule which allocates the next patient to the arm $j$ which minimized the distance between the current Gittins index and the asymptotical targeted Gittins index (i.e., $\gamma$):
\begin{equation}
\label{eq:TGI}
\delta_{\mathcal{G}_j,\mathcal{G}_{\gamma}}=\begin{cases}
    |\mathcal{G}_j^+ -\gamma| & \text{if } \bar x_{n_j}\leq\gamma\\
    |\mathcal{G}_j^- -\gamma| & \text{if } \bar x_{n_j}>\gamma\\
\end{cases}\,,
\end{equation}
where the $\mathcal{G}_j^+$ (respectively, $\mathcal{G}_j^-$) is adopted when the current sample mean is below (above) the target. Interestingly, as $n_j$ tends to infinity, $\delta_{\mathcal{G}_j,\mathcal{G}_{\gamma}}$ tends to $|\mu_j-\gamma|$ which is the key quantity of our adopted recommendation rule.

\subsection{Trend of the allocation ratio for different response-adaptive designs}
\label{subsec:allocRatio_compets}

\begin{figure}[ht]
            \centering
            \includegraphics[width=0.9\linewidth]{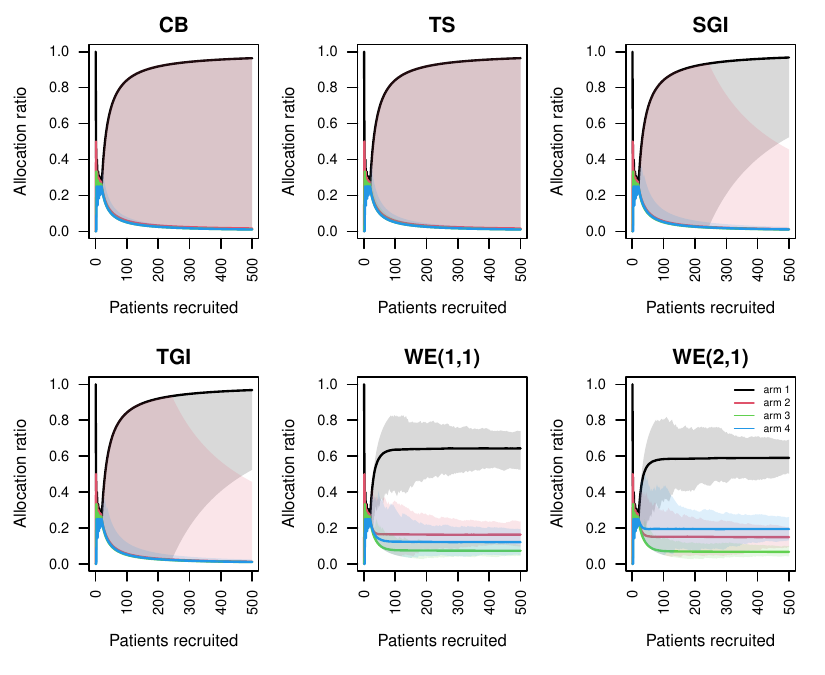}
            \caption{Allocation ratios as number of patients recruited increases, shown for each of the four arms of the trial and for different response-adaptive designs. Solid lines represent the mean allocation ratios across $10^3$ trial replicates, while the light-shaded areas show the range of variation observed across those replicates. Each replicate assumes $\bm{\mu}=(1,2,3,3.5)$, $\bm{\sigma}=(1,1,1,2)$, $\gamma=0$ and $B=5$.}
            \label{fig:allocRatio_compets}
\end{figure}

Figure \ref{fig:allocRatio_compets} illustrates the the evolution of the allocation ratios as the size of the multi-arm trial increases for the set of competing designs considered in the manuscript. The solid lines represent the mean allocation ratios across $10^3$ trial replicates, ensuring that the observed behaviour reflects a consistent trend across many independent trial realizations rather than a random feature of a single run. The light-shaded areas show the range of variation observed across those replicates. \\
The class of WE($p$,$\kappa$) designs introduced in our manuscript shows higher variability in the allocation proportions when the number of patients is small, reflecting the greater uncertainty and instability in early posterior estimates. This variability naturally decreases as more patients are recruited. In contrast, the competitor designs tend to concentrate most allocations on a single arm, either the best (black) or the second best (red) arm.

\subsection{Additional details on the simulation study with single endpoint}

\subsubsection{Burn-in size and type-I error rate}
\label{SM:burnin}

\begin{figure}[ht]
    \centering
    \includegraphics[width=0.95\textwidth]{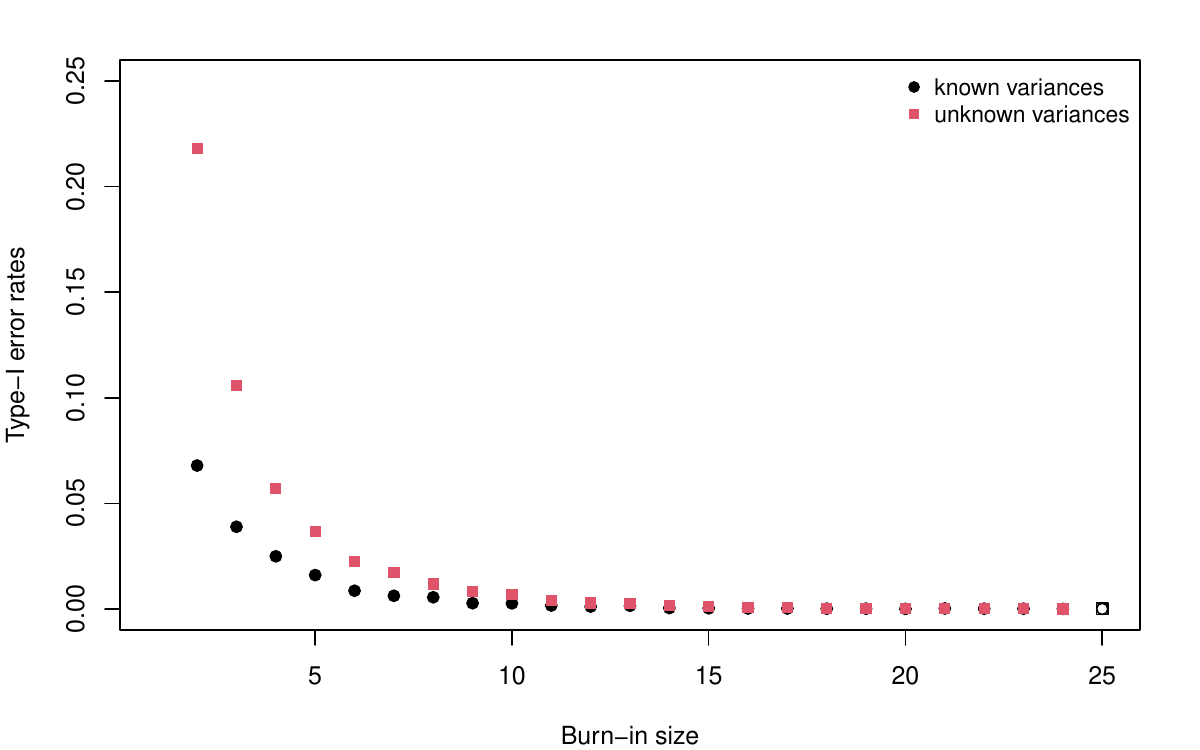}
    \caption{Type-I error rates under $H_0:\,\mu_j=\gamma$ and $\eta_\alpha=0.95$ when the design WE($2$,$1$) is used in combination of different burn-in sizes, in the case of either known or unknown variances. The empty shapes in correspondence of burn-in size equal to $25$ correspond to the extreme case of full burn-in (i.e. lack of adapting phase). Other parameters are set as follows: $K=4$, $N=100$, $\bm\sigma=(2,2,2,4)$ and $\gamma=0,\,\forall j$.}
    \label{fig:burnin}
\end{figure}

Figure \ref{fig:burnin} shows how the type-I error rate decreases as the burn-in size ($B$) increases. We choose the scenario where all true means are equal to the target: this scenario can be considered challenging for our proposed design, as it is generally associated with a high inflation of the Type I error rate. The drop is surprisingly rapid when passing from $B=2$ to $B=5$, meaning that an equal allocation of $20\%$ of the patients to each arm prior to the adaptive allocation may be useful to stabilize the algorithm and reduce the its dependence from the initial values. This is even more evident in the case of unknown variances since an additional parameter must be estimated in each arm. 

For the full simulation study, we adopt $B=5$ which seems to offer a good balance between protecting patients from default allocations to sub-optimal arms and control of type-I error rate. Notably, in the considered example, the choice $B=5$ leads to a small difference between type-I error rates of the known and unknown cases. 

\subsubsection{Choice of the null scenarios used for calibration of cut-off probabilities}

Recall that we focus on the set of null scenarios $$\mathcal{S}_0=\bigl\{(\mu_1^{(s)},\dots,\mu_K^{(s)}):\,\mu_j^{(s)}-\gamma=c_s,\forall j=1,\dots,K,\,c_s\geq0,\,s=1,\dots,S_0\bigr\}$$ for the calibration of the cut-off probabilities. Our strategy consists in considering values of $c_s$ from $0$ to $40$ increasing in a quadratic way. This reflects the interest in having a larger number of scenarios where $c_s$ is closer to $\gamma=0$, which are frequently associated with high inflation or deflation of the type-I error rates. Indeed, as $c_s$ increases, the type-I error rate tends to stabilize.

In practice, we first consider a grid of evenly spaced values from $0$ to $\sqrt{40}$; then, we take the square of these values to obtain the final grid of values for $c_s$.

\subsubsection{How $p$ and $\kappa$ affect operating characteristics}

\begin{figure}[ht]
    \centering
    \includegraphics[width=0.95\textwidth]{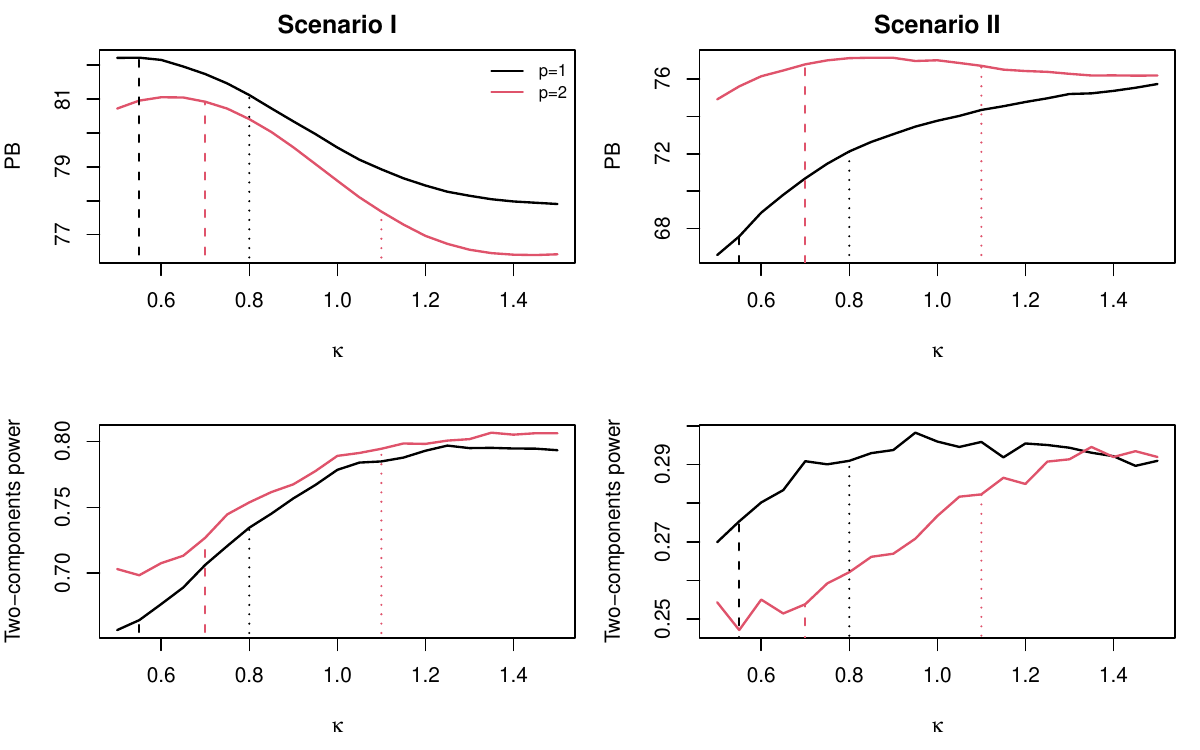}
    \caption{Average percentage of experimentation of the best arm during the trial (PB, i.e. patient benefit measure) and two-components power (cut-off values based on average type-I error rate control) for the WE design for different values of $\kappa$, under either $p=1$ or $p=2$. Dashed and dotted lines highlights the values of the penalization parameter $\kappa$ chosen using patient benefit and power criteria. Scenario I is characterized by $\bm\mu=(1.91,-3.36,-0.37,3.99)$, while Scenario II by $\bm\mu=(1.13,-3.48,-3.57,0.34)$. Both scenarios consider $\bm\sigma=(2,2,2,4)$ and $\gamma=0$.}
    \label{fig:OCwithKappa}
\end{figure}

Figure \ref{fig:OCwithKappa} illustrates how the parameter $\kappa$ has a key role in regulating the trade-off between patient benefit (individual objective) and power requirements (collective objective). Notably, in Scenario I, the power tends to increase and the patient benefit to decrease with $\kappa$ as result of the more spread allocation of patients in the various treatment arms. Moreover, if one compares the operating characteristics under the two different choices of $p$, the case of $p=2$ uniformly leads to higher power but lower patient benefit than the case of $p=1$. A reasonable explanation dwells in the fact that a high $p$ gives more chance to the arm with highest variance to be explored: while this generally helps improving the power, it may slightly reduce the exploitation of the best treatment when the most variable arm is suboptimal as in Scenario I. In contrast, Scenario II tells a partially different story: WE designs with lower $p$ are uniformly (across $\kappa$'s) dominated by the ones with higher $p$ in terms of patient benefit since the best treatment is also the one with most variable response; unsurprisingly, the opposite tends to happen in terms of power. The only unexpected behaviour is represented by the patient benefit increasing with $\kappa$ when $p=1$, which is in net contrast with the value $\kappa=0.55$ optimizing the patient benefit criterion. Once again, it is important to remind that this value of $\kappa$ was selected by averaging across a large number of possible configurations, and the ones like Scenario II can be regarded as outliers.

\subsubsection{Sensitivity analysis under model misspecification}
\label{subsec:misspecif_SM}

\begin{figure}[ht]
        \centering
        \includegraphics[width=0.75\linewidth]{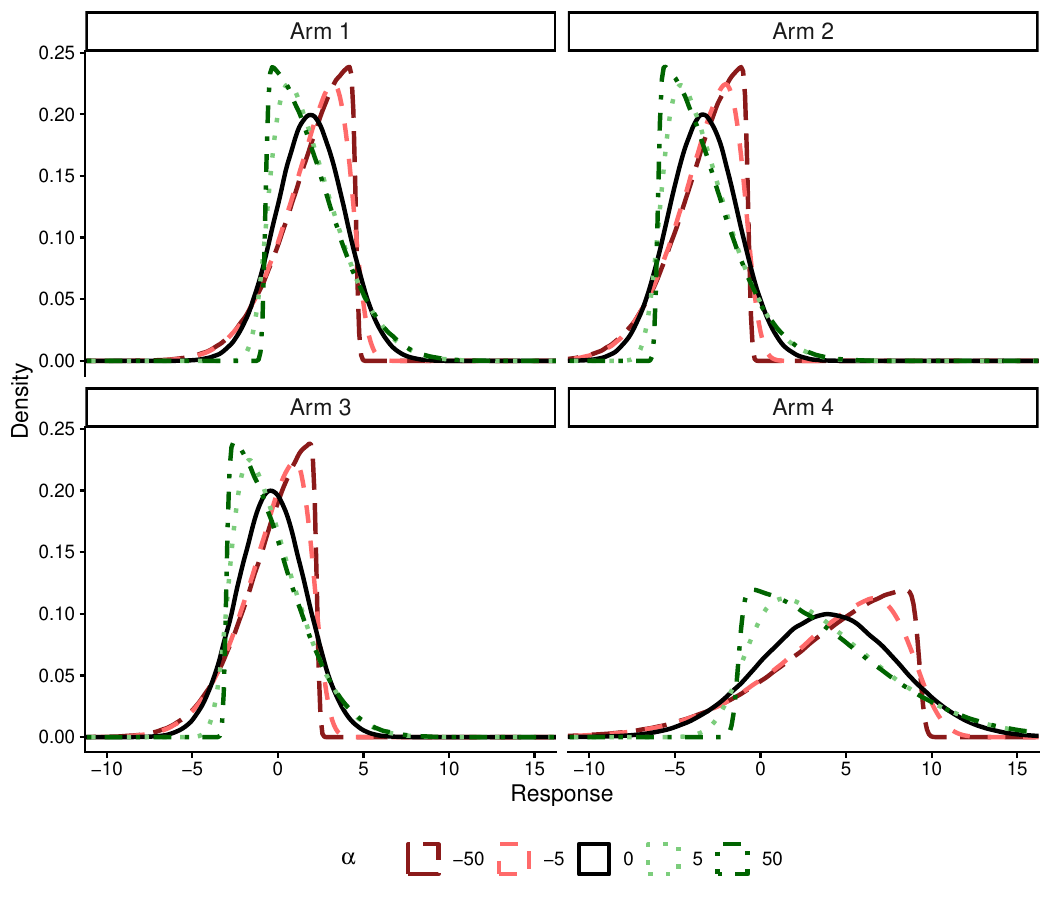}
        \caption{Skew-normal response densities for the four treatment arms. The distributions are parameterized via location and scale parameters, and calibrated to match the mean vector $\bm\mu=(1.91,-3.36,-0.37,3.99)$ and standard deviation vector $\bm\sigma=(2,2,2,4)$. A common asymmetry parameter is assumed across arms, $\alpha_j=\alpha\in\{-50,-5,0,5,50\},\,\forall j$, with $\alpha=0$ (\textit{solid line}) corresponding to the symmetric (normal) case.}
        \label{fig:dens_misspec}
\end{figure}

We investigate the performance of the WE($p$, $\kappa$) design under model misspecification. In particular, we consider a 4-armed trial with outcomes generated from a skew-normal distribution, $SN(\xi_j,\omega_j,\alpha_j)$, $j=1,\dots,4$, where $\xi_j$ and $\omega_j$ denote location and scale parameters, and $\alpha_j\in\mathbb R$ controls the degree of asymmetry \citep[p.~24]{azzalini2013skew}. The location and scale parameters are chosen so that the resulting means and standard deviations match those considered in Scenario I of the illustration Table 1 (Section 5.6 of the main text). We assume a common asymmetry parameter across arms, $\alpha_j=\alpha,\,\forall j$, and consider a set of asymmetry levels $\alpha\in\{-50,-5,0,5,50\}$, ranging from strong left-skewness to strong right-skewness, with $\alpha=0$ corresponding to the symmetric (normal) case. This setup allows us to assess the impact of distributional misspecification relative to the correctly specified normal setting. Figure \ref{fig:dens_misspec} shows the response densities for each arm across different $\alpha$.

\begin{figure}[ht]
        \centering
        \includegraphics[width=0.75\linewidth]{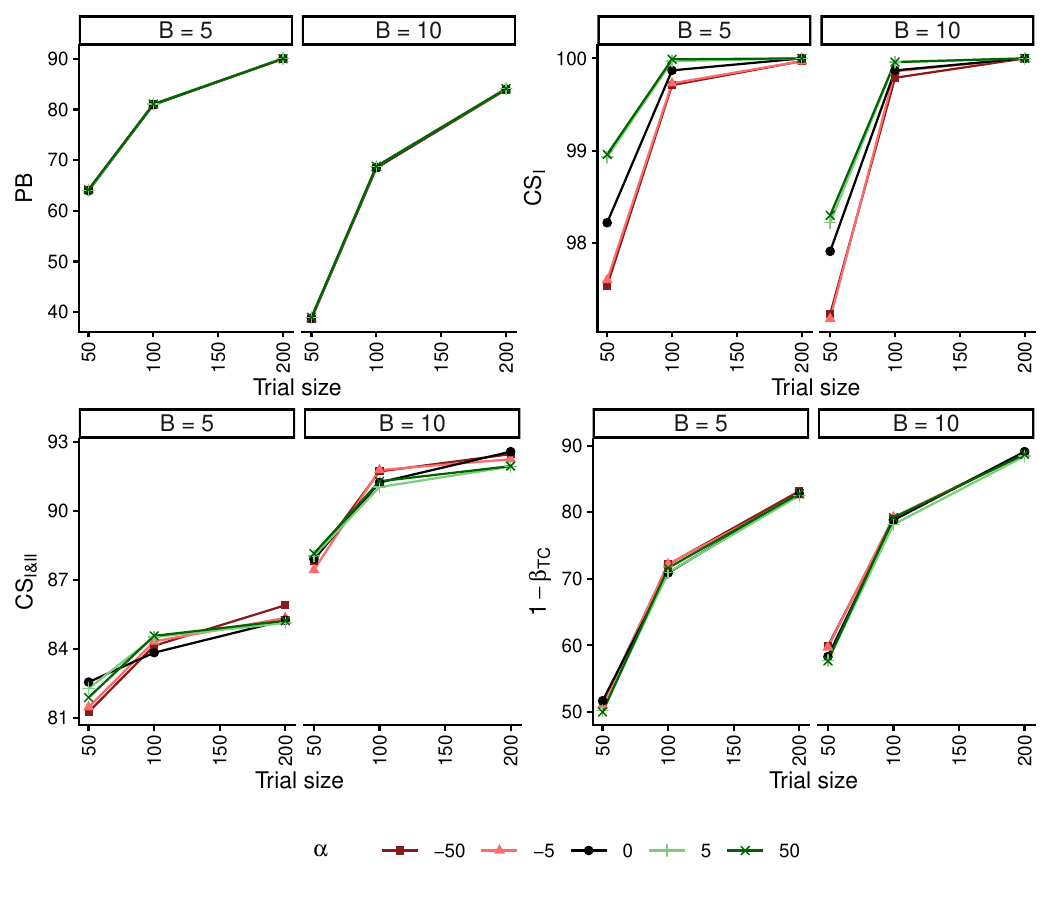}
        \caption{Average percentage of experimentation of the best arm during the trial (patient benefit, PB; \textit{top left}), percentage of correct selection of the best arm ($CS_I$; \textit{top right}), percentage of correct selection of the two best arms ($CS_{I\&II}$; \textit{bottom left}), and two-components power ($1-\beta_{TC}$; \textit{bottom right}, cut-off probability $\eta=0.913$) under the design WE(2, 0.7) with target $\gamma=0$. The treatment responses follow the skew-normal distributions described in Figure \ref{fig:dens_misspec}, with $\alpha=0$ (\textit{black dots}) corresponding to the symmetric (normal) case. Results are reported across different trial sizes $N\in\{50,100,200\}$ and burn-in sizes $B\in\{5,10\}$. Results are based on $10^4$ Monte Carlo replicates.}
        \label{fig:OC_misspec}
\end{figure}

We further consider a range of trial configurations with trial size $N\in\{50,100,200\}$ and burn-in size $B\in\{5,10\}$. The operating characteristics, reported in Figure \ref{fig:OC_misspec}, are largely insensitive to the degree of skewness. When the burn-in size is smaller, slight deviations from the symmetric case are observed in the percentage of correct identification of the best arm ($CS_I$) and of the two best arms ($CS_{I\&II}$). Overall, these results suggest that the WE($p$, $\kappa$) design is robust to deviations from normality, even under substantial skewness.

\subsection{Unknown variance case with single endpoint} 
\label{subsec:simUnknown}

When the variance of the response is supposed to be unknown, we rely on the unbiased sample variance $\frac{1}{n_j-1}\sum_{i=1}^{n_j}\,(x_{i,\,j}-\bar x_{n_j})^2$ ($j=1,\dots,K)$ as plug-in estimate of the true one. This requires a minimum burn-in of $B=2$ patients per arm to obtain a first estimate of the arm-specific variances when response-adaptive designs are used. However, in practice, a higher burn-in may be needed to achieve better operating characteristics when the variances are supposed to be unknown (cfr. Section \ref{SM:burnin}).
Here, we adopt the same setting as in Section 5 of the main text, that the scenarios are generated by drawing the mean and standard deviation parameters independently as $\mu_j\sim Unif[-4,4]$ and $\sigma_j\sim Unif[2,4]$, $\forall j=1,\dots,K$. We run a simulation study over $S_1=500$ alternative scenarios.

\subsubsection{Type-I error rate control and cut-off probabilities} 
\label{SM:typeIcontrolSim}

\begin{figure}[ht]
    \centering
    \includegraphics[width=0.95\textwidth]{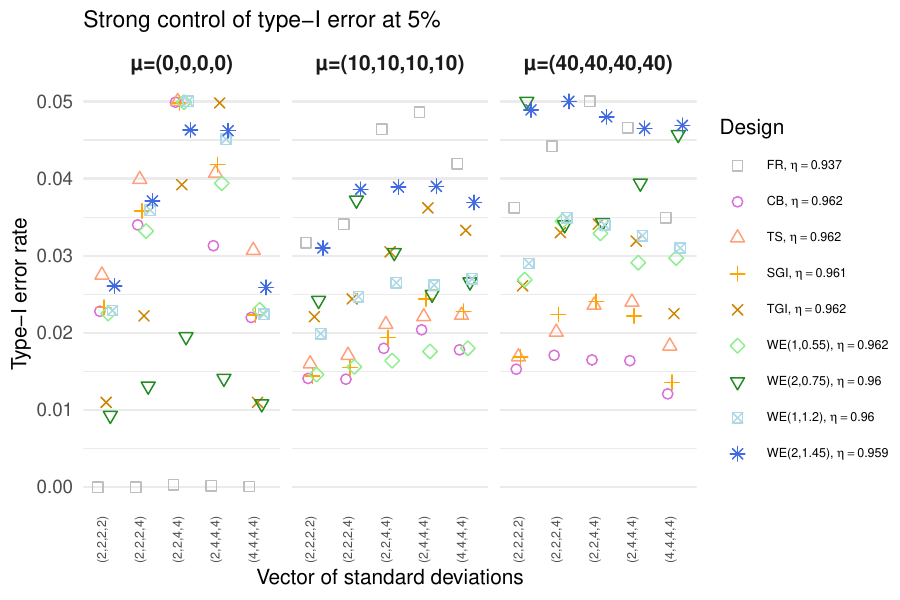}
    \caption{Individual type-I error rates in correspondence of the null scenarios in $\mathcal{S}'_0$ for several designs in the case of unknown variances. The target mean is set to $\gamma=0$ for all the scenarios. Design-specific cut-off probabilities are calibrated under strong control of the type-I error rate at level $\alpha=5\%$.}
    \label{fig:alphaStrong_DiffUnknown}
\end{figure}

\begin{figure}[ht]
    \centering
    \includegraphics[width=0.95\textwidth]{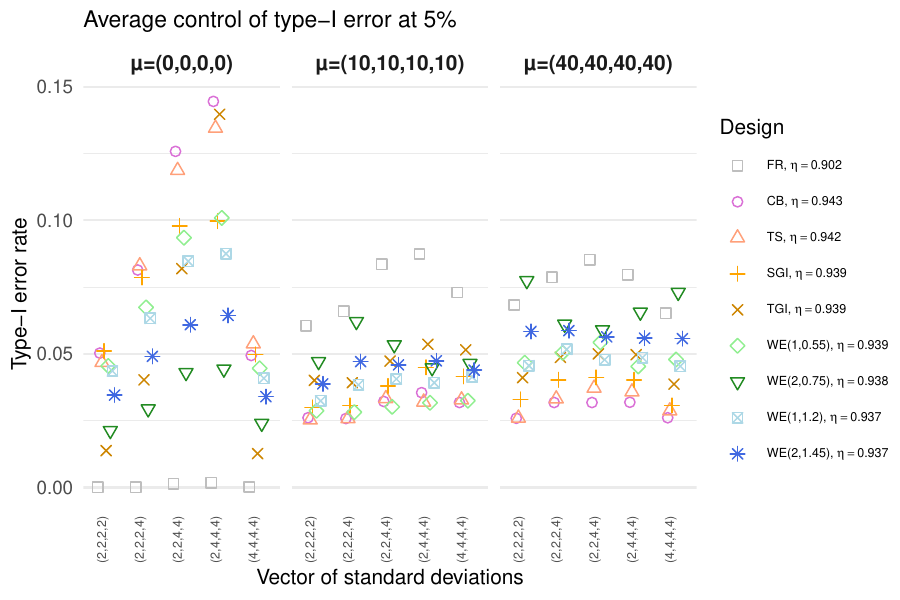}
    \caption{Same as Figure \ref{fig:alphaStrong_DiffUnknown}, but with design-specific cut-off probabilities are calibrated under average control of the type-I error rate at level $\alpha=5\%$.}
    \label{fig:alphaMean_DiffUnknown}
\end{figure}

We consider the set of null scenarios defined by the Cartesian product of the mean and variance configurations $\mathcal{S}'_0=\mathcal C \times \mathcal D$, where $$\mathcal{C}=\Bigl\{(\mu_1^{(s)},\dots,\mu_K^{(s)}):\,\mu_j^{(s)}-\gamma=c_s,\,\forall j,\,s=1,2,3\Bigr\}$$ with $c_s=\{0,\,10,\,40\}$, and $$\mathcal{D}=\Bigr\{(\sigma_1^{(r)},\dots,\sigma_K^{(r)})\,:\sigma_j^{(r)}=d_{jr},\,\forall j,\,r=1,2\Bigr\}\,,$$ with $d_{jr}=\{2,\,4\}$. Figure \ref{fig:alphaStrong_DiffUnknown} and \ref{fig:alphaMean_DiffUnknown} show the type-I error rates for each considered design in correspondence of the scenarios in $\mathcal{S}'_0$ when cut-off probabilities are calibrated - respectively - under strong and average control of the type-I error rate at level $\alpha=5\%$.

Interestingly, most of the design are associated with an inflation of the type-I error rate in correspondence when arms have different variances, and this is even more evident in the case of $\bm\sigma=(2,2,4,4)$. The magnitude of this inflation strongly depends on the particular vector of means: for example, when $c_1=0$, most of the response-adaptive designs are associated with higher type-I error rates with respect to the case of $c_2=10$ or $c_3=40$. An interesting exception is represented by WE($2$, $0.75$), which is associated with higher type-I error rates when $c_3=40$ and all the variances are equal. Finally, FR is associated with higher type-I error rates when $c_s$ becomes large.

\subsubsection{Selection of $\kappa$} 

\begin{figure}[ht]
    \centering
    \includegraphics[width=1\textwidth]{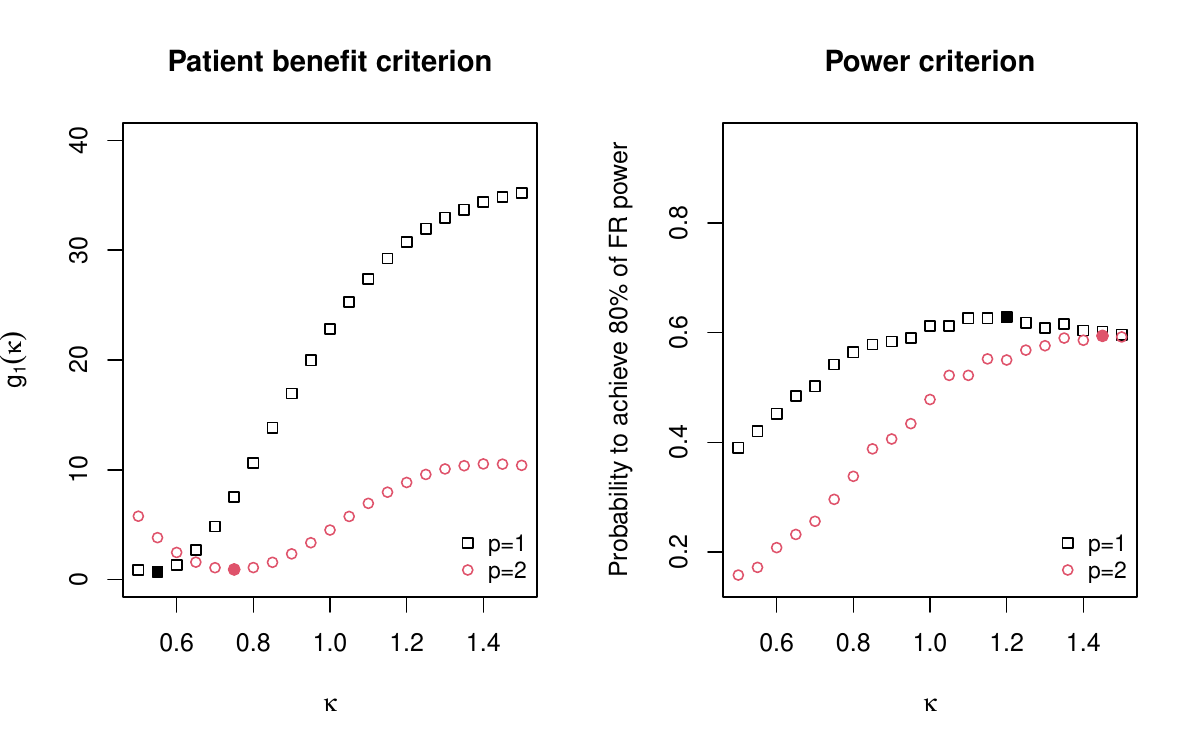}
    \caption{Values of the objective function $g_1$ (\textit{left panel}) and the probability to achieve at least $80\%$ of the power of FR (\textit{right panel}) for several values of $\kappa$, and for both $p=1$ (\textit{black square} and $p=2$ (\textit{red circles}) in the case of unknown variances. Filled shapes (either circles or squares) corresponds to the robust optimal values of $\kappa$. The two-components power is considered, along with design-specific cut-off probabilities calibrated to achieve a $5\%$ type-I error rate on average over the set of scenarios considered in Section \ref{SM:typeIcontrolSim}.}
    \label{fig:kappaSelection_DiffUnknown}
\end{figure}

Figure \ref{fig:kappaSelection_DiffUnknown} illustrates the result of the optimization of the patient benefit and power criteria for $p=1$ and $p=2$. The probability to achieve $80\%$ of FR power does not achieve the threshold $\xi=90\%$, therefore we choose $\kappa$ in order to maximize this probability. The proposed values of $\kappa$ are $\{0.55,\,1.2\}$ for $p=1$ and $\{0.75,\,1.45\}$ for $p=2$.

\subsubsection{Numerical results}

\begin{figure}[ht]
    \centering
    \includegraphics[width=1\textwidth]{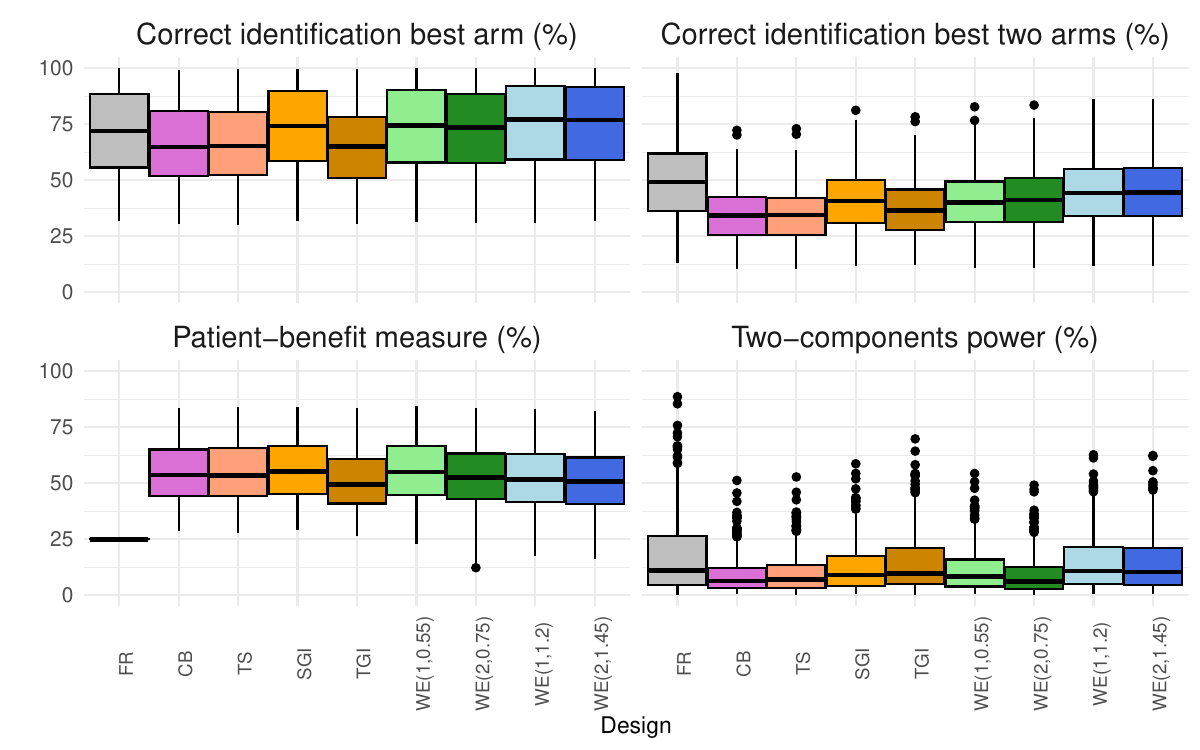}
    \caption{Operating characteristics across $S_1=500$ randomly generated scenarios for the considered set of designs. Each scenario is characterized by a vector of mean responses. For all scenarios, we assume $K=4$, $N=100$, $\bm \sigma=(2,2,2,4)$ (unknown) and $\gamma=0$, for $j=1,\dots,4$. Burn-in size is fixed to $B=5$ for all the response-adaptive designs.}
    \label{fig:OC_DiffUnknown}
\end{figure}

Figure \ref{fig:OC_DiffUnknown} shows the results of the simulation study in the case of unknown variances. As anticipated, the uncertainty about the estimates of the unknown variances is reflected in poorer performance of all the methods. Similarly to the case of known variances (cfr. main text), WE designs with $\kappa$ optimizing the power criterion - i.e. WE($1$, $1.2$) and WE($2$, $1.45$) - are associated with high percentage of identification of the best arm at the end of the trial. TGI, WE($1$, $1.2$) and WE($2$, $1.45$) exhibit similar distributions of the two-components power, but TGI obtains lower performance in terms of identification of the best arm and patient benefit. Surprisingly, in few random scenarios WE designs may perform even poorer than FR in terms of average percentage of experimentation of the best arm (i.e. patient benefit measure): specifically, this occurs in $14$ scenarios for WE($2$, $0.75$) and only in $2$ scenarios for WE($1$, $0.55$). However, it is interesting to notice that these scenarios are all characterized by the mean of the best treatment being relatively far from the target and close to the one of the other treatments (the median two-components power under FR is $3.1\%$ in this subset of scenarios).
Overall, WE(1,1.2) achieves slightly lower power than FR but with more satisfactory performance in terms of correct identification of the best arm and, above all, patient benefit. 

Now, we assess the operating characteristics of the considered designs in two specific scenarios:
\begin{enumerate}[{i}]
    \item Scenario Ibis, characterized by $\bm\mu=(1.91, -3.36, -0.37, 3.99)$ and $\bm\sigma=(3.48, 2.16, 2.91, 4)$, with the worst treatment arm having the highest variability but the second worst having the lowest variability;
    \item Scenario IIbis, characterized by $\bm\mu=(1.13, -3.48, -3.57,  0.34)$ and $\bm\sigma=(3.28, 2.13, 2.11, 3.08)$, with the best two treatment arms having the highest variability.
\end{enumerate}
Table \ref{tab:simUniv_unknownVar} illustrates the operating characteristics of the considered designs under Scenario Ibis and IIbis. In both cases, WE(1,1.2) and WE(2,1.45) perform well in terms of correct identification of the best arm at the end of the trial and power: in Scenario I, their two-components power is $11$ points lower than FR, but high a remarkable gain in terms of patients benefit, while in Scenario II they both achieve the highest two-components power, along with SGI. Interestingly, SGI results in slightly better operating characteristics than WE(1,0.55) and WE(2,0.75) in the two considered scenarios. TGI achieves relatively high conditional power (higher than FR), but performs poorly in terms of patient benefit measure, with a corresponding high standard error. 
\begin{table}[ht]
\centering

\textbf{Scenario Ibis (unknown variances):} $(\mu_j,\,\sigma_j)=\bigl\{(1.91,\,3.48),\,(-3.36,\,2.16),\,(-0.37,\,2.91),\,(3.99,\,4)\bigr\}$, $\gamma=0$ \\ 
\begin{tabular}{lccccc}
  \midrule
  Design & PB (s.e.) & CS$_{I}(\%)$ & CS$_{I\&II}(\%)$ & $1-\beta_C$ & $1-\beta_{TC}$ \\  
  \hline
FR & 24.99 (0.04) & 94.49 & 88.81 & 0.57 & 0.50 \\ 
  CB & 74.31 (0.24) & 90.40 & 59.71 & 0.37 & 0.22 \\ 
  TS & 75.10 (0.23) & 89.01 & 58.52 & 0.37 & 0.22 \\ 
  SGI & 74.42 (0.17) & 95.80 & 66.58 & 0.50 & 0.33 \\ 
  TGI & 70.33 (0.24) & 86.60 & 59.59 & 0.58 & 0.34 \\ 
  WE(1,0.55) & 74.23 (0.18) & 95.88 & 66.73 & 0.44 & 0.29 \\ 
  WE(2,0.75) & 69.93 (0.22) & 93.98 & 66.23 & 0.39 & 0.26 \\ 
  WE(1,1.2) & 71.95 (0.16) & 97.80 & 71.11 & 0.54 & 0.39 \\ 
  WE(2,1.45) & 69.84 (0.17) & 97.04 & 71.46 & 0.54 & 0.39 \\ 
  \hline \vspace{-0.2cm}\\
\end{tabular}

\textbf{Scenario IIbis (unknown variances):} $(\mu_j,\,\sigma_j)=\bigl\{(1.13,\,3.28),\,(-3.48,\,2.13),\,(-3.57,\,2.11),\,(0.34,\,3.08)\bigr\}$, $\gamma=0$ \\ 
\begin{tabular}{lccccc}
  \midrule
  Design & PB (s.e.) & CS$_{I}(\%)$ & CS$_{I\&II}(\%)$ & $1-\beta_C$ & $1-\beta_{TC}$ \\  
  \hline
FR & 25.05 (0.04) & 78.25 & 78.03 & 0.23 & 0.18 \\ 
  CB & 60.26 (0.35) & 71.87 & 58.59 & 0.30 & 0.17 \\ 
  TS & 60.09 (0.35) & 71.00 & 57.40 & 0.30 & 0.17 \\ 
  SGI & 62.85 (0.27) & 80.72 & 67.42 & 0.35 & 0.24 \\ 
  TGI & 54.69 (0.36) & 75.43 & 64.86 & 0.32 & 0.21 \\ 
  WE(1,0.55) & 62.67 (0.29) & 80.08 & 66.59 & 0.31 & 0.21 \\ 
  WE(2,0.75) & 60.95 (0.29) & 79.39 & 66.57 & 0.24 & 0.16 \\ 
  WE(1,1.2) & 63.60 (0.23) & 85.40 & 73.56 & 0.34 & 0.25 \\ 
  WE(2,1.45) & 62.69 (0.23) & 84.83 & 73.46 & 0.33 & 0.24 \\ 
   \hline
\end{tabular}
\caption{Operating characteristics of the WE($p$,$\kappa$) design for different combinations of $\kappa$ and $p$, FR design, CB, SGI and TGI under Scenario Ibis and IIbis (unknown variances). Both $1-\beta_C$ and $1-\beta_{TC}$ are based on the design-specific cut-off probabilities $\eta_{0.05}^\text{ave}$ shown in Figure \ref{fig:alphaMean_DiffUnknown}. Results are based on $10^4$ replicated trials.}
\label{tab:simUniv_unknownVar}
\end{table}

\subsection{Type-I error rate control for illustration with co-primary endpoints}

\begin{figure}[ht]
    \centering
    \includegraphics[width=0.8\textwidth]{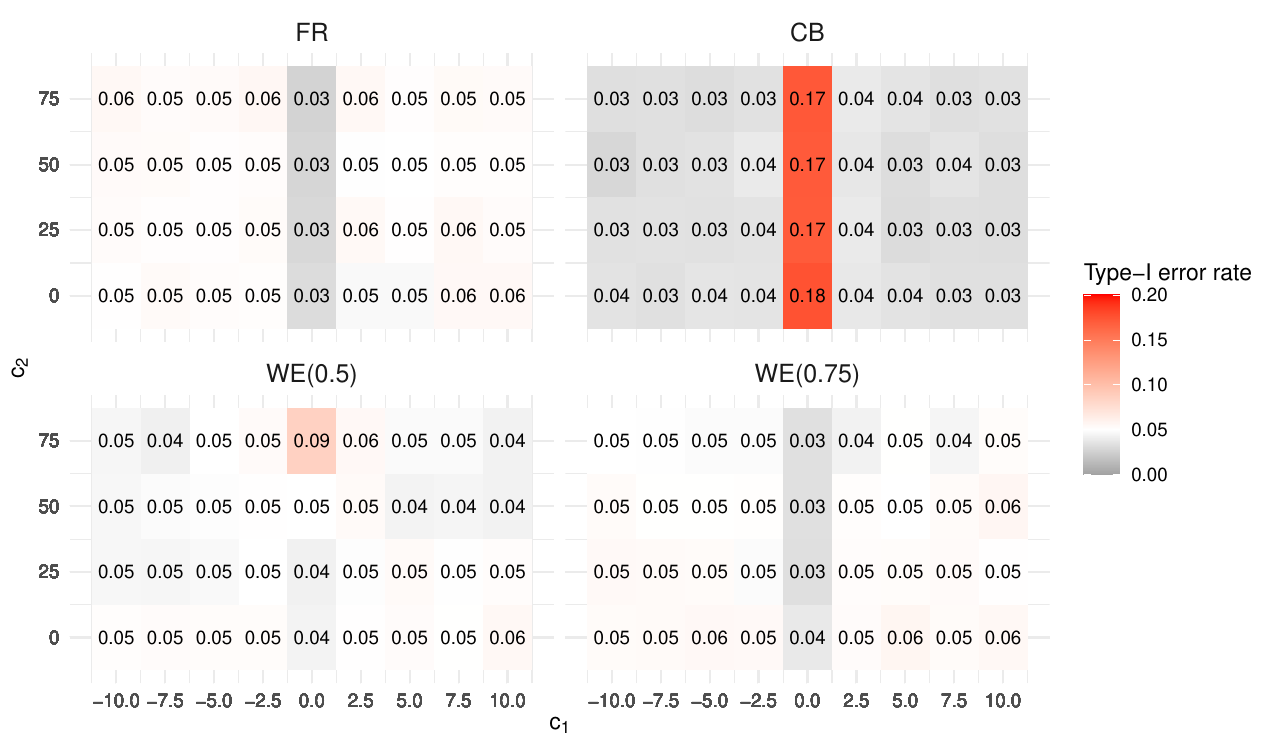}
    \vspace{-0.4cm}
    \caption{Individual type-I error rates in correspondence of several null scenarios (combinations of $c_1$ and $c_2$) for the considered designs under average control of type-I error rate at $\alpha=5\%$. Cut-off probabilities are $0.911$ (FR), $0.918$ (CB), $0.898$ (WE(0.5)),  $0.904$ (WE(0.75)).}
    \label{fig:alphaMean_bivar}
\end{figure}

Design-specific cut-off probabilities are calibrated under the set of null scenarios $$\mathcal{S}''_0=\bigl\{(\bm\mu_1^{(s)},\dots,\bm\mu_K^{(s)}):\,\bm\mu_j^{(s)}-\bm\gamma=(c_{s1},\,c_{s2})',\,\forall j=1,\dots,K,\,c_{s1}\in\mathbb{R},\,c_{s2}>0,s=1,\dots,36 \bigr\}$$ to achieve an average control of the type-I error rate at a level $\alpha=5\%$. Figure \ref{fig:alphaMean_bivar} shows the resulting individual type-I error rates across the 36 considered combinations $\bm c_s=(c_{s1},\,c_{s2})'$. 
Notably, FR and WE designs are associated with individual type-I error rates which do not exceed $6\%$, with the exception of WE(0.5) which registers a $9\%$ type-I error rate under the configuration $\bm c_s=(0,\,75)'$. FR and WE(0.75) have more conservative type-I error rates when the mean of PDM endpoint equals the target in all the treatment arms. On the contrary, there is a large inflation of type-I error rate of CB in these scenarios, while elsewhere it is below $5\%$.

\begin{figure}[ht]
    \centering
    \includegraphics[width=0.8\textwidth]{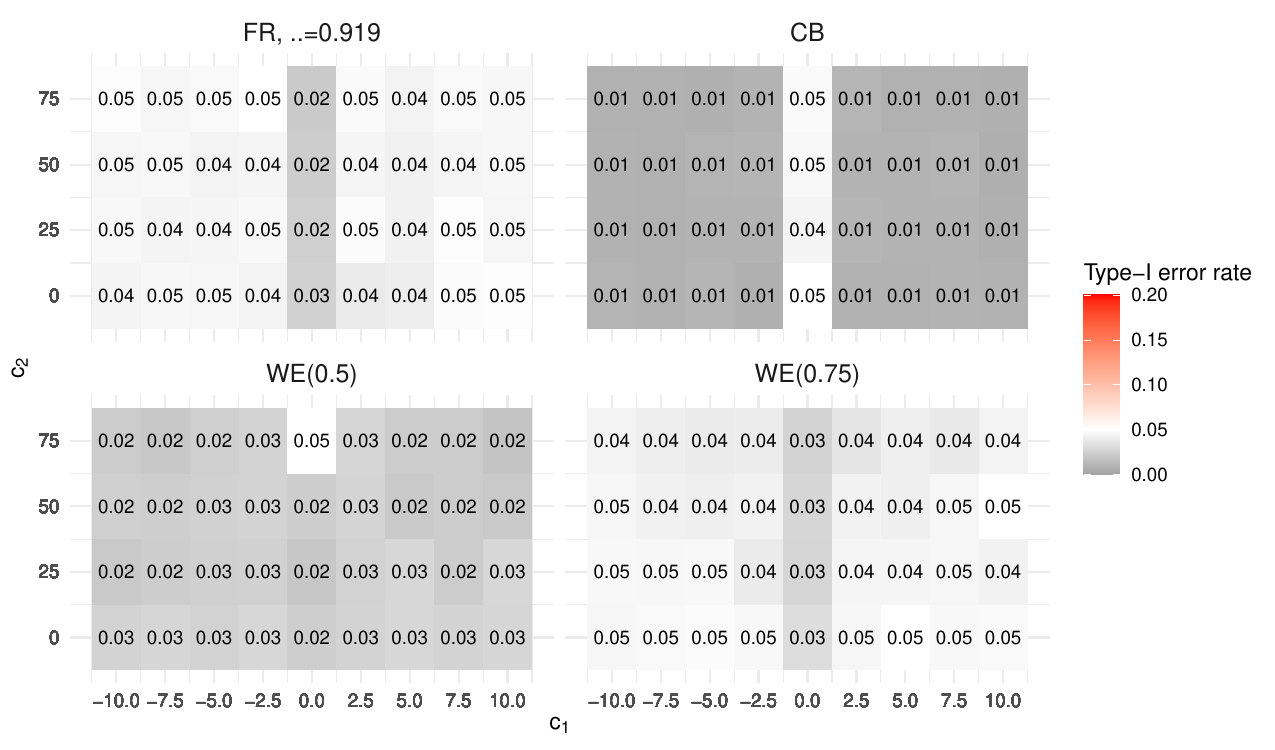}
    \caption{Individual type-I error rates in correspondence of several null scenarios (combinations of $c_1$ and $c_2$) for the considered designs under strong control of type-I error rate. Cut-off probabilities are $0.919$ (FR), $0.964$ (CB), $0.928$ (WE(0.5)),  $0.912$ (WE(0.75)).}
    \label{fig:alphaStrong_bivar}
\end{figure}

Figure \ref{fig:alphaStrong_bivar} shows the same type of analysis under strong control of type-I error rate. As expected, type-I errors do not exceed $5\%$, with scenarios-specific inflation (or deflation) as in Figure \ref{fig:alphaMean_bivar}.

\end{document}